\documentclass[apj,iop,preprint]{emulateapj}

\usepackage{times,amsmath,amsfonts,amssymb}

\newcommand{\dif}{\mbox{$\mathrm{d}$}}

\newcommand\Rsq{{\mathbb{R}^2}}

\slugcomment{To appear in ApJ}

\shorttitle{Optimal Linear Image Combination}
\shortauthors{Rowe, Hirata, \& Rhodes}

\begin{document}


\title{Optimal Linear Image Combination}


\author{Barnaby Rowe,\altaffilmark{1,2} Christopher Hirata,\altaffilmark{3} and Jason Rhodes\altaffilmark{1,2}}
\email{browe@caltech.edu; chirata@tapir.caltech.edu; jason.d.rhodes@jpl.nasa.gov}

\altaffiltext{1}{Jet Propulsion Laboratory, California Institute of Technology, 4800 Oak Grove Drive, Pasadena, CA, 91109}
\altaffiltext{2}{California Institute of Technology, 1200 E California Boulevard, Pasadena, CA, 91106}
\altaffiltext{3}{Department of Astrophysics, California Institute of Technology, M/C 350-17, 1216 E California Boulevard, Pasadena, CA, 91106}

\begin{abstract}
A simple, yet general, formalism for the optimized linear combination of astrophysical images is constructed and demonstrated.  The formalism allows the user to combine multiple undersampled images to provide oversampled output at high precision. The proposed method is general and may be used for any configuration of input pixels and point spread function; it also provides the noise covariance in the output image along with a powerful metric for describing undesired distortion to the image convolution kernel.  The method explicitly provides knowledge and control of the inevitable compromise between noise and fidelity in the output image. 

We present a first prototype implementation of the method, outlining steps taken to generate an efficient algorithm.  This implementation is then put to practical use in reconstructing fully-sampled output images using simulated, undersampled input exposures that are designed to mimic the proposed \emph{Wide Field InfraRed Survey Telescope} (\emph{WFIRST}).   We examine results using randomly rotated and dithered input images, while also assessing better-known ``ideal'' dither patterns: comparing results we illustrate the use of the method as a survey design tool.  Finally, we use the method to test the robustness of linear image combination when subject to practical realities such as missing input pixels and focal plane plate scale variations.
\end{abstract}





\section{Introduction}
Weak gravitational lensing (e.g. \citealp{schneider06}) is a powerful means by which to probe the distribution of matter in the Universe. As such, it has been identified as among the most promising methods by which to constrain the growth rate of matter structure in the Universe, and thereby to test models of evolving dark energy and modifications to general relativity (see, e.g., \citealp{albrechtetal06, albrechtetal09,peacocketal06}, and references therein). However, weak lensing is arguably the most technically challenging of the cosmological probes from an image analysis standpoint, due to its extreme sensitivity to any systematic errors made when recovering galaxy shape information.  Much work has gone into understanding how to accurately recover tiny, coherent gravitational shear signals from large numbers of noisy galaxy images that have been convolved with the instrumental Point Spread Function (PSF) \citep{heymansetal06step,masseyetal07step,bridleetal10,kitchingetal10}.

In the analysis of astronomical images, ensuring the adequate spatial sampling of data by pixels of finite size and spacing is a key concern.  Ideally, images should be sampled at or above the Nyquist-Shannon sampling rate for the \emph{band limit} set by the optical response of the system (see e.g.\ \citealp{marks09}), so that the full continuous image can be determined from the discrete pixel samples.  Throughout this Paper we will define a single plane wave to be $\propto { e}^{2\pi {\rm i}{\bf u}\cdot{\bf r}}$, so that ${\bf u}$ has units of cycles per arc second.  In general, if an image contains only Fourier modes whose spatial frequency ${\bf u}$ is no larger in magnitude than $u_{\rm max}$ then the Nyquist criterion demands sample spacing $P$ satisfying $P<1/(2u_{\rm max})$.  An image sampled more finely than the critical rate $1/(2u_{\rm max})$ is referred to as \emph{oversampled} (and one sampled at the critical rate is \emph{critically sampled}). Since it is possible to reconstruct the entire original image from oversampled data, operations such as interpolation/regridding, rotation, and translation can be carried out with no pixelization artifacts. This makes oversampled data the preferred input for most precision image analysis applications, including weak lensing.

The ideal location for weak lensing observations is a space-based telescope, where one is free from the blurring effects of the Earth's atmosphere and can achieve a level of stability of the optics and hence the PSF that is impossible from the ground. Weak lensing is thus a key project for proposed space-based imaging missions such as the \emph{Wide Field InfraRed Survey Telescope} (\emph{WFIRST}: \citealp{blandfordetal10}; the reference design is based on the \emph{Joint Dark Energy Mission/JDEM-Omega} concept: \citealt{gehrels10}) and \emph{Euclid} \citep{refregieretal10}. However, in both cases, practical design considerations prohibit oversampling at the native pixel scale of the system.  The optics of a space telescope deliver a PSF that preserves Fourier modes out to $u_{\rm max}=D/\lambda$ where $D$ is the outer diameter of the primary mirror and $\lambda$ is the wavelength of observation; the high spatial frequencies may be suppressed by e.g. charge diffusion, but in most cases $u_{\rm max}$ is still large.  Indeed, preserving the high spatial frequency components of the image is a major reason to choose a space-based platform. If one is to oversample the image at the native pixel scale, then, one is forced to choose very small pixels. However, there are competing considerations that drive one to larger pixels, including (i) the desire for large field of view within engineering or cost constraints on the number of detectors; and (ii) the high read noise of near-infrared detectors, which results in increased photometric errors as light from an object is spread over more pixels. For these reasons, both {\emph{JDEM-Omega} and \emph{Euclid} chose pixels larger than $1/(2u_{\rm max})$. Images from such systems, sampled at the native pixel scale, are \emph{undersampled}.  Undersampled images suffer from aliasing, in which each Fourier mode of the observed image contains contributions from several Fourier modes of the original image: in this case the original image cannot be unambiguously reconstructed. Operations such as interpolation, translation by a non-integer number of pixels, etc. are not mathematically possible on undersampled data, which represents a major difficulty for image analysis.

Although proposed space-based dark energy missions are unlikely to oversample their imaging data on a given single image, typical design concepts allow accurate pointing capability (e.g.\ \citealp{gehrels10}).  What is required is thus a means of reconstructing oversampled images from two or more undersampled, dithered images.  The simplest technique to improve sampling is the ideal subpixel dither: for example, one may take 4 exposures whose relative offsets are $(0,0)$, $(\frac12,0)$, $(0,\frac12)$, and $(\frac12,\frac12)$ pixels (a ``$2\times 2$'' dither pattern). With the use of these multiple exposures, where the observed image depends on different linear combinations of the aliased modes, one may use linear algebra techniques to separately ``solve out'' the aliased Fourier modes \citep{lauer99, lauer99psf} so long as the native pixel scale $P<1/u_{\rm max}$. The linear algebra techniques allow us to analyze any subpixel dither pattern: notable examples include the ``$3\times 3$'' pattern (9 exposures); subpixel dithers with diagonal grids such as the ``$\sqrt2\times\sqrt2$'' pattern [where the relative offsets are $(0,0)$ and $(\frac12,\frac12)$ pixels] or the ``$\sqrt5\times\sqrt5$'' pattern (see Section \ref{sect:rt5xrt5}); or $N$ exposures with arbitrary offsets. Given any such subpixel dither pattern, one can determine whether the linear system for the Fourier modes of the input image can be solved. The output image from this approach has the same PSF as the input images.

However, it is likely that \emph{WFIRST} -- or any wide-angle survey that attempts to fill in gaps between detectors -- will end up with a survey pattern that contains large slews (of order a detector width) between each exposure.  Under such circumstances, the different input images that are being used 
to build the final, oversampled output image will be sampled on different grids whose phase cannot be controlled, will have different field 
distortions, and will have ``holes'' due to cosmic rays and detector effects.  Moreover, the input PSFs will be 
different even if the telescope optics are perfectly stable, because each observation of a galaxy will be made in a different part of the field.  The 
Fourier space/linear algebra techniques are not well-equipped to handle such situations efficiently.   There are stable, commonly-used image combination algorithms such as \textsc{Drizzle} 
\citep{fruchterhook02} with mature accompanying scripts (e.g.\ \citealp{koekemoeretal02}) that will produce an output image for such inputs.  However, these algorithms do not give full control of the PSF in the output image, or even guarantee that this PSF is constant across the image of a galaxy\footnote{Special cases such as the $2\times 2$ ideal subpixel dither with identical input PSFs may be exceptions because of special symmetries, e.g. translation by half-pixels. However, our interest is in the combination of images with no such symmetries.}, which should be seen as a prerequisite for high-precision shape measurement for a space weak lensing mission.  Recently \citet{fruchter11} proposed a new method (\textsc{iDrizzle}: iterative \textsc{Drizzle}) that tackles some of these issues using an iterative application of \textsc{Drizzle} commands.  It will be interesting in the future to compare results using this algorithm to those of this Paper.

A third approach to combining images would be to adapt methods from Cosmic Microwave Background (CMB) experiments.  CMB experiments typically have only a few pixels in their focal plane (the \emph{Wilkinson Microwave Anisotropy Probe/WMAP} had a total of 10 differencing assemblies: e.g.\ \citealp{jarosiketal11}, and references therein) and use a large number of scans across the sky to build up an output map.  Mapmaking for CMB experiments is usually done by taking the vector ${\bf d}$ of time-ordered data from a small number of detectors, and writing it in the form:
\begin{equation}
{\bf d} = {\bf Ms}+{\bf \eta},
\label{eq:d}
\end{equation}
where ${\bf s}$ is a vector of temperatures in pixels on the sky, ${\bf \eta}$ is the noise, and ${\bf M}$ is the $n_d\times n_p$ mapping matrix  
($n_d$ being the number of samples of data and $n_p$  the number of pixels in the output map).  In its simplest form ${\bf M}$ is a sparse matrix 
containing $M_{ij}=1$ if the $i$th sample is collected when looking at the $j$th pixel, and 0 otherwise, but many generalizations are possible (e.g.\ 
for differential experiments, methods that mask bright or variable sources so they do not contaminate the data processing, etc. see \citealp{hinshawetal07}).  Taking the known noise covariance matrix ${\bf N}=\langle {\bf \eta \eta}^{\rm 
T}\rangle$, one may obtain a least-squares solution for the sky map ${\bf s}$: this is $({\bf M}^{\rm T}{\bf N}^{-1}{\bf M})^{-1}{\bf M}^{\rm T}{\bf N}^{-1}{\bf d}$.

However, this method (as written here and in most of the literature) does not trivially incorporate a variable PSF (beam), requires the output ${\bf s}$ to have pixels {\em much} smaller than the beam, and then requires at least several samples per output pixel -- this is a luxury that is not present in optical astronomy with sub-arcsec PSF sizes.  This problem can be solved by allowing ${\bf M}$ to contain interpolation coefficients, but an implementation of this approach is not yet public, and is typically unnecessary for the arcminute beam sizes commonly seen in CMB experiments.

In this Paper, we take the first steps toward developing a general linear algorithm that attempts to combine the best features of these approaches.  As in the Fourier/linear algebra approach \citep{lauer99,lauer99psf}, we desire a well-controlled final PSF.  As is the case for \textsc{Drizzle} \citep{fruchterhook02}, we want to be able to use arbitrarily placed samples if possible.  Sometimes, reconstructing a fully-sampled image with a controlled PSF is impossible given the positions of the samples provided; unlike \textsc{Drizzle}, we desire a method that alerts the user when this happens.  Finally, as in the CMB approach, we aim to do something that minimizes output noise using some variant of the least-squares method.

In Section \ref{sect:linform} we derive a linear formalism that can meet these aims, and in Section \ref{sect:imp} describe a prototype implementation of the method called \textsc{Imcom}, which is available freely from the authors on request.  In Section \ref{sect:example} we illustrate the process of combining multiple images using \textsc{Imcom} via a detailed worked example, before turning in Section \ref{sect:wfirst} to examine a set of realistic potential observing scenarios for a space-based dark energy mission such as \emph{WFIRST}.  While it should be stressed that the dither scenarios and PSF patterns used in these tests do not necessarily reflect any firm plans for the \emph{WFIRST} mission, they are designed to approximate such a mission in a broad sense.  In Section \ref{sect:resources} we discuss the computing resources necessary to process imaging data from a dark energy mission using the \textsc{Imcom} technique, before drawing conclusions in Section \ref{sect:conc}.




\section{Linear Formalism}\label{sect:linform}


The problem is to combine several undersampled input images into a single oversampled image.  The input images may be written as a vector of intensities 
$I_i$ of length $n$, where $n$ is the total number of usable pixels.  For example, if the inputs are four $128 \times 128$ pixel images with no defects, then $n=4(128)^2$.  The 
inputs are sampled at their pixel centers $\{{\bf r}_i\}_{i=1}^n$.  We also suppose that the PSF (including jitter, optics, and detector response) at separation ${\bf 
s}$ is $G_i({\bf s})$.  That is, the intensities are given by
\begin{equation}
I_i = \int_\Rsq f({\bf r}') G_i({\bf r}_i-{\bf r}') \dif^2{\bf r}' + \eta_i,
\label{eq:I}
\end{equation}
where $\eta_i$ is the noise with $\langle\eta_i\rangle=0$ and some covariance matrix $N_{ij}=\langle\eta_i\eta_j\rangle$. The formalism can rigoroursly treat any general noise covariance matrix $N_{ij}$, although in most cases $N_{ij}$ is close to diagonal (with small off-diagonal correlations due to, e.g., corrections for inter-pixel capacitance, electronic $1/f$ or flicker noise, etc.).  The function $f({\bf r}')$ describes the physical image on the sky.

We note here that the PSF $G_i({\bf r})$ is assumed as being fully specified, in advance, at the position centered on each input pixel $\textbf{r}_i$, in each contributing exposure individually.  This knowledge must be expected to come from either a well-motivated optical model or the image data $I_i$, or both.  Of course, images of point sources (stars) within $I_i$ will be subject to the same degree of undersampling as the rest of the $I_i$ pixels.  This issue can be countered if adopting a fully parametric model and fitting for $G_i({\bf r})$ from stellar images (e.g.\ \citealp{maetal08}).  However, the use of a non-parametric model that potentially contains defects due to aliasing will undermine the success of any image combination.  The problem of recovering $G_i({\bf r})$ from undersampled stellar images is discussed further in Section \ref{sect:conc}.

The output is to be a synthesized, fully sampled image $H_\alpha$ on a grid of pixel centers ${\bf R}_\alpha$, where $\alpha=1, \dotsc, m$.  For example, if the synthesized image is to be 256$\times$256 then $m=256^2$.  We wish for $H_\alpha$ to be as close as possible to the \emph{target image} $J_{\alpha}$, which is defined as:
\begin{equation}\label{eq:Ja}
J_\alpha \equiv \int_\Rsq f({\bf r}') \Gamma({\bf R}_\alpha-{\bf r}') \dif^2{\bf r}'.
\end{equation}
Here $\Gamma$ is the desired effective PSF of the synthesized image, a free choice for the user.  A well-motivated choice for $\Gamma$ might be $G_i({\bf r})$ itself (if it is approximately constant), preserving the input PSF as much as possible.   This is the approach taken in later Sections, but in general $\Gamma$ may be freely chosen to additionally filter the output image as desired (see, e.g., Section \ref{sect:scalev}).

\subsection{Notation}
Fourier transform pairs are written with Fourier space in units of cycles per arcsecond rather than radians per arcsecond.  That is,
\begin{equation}
\tilde{f}({\bf u})  =   \mathcal{F} \left[ f({\bf r}) \right] ({\bf u}) = \int_\Rsq f({\bf r}) e^{-2\pi \textrm{i}{\bf u}\cdot{\bf r}} \dif^2{\bf r}
\end{equation}
for the forward Fourier transform, and
\begin{equation}
f({\bf r})  =  \mathcal{F}^{-1} \left[ \tilde{f}({\bf u}) \right] ({\bf r}) = \int_\Rsq \tilde{f}({\bf u}) e^{2\pi \textrm{i}{\bf u}\cdot{\bf r}} \dif^2{\bf u}
\end{equation}
for the inverse Fourier transform.

The Einstein summation convention is {\em not} used: all summations over indices will be either be written explicitly or, for brevity, as implied matrix multiplications upon matrix and vector objects in bold face.  Latin indices are used for input vectors and in derived objects where reference is made to input pixel locations, and Greek indices are similarly used to refer to output pixel locations.

\subsection{Linear Solutions}
We have defined the fully-sampled output $H_\alpha$ in relation to the target image $J_{\alpha}$ of equation \eqref{eq:Ja}, to which we intend $H_{\alpha}$ to be as close an approximation as possible.  We then desire $H_{\alpha}$ to be linearly related to the input images $I_i$ for several reasons.  Most importantly, it ensures that the final PSF in $H_\alpha$ is independent of image brightness, so the stars that were not used in the PSF determination are tracers of how well the image synthesis worked.  Linearity simplifies the analysis of 
systematics.  Furthermore, it dramatically reduces the space of possible algorithms and allows a systematic search of the space that remains.  We further note that 
some pre-existing algorithms (e.g. \textsc{Drizzle}: \citealp{fruchterhook02}) are also linear and hence are included in our search space.
It should be made clear, however, that detectors usually display some level of non-linear behavior (e.g.\ non-linear gain), changing the PSF for bright objects.  Corrections for this non-linearity are typically applied in image pre-processing at the raw pixel level, and we assume that any such non-linearity will have been duly corrected in $I_i$ before starting to combine images.

A general linear mapping from $I\rightarrow H$ is:
\begin{equation}
H_\alpha = \sum_i T_{\alpha i} I_i,
\end{equation}
where $T_{\alpha i}$ is an $m\times n$ matrix.  We now need to construct an appropriate objective function for the performance of the matrix $T_{\alpha i}$ and extremize it to find the optimal linear transformation we desire.

\subsection{Objective Function} \label{sect:obj}

We note that for a matrix $T_{\alpha i}$, an error map can be constructed:
\begin{eqnarray}\label{eq:R}
Z_\alpha & \equiv & H_\alpha-J_\alpha \nonumber \\ 
& = &  \int_\Rsq  f({\bf r}') L_{\alpha} ({\bf R}_{\alpha} - {\bf r'}) \dif^2 {\bf r'}   +   \sum_i T_{\alpha i} \eta_i \nonumber
\end{eqnarray}
where we have defined the leakage function $L_{\alpha}$ as
\begin{equation}\label{eq:L}
L_{\alpha} ({\bf R}_{\alpha} - {\bf r'}) \equiv  \sum_i T_{\alpha i} G_i({\bf r}_i-{\bf r}') - \Gamma({\bf R}_\alpha-{\bf r}').
\end{equation}
The leakage is simply the difference between the desired PSF $\Gamma$ and its actual reconstructed counterpart in $H_{\alpha}$.  We will seek to minimize this difference.
Taking equation \eqref{eq:R}, we may then separate out the deterministic part,
\begin{equation}
\langle Z_\alpha\rangle = \int_\Rsq f({\bf r}') L_\alpha({\bf r}' - {\bf R}_\alpha) \dif^2{\bf r}',
\end{equation}
from the stochastic contribution to the residuals as specified by the noise covariance,
\begin{equation}
\Sigma_{\alpha\beta} = {\rm Cov}(Z_\alpha,Z_\beta) = \sum_{ij} T_{\alpha i}T_{\beta j}N_{ij}.
\label{eq:Cr}
\end{equation}
We will also seek to minimize the noise variance $\Sigma_{\alpha \alpha}$ for each output pixel value $H_{\alpha}$.
Noisy output images will be those that combine input pixels from $I_i$ with large positive and negative weights in $T_{\alpha i}$.  Such solutions will tend to be unstable, which gives a further practical reason for wishing to minimize $\Sigma_{\alpha \alpha}$ where possible,

This now allows the construction of an objective function $W$, i.e.\ a function to be minimized in solving for $T_{\alpha i}$.  We will choose a function that is a simple sum over values of $\alpha$, so that if we wish to we may write $W=\sum_\alpha 
W_\alpha$, where $W_\alpha$ depends only on $L_\alpha({\bf s})$ and $\Sigma_{\alpha\alpha}$. In other words, we choose to optimize each output pixel independently: one minimizes $W$ by minimizing each $W_\alpha$. 

This choice is made both for computational simplicity, and also so that the 
optimization 
does not mangle good regions of the synthesized image to try to clean up bad regions (e.g.\ the edges of frames or the occasional intersection of multiple cosmic ray 
tracks).  We desire for $W_\alpha$ to be a quadratic function, so that linear algebra methods can be used to solve for $T_{\alpha i}$, and for it to be 
translation-invariant (i.e. we care only about the normalization and shape of the leakage function $L_\alpha$, not about its position on the sky). The most general such function is
\begin{equation}
W_\alpha = U_{\alpha} + \kappa \Sigma_{\alpha\alpha},
\label{eq:Wa}
\end{equation}
where $\kappa>0$ is a Lagrange multiplier, and we have defined the \emph{leakage objective} $U_{\alpha}$ as
\begin{equation}\label{eq:Ulong}
U_{\alpha} = \int_\Rsq  \! \! \int_\Rsq  \Upsilon({\bf s}'-{\bf s}) L_\alpha({\bf s}) L_\alpha({\bf s}') \dif^2{\bf s} \, \dif^2{\bf s}' ,
\end{equation}
where $\Upsilon({\bf s})$ is a real symmetric positive-definite kernel, i.e. $\Upsilon(-{\bf s})=\Upsilon({\bf s})$, $\Im\Upsilon({\bf s})=0$, 
and $\Upsilon({\bf s})>0$.  We note that for the simplest choice of $\Upsilon({\bf s})$, the Dirac delta function $\delta({\bf s})$, this expression becomes
\begin{equation}
U_{\alpha} = \int_\Rsq  \left[L_{\alpha} ({\bf s})\right]^2 \dif^2 {\bf s}.
\end{equation}
The leakage objective $U_{\alpha}$ is thus, in the case $\Upsilon ({\bf s})= \delta({\bf s})$, a very simple, direct, quadratic measure of the local leakage at the point ${\bf R}_{\alpha}$.  Returning to the general case, equation \eqref{eq:Ulong} may also be written conveniently in Fourier space:
\begin{equation}
U_\alpha = \int_\Rsq \tilde\Upsilon({\bf u}) |\tilde L_\alpha({\bf u})|^2 d^2{\bf u}.
\label{eq:Ua-fourier}
\end{equation}
It will be seen this expression for $U_{\alpha}$ is computationally convenient.

The role of the Lagrange multiplier $\kappa$ in equation \eqref{eq:Wa} merits some discussion here.  It can be seen that $\kappa$ may be used to weight the relative importance of the two terms contributing to the objective function $W_\alpha$.  These relate to two key properties of the output image:
\begin{itemize}
\item The \emph{fidelity} of the ensemble mean $H_{\alpha}$ to the target image of equation \eqref{eq:Ja}, on which $U_{\alpha}$ depends.
\item The noise variance in the output image, $\Sigma_{\alpha \alpha}$.
\end{itemize}
To ensure fidelity we naturally wish to emphasize the minimization of $U_{\alpha}$ relative to $\Sigma_{\alpha \alpha}$, implying $\kappa \ll 1$, and vice versa if we wish to suppress noise.  Thus the minimization of $W_\alpha$ to solve for $T_{\alpha i}$, subject to the freely-chosen Lagrange multiplier constraint $\kappa$, gives the user control of the balance between image fidelity and noise at each output pixel $H_{\alpha}$.

\subsection{Minimization of the Objective Function}\label{sect:minobs}
We want to minimize $W_\alpha$ over the space of $T_{\alpha i}$.  To do this, we first take the Fourier transform of equation \eqref{eq:L}:
\begin{eqnarray}
\tilde L_\alpha({\bf u}) & = & \left[
\sum_i T_{\alpha i} e^{-2\pi \textrm{i}{\bf u}\cdot{\bf r}_i} \tilde G_i^\ast({\bf u})
  - e^{-2\pi \textrm{i}{\bf u}\cdot{\bf R}_\alpha} \tilde \Gamma^\ast({\bf u}) \right] \nonumber \\
 & \times & e^{2\pi \textrm{i}{\bf u}\cdot {\bf R}_\alpha} 
\label{eq:tL}
\end{eqnarray}
The leakage objective is then
\begin{eqnarray}
U_\alpha & = & \int_\Rsq \left| \sum_i T_{\alpha i} e^{-2\pi \textrm{i}{\bf u}\cdot{\bf r}_i} \tilde G_i^\ast({\bf u})  - e^{-2\pi \textrm{i}{\bf u}\cdot{\bf R}_\alpha} \tilde \Gamma^\ast({\bf u}) \right|^2  \nonumber \\
& \times & \tilde{\Upsilon}({\bf u})  \dif^2{\bf u} .
\label{eq:obj1}
\end{eqnarray}
Recalling that $T_{\alpha i}$ is real, and using equations \eqref{eq:Cr} \& \eqref{eq:Wa}, this expands to give the combined objective function:
\begin{equation}
W_\alpha = \sum_{ij} \left( A_{\alpha ij} + \kappa N_{ij} \right) T_{\alpha i} T_{\alpha j} + \sum_i B_{\alpha i}T_{\alpha i} + C_\alpha,
\label{eq:obj2}
\end{equation}
where the coefficients are given by the system matrices:
\begin{eqnarray}
A_{\alpha ij} &=& \int_\Rsq \tilde\Upsilon({\bf u}) e^{2\pi \textrm{i}{\bf u}\cdot({\bf r}_j-{\bf r}_i)} \tilde G_i^\ast({\bf u}) \tilde G_j({\bf u}) d^2{\bf u}, \label{eq:Aaij} \\
B_{\alpha i} &=& -2 \int_\Rsq \tilde\Upsilon({\bf u})
  e^{2\pi \textrm{i}{\bf u}\cdot({\bf r}_i-{\bf R}_\alpha)} \tilde G_i({\bf u}) \tilde\Gamma^\ast({\bf u}) d^2{\bf u}, \label{eq:Bai} \\
C_\alpha &=& \int \tilde\Upsilon({\bf u}) |\tilde\Gamma({\bf u})|^2 d^2{\bf u}. \label{eq:Ca}
\end{eqnarray}
Here $A_{\alpha i j}$ is an $n \times n$, positive definite symmetric real matrix, $B_{\alpha i}$ is a real matrix of dimension $m \times n$, and $C_\alpha$ is a positive real scalar. (The integrals can be seen to be real since the integrands in $A_{\alpha ij}$, $B_{\alpha i}$, and $C_\alpha$ transform to their complex conjugates under change of the integration variable ${\bf u}\rightarrow-{\bf u}$.)  We note that direct expansion of equation \eqref{eq:obj1} would have lead to a Hermitian matrix; however, 
since $T_{\alpha i}$ is real the contributions to $U_\alpha$ from the imaginary, antisymmetric part vanish.

Any function written in the form of equation \eqref{eq:obj2} can be immediately be minimized to give
\begin{equation}
T_{\alpha i} = -\frac12\sum_j \left[ \left( {\bf A}_\alpha + \kappa {\bf N} \right)^{-1} \right]_{ij} B_{\alpha j},
\label{eq:tai}
\end{equation}
where $\left({\bf A}_\alpha + \kappa {\bf N} \right)^{-1}$ is the $n\times n$ matrix inverse of $A_{\alpha ij} + \kappa N_{ij}$.  In practice it will be necessary to restrict the synthesized image pixel $\alpha$ to only depend on nearby input pixels $i$ so as to reduce the dimensionality of the 
matrix $ A_{\alpha i j}$.  If this is done, however, $A_{\alpha i j}$ actually depends on $\alpha$ which is not true if the whole image is used.  
Since the whole image may have $n \sim 10^7$, this reduction of dimensionality is required in order to make the approach computationally feasible. These issues are discussed further in Section \ref{sect:imp}.

We note that equation \eqref{eq:tai} is a general result and can be used in principle for any system of non-ideal dithers, rotated images, and even images with defects (the ``lost'' pixels are simply removed from the system matrices).  Some of these cases will be explored in Sections \ref{sect:example} \& \ref{sect:wfirst}.

The noise variance in each output image pixel is given by the diagonal terms in the full covariance of equation \eqref{eq:Cr}.
As discussed in Section \ref{sect:obj}, the noise properties can be adjusted to using the Lagrange multiplier $\kappa$.  In Appendix \ref{app:asym} we derive in detail the limiting behavior of output images across the range of $\kappa$ values.  These results make it clear that, in setting $\kappa$, some compromise is desired between the limiting cases of maximum fidelity and minimum noise.
In some cases, even if $\kappa$ is taken to zero (i.e.\ the optimization is sensitive only to mean output image fidelity and not to noise) the leakage function will not approach zero.
In these cases, reconstructing an unbiased synthesized image is {\em not mathematically possible} using a purely linear combination of the input pixels, and $\lim_{\kappa\rightarrow0^+}U_\alpha$ represents a fundamental limitation.

The choice of $\kappa$ may be made globally or from pixel to pixel.  Setting a single value of $\kappa$ is certainly the simplest approach, but it is advantageous to allow the user to specify the properties of their input image in terms of key deliverables such as the noise variance $\Sigma_{\alpha \alpha}$ or leakage objective $U_\alpha$.
One strategy for image combination would therefore be to regard $\Sigma_{\alpha\alpha}$ as fixed (using this to set the value of 
$\kappa$ from pixel to pixel) and then obtain the highest-fidelity mean image consistent with this noise value.   The opposing strategy would be to set a tolerance on the image fidelity by requiring $U_{\alpha}$ to be less than some threshold value, and then finding the least noisy reconstruction consistent with this threshold.  We now consider these strategies and give more detail about practical issues in the implementation of the algorithm.

\section{Implementation}\label{sect:imp}

In this Section  we describe a prototype implementation of the image combination method presented in Section \ref{sect:linform}, describing some of the numerical techniques employed and issues raised in finding solutions for $T_{\alpha i}$.  We have named this prototype package {\sc Imcom} (IMage COMbination), and plan to make it available to the public either in its current version or as part of a more developed software package at a later date.  The code is written in {\sc Fortran 95}, supports multi-threading via {\sc OpenMP} (see, e.g., \citealp*{chapman07}), and uses only public software libraries, some of which are available in specific hardware-optimized versions: CFITSIO \citep{pence99}, FFTW3 \citep{frigojohnson05}, BLAS \citep{blackfordetal02} \& LAPACK \citep{andersonetal99}.

\subsection{{\sc Imcom} Program Outline and Inputs}\label{sect:outline}

The {\sc Imcom} code solves for $T_{\alpha i}$ by varying $\kappa$ on an output pixel-by-output pixel basis as described in Section \ref{sect:minobs}, so that $\kappa = \kappa_{\alpha}$.   The user may specify a maximum tolerance value of either the noise variance or leakage objective: these we label $\Sigma_{\alpha \alpha}^{\textrm{max}}$ and $U_{\alpha}^{\textrm{max}}$, respectively.  The algorithm may then be used to solve for $\kappa_\alpha$ so that $\Sigma_{\alpha \alpha} \le \Sigma_{\alpha \alpha}^{\textrm{max}}$ or
$U_{\alpha} \le U_{\alpha}^{\textrm{max}}$, while simultaneously minimizing the corresponding $U_\alpha$ or $\Sigma_{\alpha \alpha}$, respectively.  This procedure leaves the user free to specify an appropriate compromise between noise and fidelity for the application in question. 

In practice it is desirable to have an iterative solution stop within finite time, and this can be achieved by specifying an acceptable range for the solutions, $\Delta \Sigma_{\alpha \alpha}^{\textrm{max}}$ and $\Delta U_{\alpha}^{\textrm{max}}$.  The algorithm can then be instructed to finish iterating once {$\Sigma_{\alpha \alpha}^{\textrm{max}} - \Delta \Sigma_{\alpha \alpha}^{\textrm{max}} < \Sigma_{\alpha \alpha} \le \Sigma_{\alpha \alpha}^{\textrm{max}}$} or $U_{\alpha}^{\textrm{max}} - \Delta U_{\alpha}^{\textrm{max}} < U_\alpha \le U_{\alpha}^{\textrm{max}}$.  This then allows total control over either the noise or fidelity of the output.

Apart from these tolerances on the properties of the output image, there are other choices that the user needs to make before the problem is fully specified.  These we refer to as ``soft inputs'', since the user has some freedom in the choice of their functional forms or values:
\begin{enumerate}
\item[i)] The synthetic convolution kernel $\Gamma({\bf r})$ of the desired output image.
\item[ii)] The desired output sampling locations ${\bf R}_{\alpha}$.
\item[iii)] A choice of objective kernel $\Upsilon ({\bf r})$ for evaluating $U_{\alpha}$. The simplest choice, made throughout this paper, is a Dirac delta function so that $\tilde{\Upsilon}({\bf u}) = 1$.
\item[iv)] The total range of search for solutions that match the input tolerances on leakage or noise: $\kappa_{\textrm{min}} \le \kappa_{\alpha} \le \kappa_{\textrm{max}} $.
\item[v)] The maximum allowed number of interval bisections $n_{\textrm{bis}}$ of this $\kappa$ range in the search for solutions.
\end{enumerate}
There are additional, low-level, implementation-specific parameter choices that affect the degree of numerical approximation made when calculating $A_{\alpha i j}$ and $B_{\alpha i}$.  These are described in Section \ref{sect:randsys}, and the values used are tabulated (along with fixed choices of the soft inputs) in Table \ref{tab:params}.

The algorithm also requires input values and functions which are \emph{not} chosen, and are determined by the images in question.  These ``hard inputs'' are assumed to be fully known in advance, and are simply:
\begin{enumerate}
\item[i)] The input images arranged into a continuous vector $I_i$ and corresponding input pixel center positions ${\bf r}_i$.
\item[ii)] The PSF at the position of each pixel center $G_i({\bf r})$.
\item[iii)] The noise covariance of the image pixels $N_{ij}$.
\end{enumerate}
Once all of these inputs are given the problem is fully specified, and the algorithm can begin constructing the optimal linear transformation $T_{\alpha i}$.

\subsection{System Matrices from the Fast Fourier Transform}\label{sect:sysfft}
The first significant computational task in the algorithm is the calculation of the system matrices $A_{\alpha i j}$ and $B_{\alpha i}$, and the scalar $C_\alpha$, given in equations \eqref{eq:Aaij}-\eqref{eq:Ca}.   We assume that $G_i({\bf r})$ and $\Gamma({\bf r})$ will be submitted to the software as fully-sampled, discrete images, although if some sufficiently accurate analytic form is known this could also be used instead.

A numerical approximation to the integrals of equations \eqref{eq:Aaij}-\eqref{eq:Ca} might then be calculated directly for each $\alpha,i,j$ element in turn. This approach begins by constructing discrete representations of $\tilde{G}_i({\bf u})$ and $\tilde{\Gamma} ({\bf u})$ using the Discrete Fourier transform (DFT: see, e.g., \citealp{pressetal92}) of the input $G_i({\bf r})$ and $\Gamma({\bf r})$, then using these functions to take a direct sum over the integrand.   However, a DFT of finite size in truth represents each PSF image as a periodic function in real space, which can causes erroneous successive maxima in the system matrices wherever the combination $({\bf r}_j - {\bf r}_i)$ exceeds the dimensions of the original PSF image; the same is true for $({\bf r}_i - {\bf R}_\alpha)$ when calculating $B_{\alpha i}$.  This problem can be overcome by zero-padding $G_i({\bf r})$ and $\Gamma({\bf r})$ to match the spatial extent of the input ${\bf r}_i$ and output ${\bf R}_\alpha$, and then taking the DFT.  The Fast Fourier Transform (FFT: e.g., \citealp{pressetal92}) algorithm performs this task highly efficiently.

Yet this schema becomes prohibitively slow in the case of large, well-sampled PSF images.  If we assume that the PSF images are better sampled than our input $I_i$ by a factor $n_{\textrm{PSF}}$ along each dimension, then the FFT of the zero-padded PSF images requires $O\left[n^2_{\textrm{PSF} }n \log{\left(n^2_{\textrm{PSF}} n \right)} \right]$ calculations.  That cost may be borne, but going on to calculate $A_{\alpha i j}$ then requires a further $O \left(n^2_{\textrm{PSF}} n^3 \right)$ calculations in total after each element is calculated in turn.  This cost is high: for a single, small, $100 \times 100$ input image we have $n=10^4$, and if we assume that $n_{\textrm{PSF}} = 3$ we arrive at an estimated $\sim 10^{13}$ floating point operations required for the calculation of $A_{\alpha i j}$.

Another approach is possible.  Inspection of equations \eqref{eq:Aaij} \& \eqref{eq:Bai} shows that we may write expressions for $A_{\alpha i j}$ and $B_{\alpha i}$ directly in terms of inverse Fourier transform operations, as follows:
\begin{equation}
A_{\alpha i j}  = \mathcal{F}^{-1} \left[ \tilde G_i^\ast({\bf u}) \tilde G_j({\bf u}) \right] \left( {\bf r}_j - {\bf r}_i \right), \label{eq:aft}
\end{equation}
and
\begin{equation}
B_{\alpha i}  =  -2 \mathcal{F}^{-1} \left[ \tilde G_i ({\bf u}) \tilde \Gamma^* ({\bf u}) \right] \left( {\bf r}_i - {\bf R}_{\alpha} \right)  . \label{eq:bft}
\end{equation}
The determination of $A_{\alpha i j}$ and $B_{\alpha i}$ thus proceeds by first constructing high-resolution lookup tables for these quantities, using the inverse DFT of the $\tilde G_i^\ast \tilde G_j $ and $\tilde G_i \tilde \Gamma^*$ arrays.  As additional zero-padding may be added to these complex functions in Fourier space, the lookup tables may be sampled at essentially arbitrary spatial resolution (physical memory constraints notwithstanding) following the inverse DFT back into real space.  The value of each element in $A_{\alpha i j}$ and $B_{\alpha i}$ can then be determined by high-order polynomial interpolation (e.g.\ \citealp{pressetal92}) of the lookup tables to the locations ${\bf r}_j -  {\bf r}_i$ and ${\bf r}_i - {\bf R}_{\alpha}$, respectively.

Assuming once more that the PSF images are zero-padded to cover the spatial extent of the input images, the lookup tables will each require $O\left[n^2_{\textrm{PSF} } n \log\left( n^2_{\textrm{PSF}} n \right) \right]$ calculations using the FFT.  A polynomial interpolation in two dimensions across regularly-gridded data is an operation of complexity $O(2n^2_{\textrm{poly}})$, where $n_{\textrm{poly}}$ is the degree of the interpolating polynomial.  Therefore, the determination of all elements of $A_{\alpha i j}$ and $B_{\alpha i}$ via lookup table requires $O(2  n^2_{\textrm{poly}} n^2)$ and $O(2 n^2_{\textrm{poly}}   n m)$ calculations, respectively.\footnote{The authors note that the calculations presented in this paper were in fact made using a polynomial interpolation schema that allows arbitrarily-spaced points along the abscissa, but at a greater relative cost of $O(n^3_{\textrm{poly}})$ operations per interpolation.}  This process will take longer than the time required to construct the lookup tables themselves, but the order $n$ reduction in complexity seen over the direct-summation approach is a significant improvement.  It is this technique that is used in the {\sc Imcom} prototype; it is noted that both approaches are trivially parallelizable, which is typically the case for calculations made throughout the method.

\subsection{Eigendecomposition of the {\bf A}$_{\alpha}$ Matrix}\label{sect:eigen}
After the calculation of the system matrices of equations \eqref{eq:Aaij}-\eqref{eq:Ca}, the next computationally challenging task is the solution of equation \eqref{eq:tai} to find $T_{\alpha i}$.  This equation may be rewritten as
\begin{equation}\label{eq:Tai}
{\bf T}_{\alpha}  \left({\bf A}_\alpha +\kappa {\bf N} \right) = -\frac{1}{2} {\bf B}_{\alpha} 
\end{equation}
in matrix form; in the case of a single, fixed $\kappa$ for all $\alpha$, this system is simply an $m$-element ensemble of transposed ${\bf M} {\bf x} = {\bf b}$ matrix-vector equations.  Here $M_{ij} = A_{\alpha ij} + \kappa N_{ij}$ will be real, symmetric and positive definite, and each solution (corresponding to a row of $T_{\alpha i}$) can be found efficiently and in parallel using the Cholesky decomposition ${\bf M} = {\bf L L}^T$ \citep{pressetal92,andersonetal99}.  This decomposition need be performed only once for constant $\kappa$; subsequently, only the back-substitution step need then be repeated $m$ times, once for each row of $-B_{\alpha i} /2$.
The full solution for constant $\kappa$ therefore requires $O(n^3 + n^2m)$ calculations.

However, as stated in Section \ref{sect:outline} the goal of this implementation is to allow the user to specify $\Sigma^{\textrm{max}}_{\alpha \alpha}$ or $U^{\textrm{max}}_{\alpha}$ for the output image, varying $\kappa$ from pixel to pixel to satisfy this requirement.
If we assume that iteratively finding such a solution requires $n_{\textrm{tries}}$ attempts, then simply solving for $T_{\alpha i }$ using Cholesky decomposition comes at a cost of $O(n_{\textrm{tries}} n^3 m)$ calculations.  We must seek to avoid fourth powers of $n$ and $m$, as even for small images each of these numbers quickly exceeds $\sim 10^3$.

The problem may also be tackled by considering the eigendecomposition of the system matrix $A_{\alpha i j}$, which is symmetric positive definite and may thus be written as
\begin{equation}\label{eq:Aeigen}
{\bf A}_\alpha = {\bf Q \Lambda Q}^{-1} =  {\bf Q \Lambda Q}^T
\end{equation}
where $Q_{ij}$ is the $n \times n$ matrix whose columns are the $n$ eigenvectors of $A_{\alpha i j}$, and the diagonal matrix $ \Lambda_{ij} = \textrm{diag}\left(\lambda_1, \lambda_2, \ldots, \lambda_n\right)$ contains the corresponding eigenvalues (e.g.\ \citealp{arfkenweber05}).  The inverse is then given simply by
\begin{equation}
{\bf A}^{-1}_\alpha =  {\bf Q \Lambda}^{-1}{\bf Q}^T.
\end{equation}
This fact can be utilized in the solution for $T_{\alpha i j}$, \emph{if we assume that the input noise covariance is diagonal} $N_{ij} = \textrm{diag}\left( \langle \eta^2_1 \rangle, \langle \eta^2_2 \rangle, \ldots, \langle \eta^2_n \rangle \right)$.  Whether this is a good or poor assumption will depend on the detectors used; off-diagonal terms might occur due to corrections for inter-pixel capacitance (e.g.\ \citealp{barronetal07}).  It should be noted that even if $N_{ij}$ is not purely diagonal no systematic error is introduced in the output image, since the calculation of the leakage objective $U_{\alpha}$ is unaffected. The method uses the input noise covariance solely to minimize the output noise, and so imperfect assumptions about $N_{ij}$ will simply cause the resulting $\Sigma_{\alpha \alpha}$ to be less than optimal. Any general $N_{ij}$ can later be propagated through to give the accurate output covariance matrix $\Sigma_{\alpha \beta}$ using equation \eqref{eq:Cr} and the solution for $T_{\alpha i}$ as determined using the diagonal $N_{ij}$ approximation.

Assuming a diagonal input noise $N_{ij}$, we may then demand that the input image $I_i$ is  pre-multiplied by a scaling matrix $\textrm{diag}\left( 1 / \langle \eta^2_1 \rangle^{1/2}, 1/\langle \eta^2_2 \rangle^{1/2}, \ldots, 1/\langle \eta^2_n \rangle^{1/2} \right)$ such that $N_{ij} = \mathbb{I}_n$, the $n \times n$ identity matrix.  We then exploit a useful invariance property of the eigenvectors of a symmetric matrix under linear combination with the identity matrix:  given equation \eqref{eq:Aeigen}, it is simple to show that
\begin{equation}
{\bf A}_\alpha + \kappa_{\alpha} \mathbb{I}_n =  {\bf Q} \left( {\bf \Lambda} + \kappa_{\alpha} \mathbb{I}_n \right) {\bf Q}^T.
\end{equation}
For our scaled image we may thus write
\begin{equation}\label{eq:Ainveigen}
\left( {\bf A}_\alpha + \kappa_{\alpha} {\bf N} \right)^{-1} = {\bf Q} \left( {\bf \Lambda} + \kappa_{\alpha} \mathbb{I}_n \right)^{-1} {\bf Q}^T,
\end{equation}
where the central matrix is purely diagonal and trivially calculated. 

The number of operations required for the initial decomposition of equation \eqref{eq:Aeigen} is $O(n^3)$, and this decomposition need only be performed once and may be parallelized (e.g.\ \citealp{andersonetal99}).
We show in the following Section \ref{sect:bisect} that, subsequently, trial values of $\Sigma_{\alpha \alpha}(\kappa_{\alpha})$ or $U_{\alpha}(\kappa_\alpha)$ may be calculated for each output pixel $\alpha$ at a cost of $O(n)$ operations per trial solution.  This implies a total cost of $O\left( n^3 +  2n^2m  + n_{\textrm{tries}} nm \right)$ for calculating $T_{\alpha i j}$, including an extra $O(n^2 m)$ overhead for necessary matrix manipulation as will be discussed in the following Section.  This represents a significant economy over the repeated use of Cholesky decompositions and justifies the use of the eigendecomposition.  This procedure also yields a figure for the condition number for the matrix $A_{\alpha i j}$, which gives a useful indication of the accuracy and stability of results regarding the input system. 


\subsection{Solution for $\kappa$ by Interval Bisection}\label{sect:bisect}
Having calculated the eigenvectors $Q_{ij}$ and eigenvalues $\Lambda_{i}$ of the system matrix $A_{\alpha ij}$, we see from equation \eqref{eq:Ainveigen} that these may by used to trivially express the matrix inverse required to calculate $T_{\alpha i}$ in equation \eqref{eq:tai}.  In Appendix \ref{app:asym}, equations \eqref{eq:SAA} \& \eqref{eq:UAA} also give expressions for $\Sigma_{\alpha \alpha}$ and $U_{\alpha}$ in terms of this matrix inverse.  Substituting equation \eqref{eq:Ainveigen} and expanding, we find
\begin{equation}
\Sigma_{\alpha \alpha} = \frac{1}{4} \sum_i \frac{N_{ii}}{\left(\lambda_i + \kappa_{\alpha}\right)^2} \left( \sum_j  B_{\alpha j}  Q_{ji} \right)^2 , \label{eq:Strial}
\end{equation}
and
\begin{equation}
U_{\alpha} = C_{\alpha} - \frac{1}{4} \sum_i \frac{\lambda_i + 2\kappa_{\alpha}}{\left( \lambda_i + \kappa_{\alpha} \right)^2}  \left( \sum_j B_{\alpha j}  Q_{ji} \right)^2 , \label{eq:Utrial}
\end{equation}
where here we have chosen to write the matrix multiplication explicitly for clarity.  These expressions may be used to generate trial values of $\Sigma_{\alpha \alpha}(\kappa_{\alpha})$ or $U_{\alpha}(\kappa_\alpha)$ directly, without first calculating $T_{\alpha i j}$.  
The object
\begin{equation}\label{eq:Pai}
P_{\alpha i} = \sum_j B_{\alpha j} Q_{ji} 
\end{equation}
 will be referred to as the projection matrix, and gives the scalar product of each row-vector ${\bf B}_\alpha$ with the successive eigenvectors $Q_{ij}$ of $A_{\alpha i j}$.  This projection matrix can be calculated in advance, immediately following the eigendecomposition of $A_{\alpha i j}$, at a cost of $O(n^2m)$ operations.
Once this is done, each trial calculation of $\Sigma_{\alpha \alpha}$ or $U_{\alpha}$ made using equations \eqref{eq:Strial} \& \eqref{eq:Utrial} comes at the low cost of $O(n)$ operations.  After each $\kappa_{\alpha}$ is found, we may then use the projection matrix once more to write the final solution for $T_{\alpha i}$ as
\begin{equation}\label{eq:Tproj}
T_{\alpha i} = -\frac{1}{2} \sum_j \frac{P_{\alpha j} Q_{ij}}{\lambda_j + \kappa_{\alpha}},
\end{equation}
at a further cost of $O(n^2m)$ operations.

The problem is then, for each output pixel $\alpha$, to find the value of $\kappa_\alpha$ that satisfies $\Sigma_{\alpha \alpha}^{\textrm{max}} - \Delta \Sigma_{\alpha \alpha}^{\textrm{max}} < \Sigma_{\alpha \alpha} \le \Sigma_{\alpha \alpha}^{\textrm{max}}$ or $U_{\alpha}^{\textrm{max}} - \Delta U_{\alpha}^{\textrm{max}} < U_\alpha \le U_{\alpha}^{\textrm{max}}$.  The simplest and most robust algorithm for finding such a solution is interval bisection \citep{pressetal92}, and this is the method we adopt here, iteratively bisecting some search range $[\kappa_{\textrm{min}}, \kappa_{\textrm{max}}]$ at intervals of $\log{\kappa}$.
If a root lies within this range the algorithm will find it, or will get as close as allowed by the user-input maximum number of bisections $n_{\textrm{bis}}$ (see Section \ref{sect:outline}, and Section \ref{sect:randout} for some example choices of these parameters).  
 
Since there is some known directionality in the problem at hand, the algorithm may be adapted somewhat.  In Appendix \ref{app:asym}, we show that the derivatives of $\Sigma_{\alpha \alpha}$ or $U_{\alpha}$ with respect to $\kappa$ are everywhere negative or positive, respectively.  Taking the case of $U_\alpha$, this means that the first trial value should be calculated using $\kappa_{\textrm{min}}$: if this solution does not satisfy $U_\alpha \le U_{\alpha}^{\textrm{max}}$ then there can be no improvement on $\kappa_{\textrm{min}}$ and this value is adopted as $\kappa_{\alpha}$.  For $\Sigma_{\alpha \alpha}$, we likewise start on $\kappa_{\max}$ before proceeding to bisect.  An output image of the interval bisection-derived $\kappa_\alpha$ is also generated, so the user may see if this range needs to be expanded.  In addition, it is also relatively fast to generate an image of the $U_\alpha$ derivative given in equation \eqref{eq:dUdk}, which allows the user to judge convergence upon the fundamentally limiting lower bound, $\lim_{\kappa\rightarrow0^+}U_\alpha$. 

The total cost of solving for $T_{\alpha i j}$ is therefore $O(n^3 + 2n^2m + n_{\textrm{tries}}nm)$ as given in Section \ref{sect:eigen}, which includes the eigendecomposition of $A_{\alpha ij}$, the calculation of the projection matrix, the $n_{\textrm{tries}}$ trial solutions to $\kappa_{\alpha}$ for each of $m$ output pixels, and the final determination of $T_{\alpha i j}$ using equation \eqref{eq:Tproj}.  It can be seen that the dominant contributions come from the eigendecomposition of $A_{\alpha ij }$ and calculating $P_{\alpha i}$, which lessens the need to seek more efficient (and potentially less robust) algorithms for the subsequent root-finding.

\section{An Illustrative Example}\label{sect:example}
We now demonstrate the use of the method by means of a worked example.  As a primary goal of this work is the analysis of survey images for a dark energy mission, we perform a simple simulation of observations made using an example design concept under consideration for the \emph{WFIRST} mission.  This particular design employs a $D = 1.3$~m primary mirror with an off-axis configuration for the secondary mirrors and detector array. This provides an unobstructed aperture, both increasing the flux throughput of the instrument and reducing the integrated light in the outer wings of the PSF.\footnote{More details regarding this family of designs can be found in the report of the Interim Science Working Group,  http://wfirst.gsfc.nasa.gov/science/ISWG\_Report.pdf}

We will simulate monochromatic observations at the near-infrared wavelength $\lambda = 1~\mu$m, close to the center of the range proposed for the \emph{WFIRST} imaging survey  \citep{gehrels10}.  It should be noted that observations may likely take place at longer wavelengths than this value, and over a broad filter band.  This will have two desirable effects, smearing together diffraction rings in the PSF and introducing a cutoff at lower spatial frequencies in the MTF.  The $\lambda = 1 \mu$m monochromatic system adopted for these examples is, in this sense, more demanding than we expect for a realistic dark energy mission.

\begin{figure*}
\plottwo{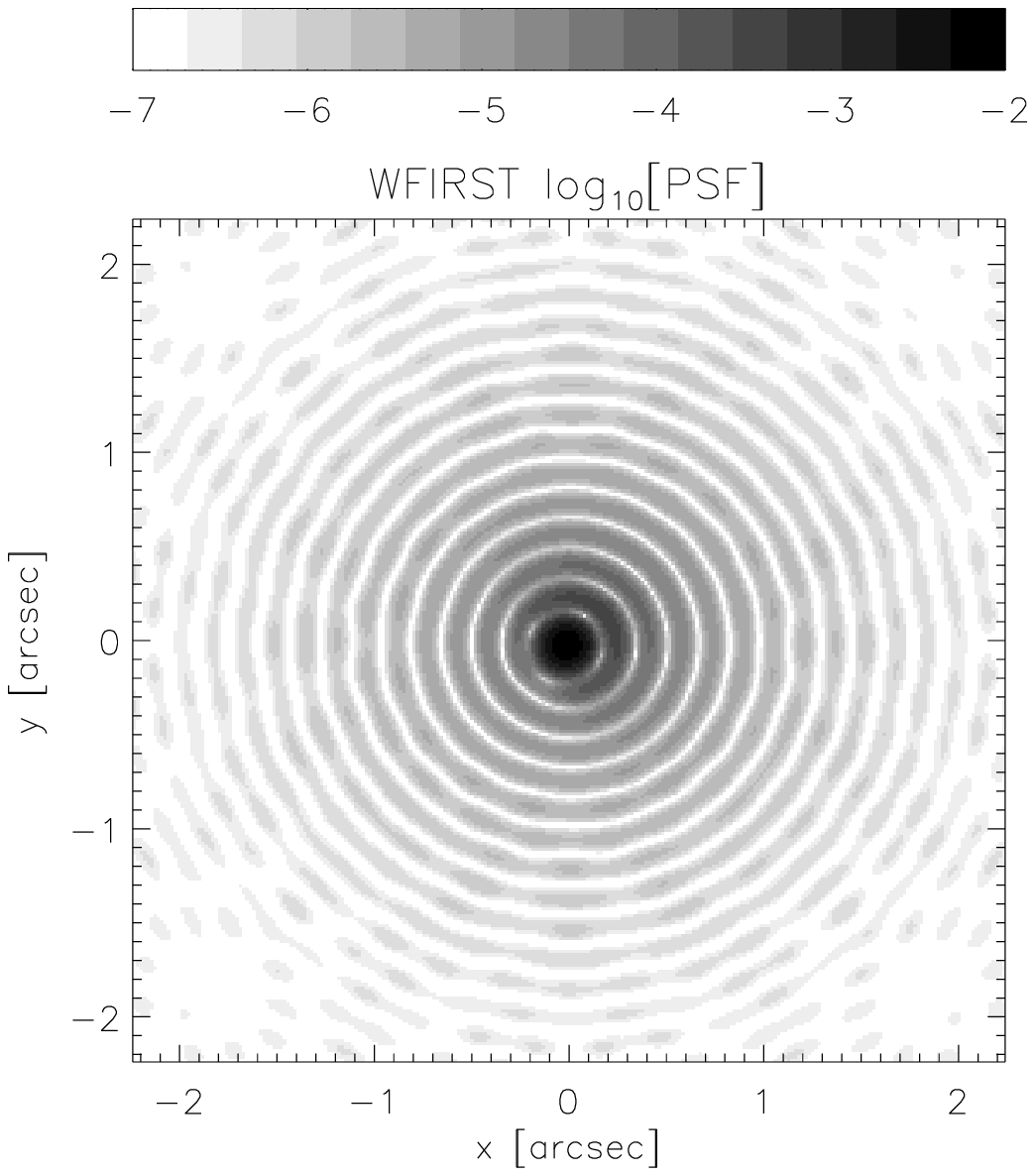}{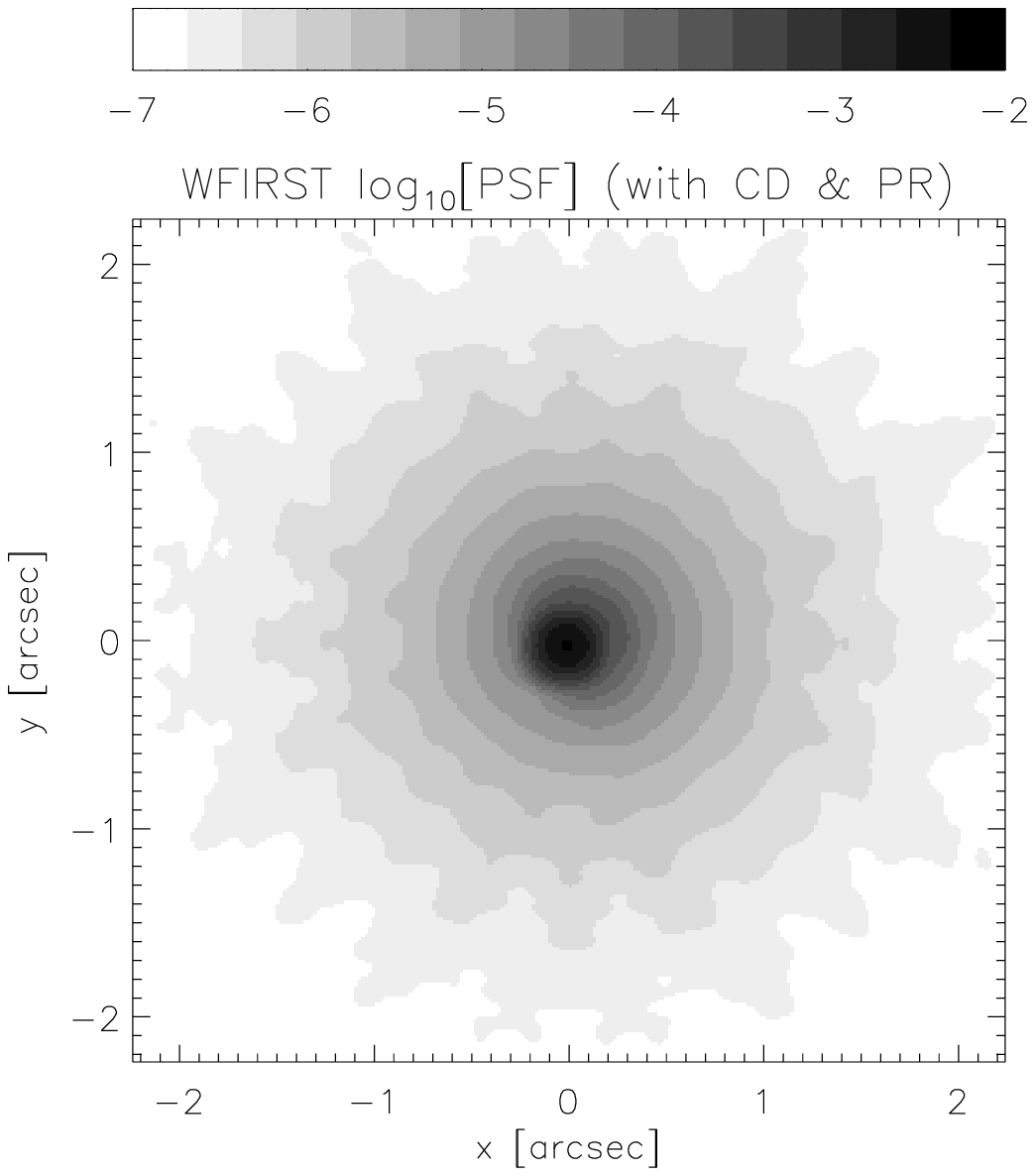}
\caption{Left panel: Simulated PSF for a \emph{WFIRST} mission concept with a 1.3~m unobstructed primary at a wavelength of $1~\mu$m. Right panel: Image of $G_i({\bf r})$ for this PSF, incorporating additional charge diffusion (CD) and the pixel response (PR) for 0.18 arcsec detectors. Both images use a logarithmic scale.\label{fig:PSF}} 
\end{figure*}
\begin{figure*}
\plottwo{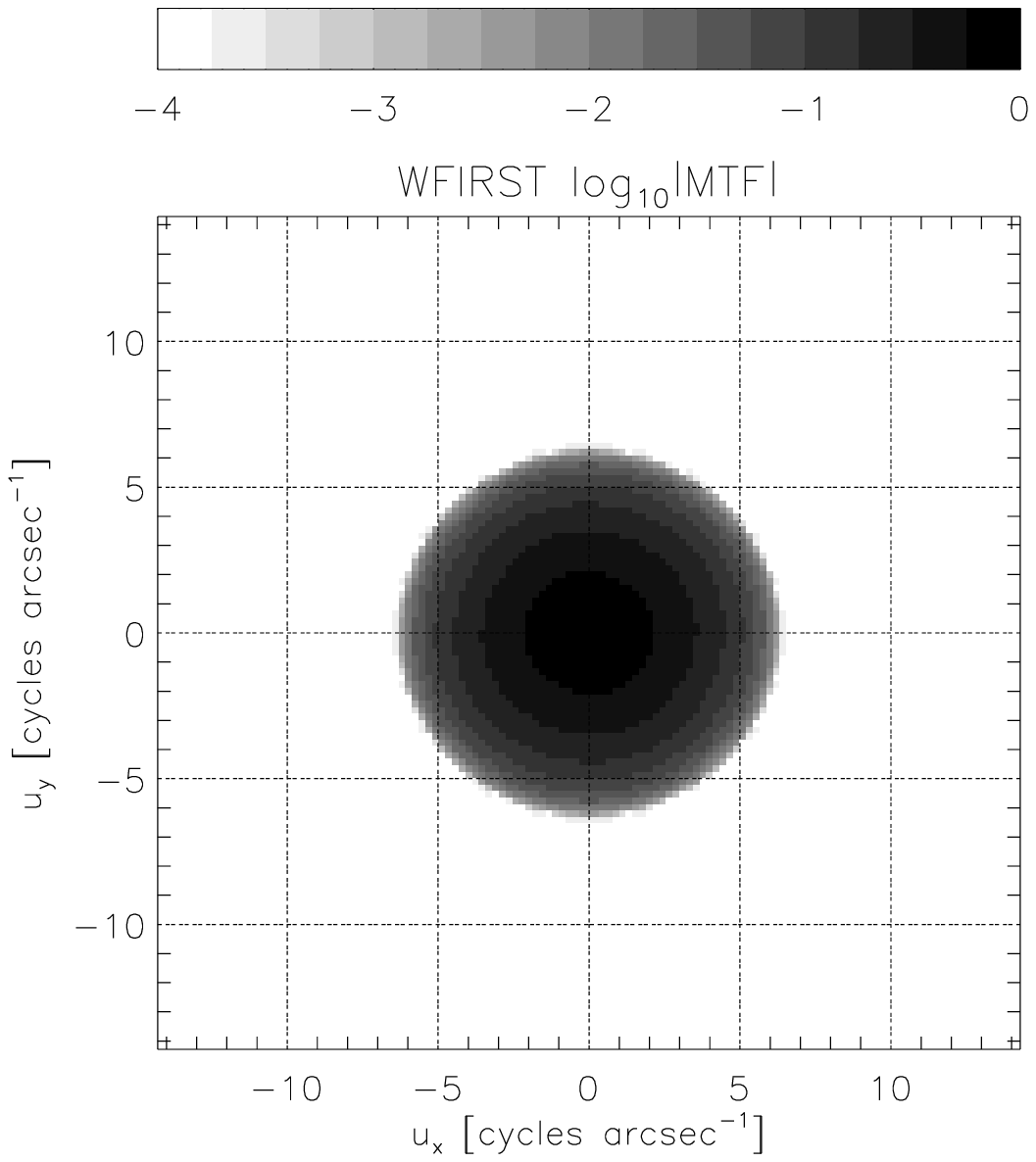}{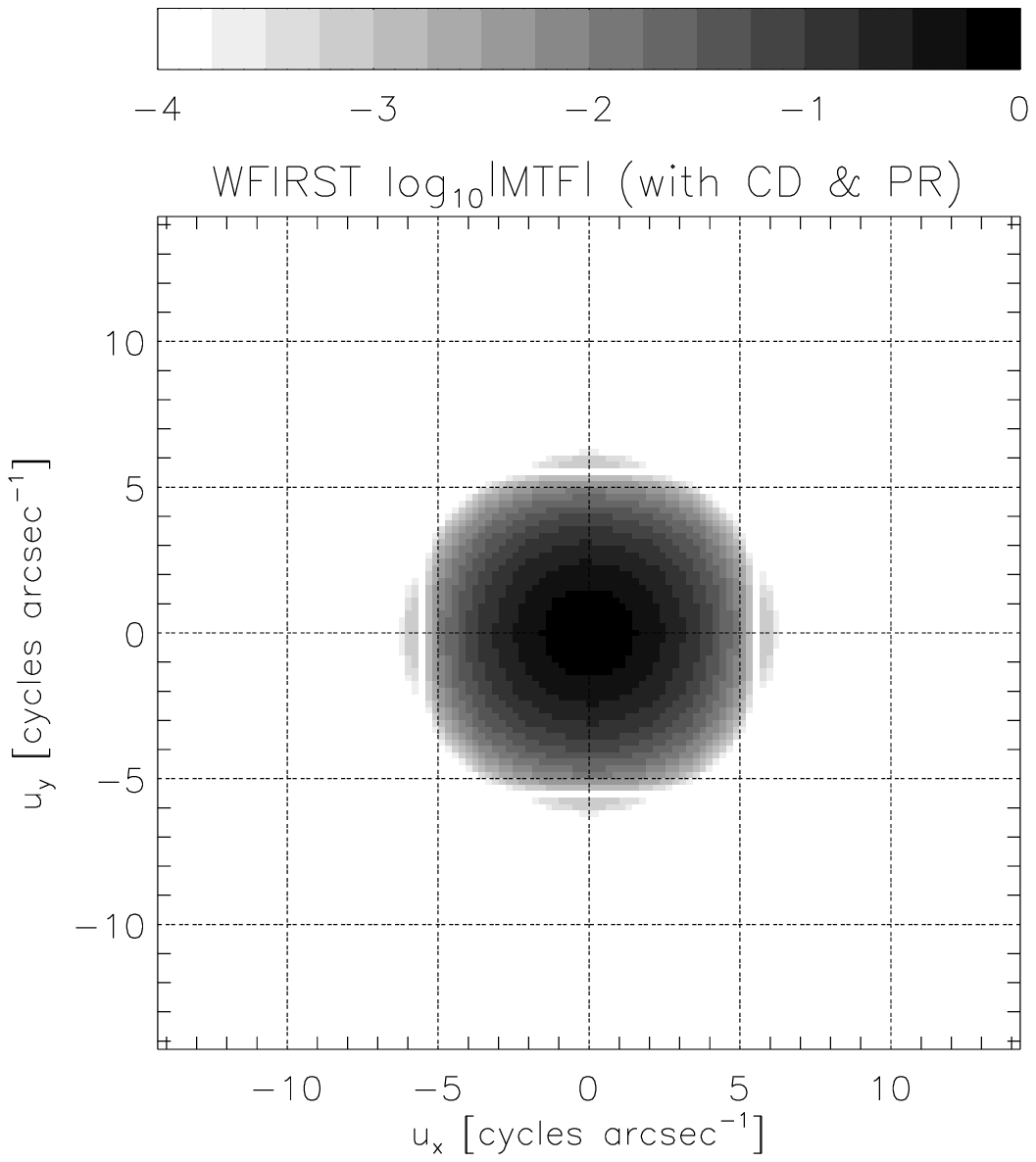}
\caption{Left panel: Magnitude of the Modulation Transfer Function (MTF) for the \emph{WFIRST} concept PSF of Figure \ref{fig:PSF}. Right panel: Image of $|\tilde{G}_i({\bf u})|$, the magnitude of the MTF corresponding to $G_i({\bf r})$; the sinc modulation of the pixel response can be clearly seen at $1/0.18 = 5 \frac{5}{9}$ cycles arcsec$^{-1}$. Both images use a logarithmic scale.\label{fig:MTF}}
\end{figure*}
In Figure \ref{fig:PSF} (left panel) we show a simulated PSF, generated using the ZEMAX software (Ed Cheng, John Lehan, David Content, \& the \emph{WFIRST} Project Office, priv.\ comm.), for this \emph{WFIRST} design concept at this wavelength. The model includes additional astigmatism to mimic aberrations due to fabrication errors, and coma to simulate possible misalignments due to thermal drift and realistic imperfections in installation and testing.  For the camera we assume an array of $18~\mu$m Teledyne Hawaii-2RG (H2RG) detectors, corresponding to an image sampling of 0.18 arcsec. For the purposes of this demonstration we model the pixel response as a simple boxcar function.

We also model the effects of charge diffusion using results from \citet{barronetal07}, who measure a projected, one-dimensional diffusion length of $l_{\textrm{CD}} = 1.87 \mu$m for a hyperbolic secant diffusion function
$I_{\textrm{CD}}(\Delta x) \propto {\textrm{sech}}{(\Delta x / l_{\textrm{CD}} )}$,
where $\Delta x$ is the distance of the collected charge from the location of the electron-hole pair.   Since we seek a de-projected charge diffusion model, some assumptions are unavoidable: for simplicity, we model the two-dimensional charge diffusion as a zero mean, circular Gaussian of width $\sigma_{\textrm{CD}} = \pi l_{\textrm{CD}} / 2$, preserving the variance (and hence diffusion length) of the function.

 Convolving the \emph{WFIRST} concept PSF with the pixel response boxcar and the charge diffusion function, we arrive at the functional form of $G_i({\bf r})$ for this demonstration, shown in Figure \ref{fig:PSF} (right panel).  For this first demonstration, $G_i({\bf r})$ will remain constant throughout the input image.

The Modulation Transfer Function (MTF) is defined as the Fourier transform of the telescope PSF. The function $\tilde{G}_i({\bf u})$ therefore represents the MTF conjugate to $G_i({\bf r})$, and the magnitude of this complex object can be seen in Figure \ref{fig:MTF} for the two corresponding PSFs of Figure \ref{fig:PSF}.  As can be seen from Figure \ref{fig:MTF}, the system is bandlimited at the fundamental frequency corresponding to $D / \lambda = 6.3026~$cycles arcsec$^{-1}$.  Therefore, a sampling interval of $\lambda / 2D = 0.079333$~arcsec in the output image $H_{\alpha}$ is the requirement for critical sampling according to the sampling theorem \citep{marks09}.  This output sampling rate is therefore of fundamental interest in demonstrating that survey design and multiple dithering strategies will allow a dark energy mission to overcome an undersampling detector array in the focal plane, and thereby utilize all available spatial information for weak lensing measurements.  Hence, $\lambda / 2D = 0.079333$~arcsec will be adopted as the desired output sampling interval for all the demonstrations in this study.

\subsection{Input Images: Random Dithers, Random Rotations}
From the \emph{WFIRST} concept described in Section \ref{sect:example} it has been possible to create a model for $G_i({\bf r})$ that shares many of the properties expected for a future dark energy mission.  However, this represents only one of the input quantities described in Section \ref{sect:outline}; not least, input images themselves are required.

\begin{figure*}
\plottwo{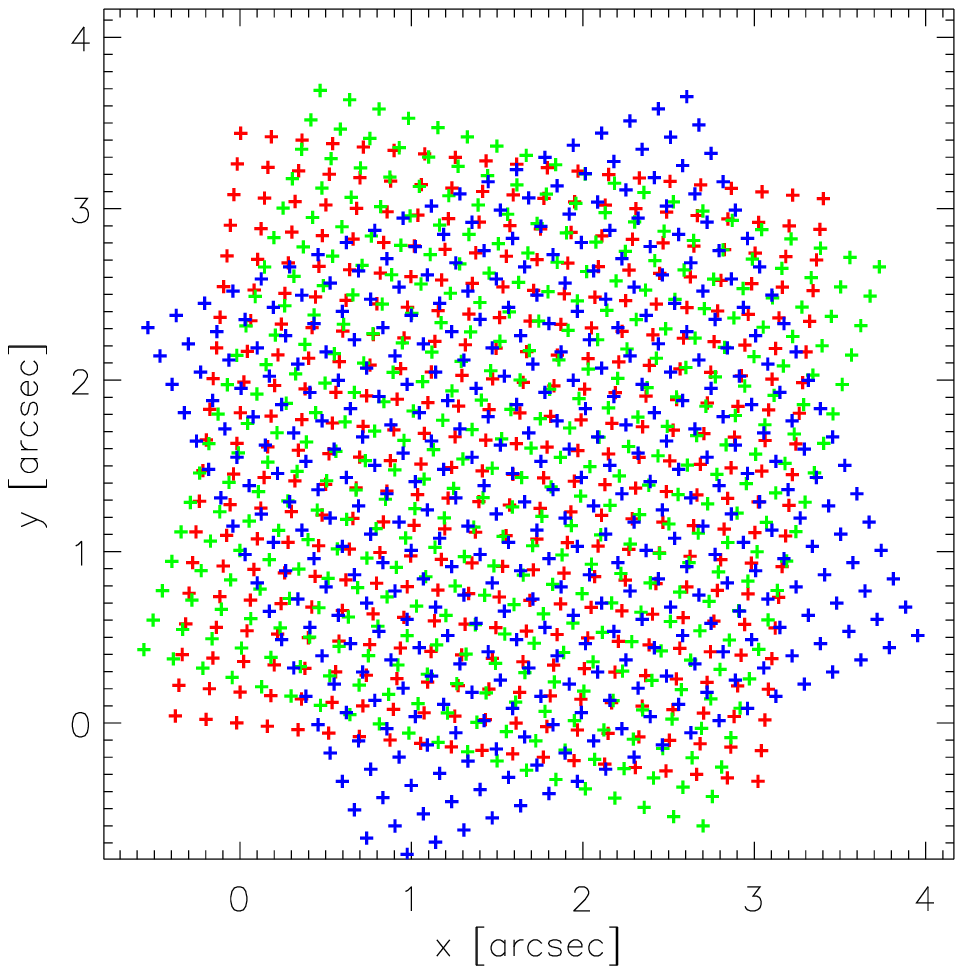}{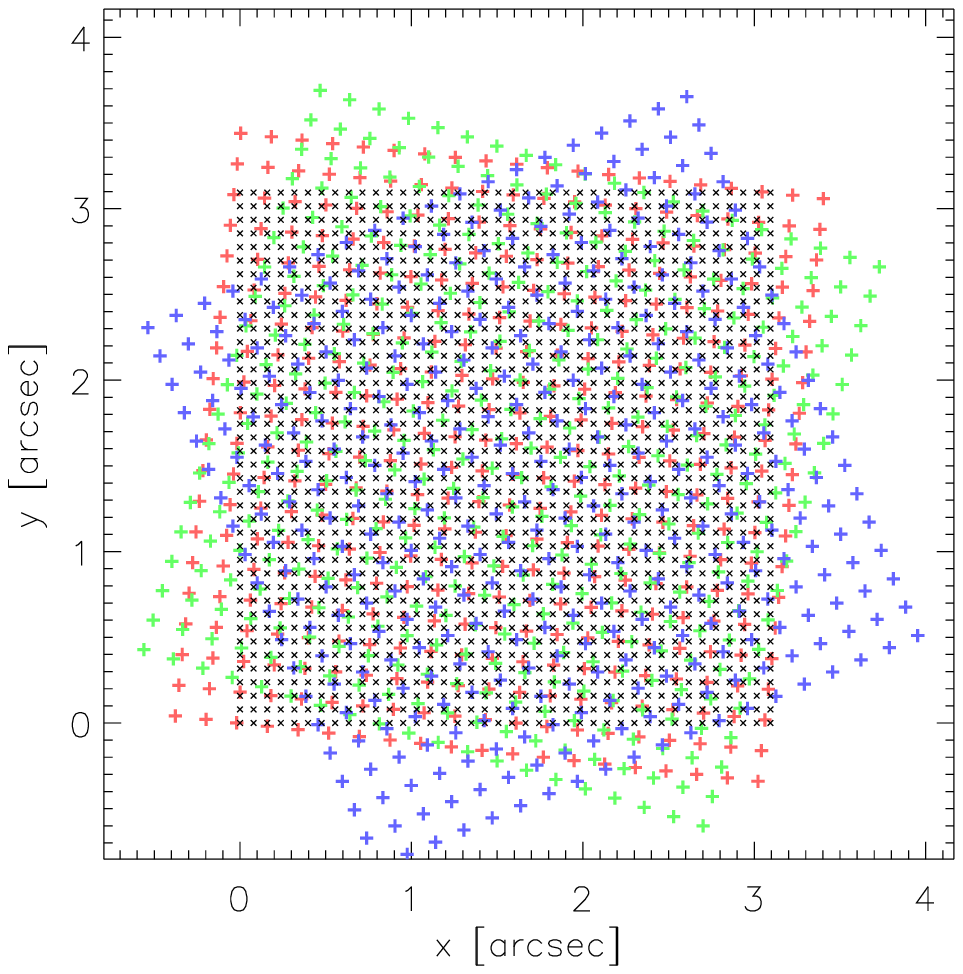}
\caption{Left panel: Input pixel locations ${\bf r}_i$ for three randomly dithered and rotated exposures of the same camera. Red, green and blue crosses represent each of the three exposures. Right panel: Output image sampling locations ${\bf R}_{\alpha}$ (black crosses), placed at a spatial interval of $0.079333$ arcsec so as to fully sample the \emph{WFIRST} concept image plane.\label{fig:xyrand}}
\end{figure*}
As a demonstration of the technique's flexibility we will attempt to reconstruct a fully-sampled output image using a set of three, randomly dithered and rotated images.  The spatial configuration ${\bf r}_i$ of the three dithered exposures can be seen in Figure \ref{fig:xyrand}, each of dimension $20 \times 20$ pixels$^2$.  Also shown are the corresponding sample locations ${\bf R}_{\alpha}$ of the desired, fully-sampled output image.   It should be stated from the outset that these three dither patterns are not expected to be sufficient to achieve full sampling, but are being used deliberately to show the behavior of the algorithm when unable to recover certain output pixels.  The user may of course use images with any pixel locations, there being no requirements of regularity or homogeneity in the formalism. 

\begin{figure}
\epsscale{0.8}
\plotone{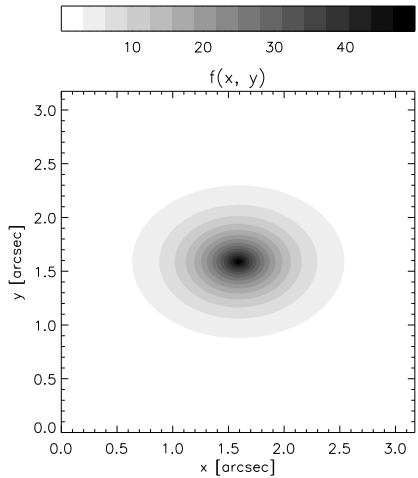}
\caption{Surface brightness $f({\bf r})$ used in Sections \ref{sect:example} \& \ref{sect:wfirst}: a simple, idealized model of an elliptical galaxy.\label{fig:fgal}}
\epsscale{1.0}
\end{figure}
For the physical image on the sky $f({\bf r})$ we choose a single galaxy, modeled as a simple elliptical, exponential profile.  The profile is given an axis ratio $a/b = 3/\sqrt{5}$, a semi-major axis scale length (aligned along the $x$ direction) of $a = 3 / (4 \sqrt{5})$ arcsec, and a central peak surface brightness of $50$ (in arbitrary units).  This model galaxy is shown in  Figure \ref{fig:fgal} in high-resolution. 

This image is then convolved with the $G_i({\bf r})$ of Figure \ref{fig:PSF} and sampled at the locations shown in Figure \ref{fig:xyrand} to generate the array of input images that make up $I_i$.  Since each image is rotated, before calculating $A_{\alpha i j}$ and $B_{\alpha i}$ the PSF $G_i ({\bf r})$ for each exposure must be rotated into a common coordinate system in which $\Gamma({\bf r})$ is defined (see Section \ref{sect:randsys} below).  In this example we choose $\Gamma({\bf r})$ to be simply the unrotated $G_i({\bf r})$ of Figure \ref{fig:PSF}, so that only the input PSFs need be rotated before solving the system.  For precise comparison of output and input images we do not add stochastic noise to $I_i$, but will nonetheless assume a unit diagonal noise covariance $N_{ij} = \mathbb{I}_n$ to illustrate the behavior of $\Sigma_{\alpha \alpha}$ in a more realistic, noisy imaging scenario. 

With $G_i({\bf r})$, $I_i$, ${\bf r}_i$ and $N_{ij}$ we have provided the three ``hard inputs'' of Section \ref{sect:outline}, but there remain choices to be made regarding the values of user-specified input quantities.  We adopt a simple Dirac delta function for $\Upsilon ({\bf r})$ so that $\tilde{\Upsilon} ({\bf u}) = 1$.  This fixed parameter choice is given in Table \ref{tab:params}, where all such implementation-specific parameter values are given.
The remaining choices that must be made relate to the process of finding solutions for $U^{\textrm{max}}_{\alpha}$ or $\Sigma^{\textrm{max}}_{\alpha}$, and so we defer a discussion of these input quantities until Section \ref{sect:randout}.  We first illustrate the system matrices $A_{\alpha i j}$, $B_{\alpha i}$, which may be calculated immediately. 

\subsection{System Matrices}\label{sect:randsys}
\begin{deluxetable*}{crl}
\tablecaption{Fixed, \textsc{Imcom} implementation-specific parameters used in calculating the results of Sections \ref{sect:example} \& \ref{sect:wfirst}. \label{tab:params}}
\tablehead{\colhead{Parameter} & \colhead{Value} & \colhead{Description}}
\startdata
$\Upsilon ({\bf r})$ & $\delta({\bf r})$ &  Kernel used for calculating leakage objective $U_{\alpha}$ (see equation \ref{eq:Ulong}). \\
$n_{\textrm{pad}}$ & 3 & Multiple by which $\tilde{G}_i({\bf u})$ and $\tilde{\Gamma}({\bf u})$ are \emph{further} zero-padded in ${\bf u}$ when calculating \\
~ & ~ &  $A_{\alpha i j}$ and $B_{\alpha i}$ (Section \ref{sect:randsys}). \\
$n_{\textrm{poly}}$ & 7 & Polynomial order used when rotating $G_i({\bf r})$, and interpolating $A_{\alpha i j}$ and $B_{\alpha i}$ from \\
~ & ~ & lookup tables (Section \ref{sect:randsys}). \\
$\kappa_{\textrm{min}}$ & $1.11C_{\alpha} \times 10^{-16} $ & Minimum $\kappa_{\alpha}$ for interval bisection search, set by machine precision \\
~ & ~ & (Sections \ref{sect:imp} \& \ref{sect:randout}). \\
$\kappa_{\textrm{max}}$ & $9.01C_{\alpha} \times 10^{15} $ & Maximum $\kappa_{\alpha}$ for interval bisection search, set by machine precision \\
~ & ~ &  (Sections \ref{sect:imp} \& \ref{sect:randout}). \\
$n_{\textrm{bis}}$ & 53 & Maximum number of interval bisections allowed in solving for each $\kappa_{\alpha}$ \\
~ & ~ & (Sections \ref{sect:imp} \& \ref{sect:randout}).  \\
\enddata
\end{deluxetable*}
Taking the input functional forms of $G_i({\bf r})$ and $\Gamma({\bf r})$, and the
pixel locations ${\bf r}_i$ and ${\bf R}_{\alpha}$, we may calculate $A_{\alpha i j}$ and $B_{\alpha i}$ using the FFT-based prescription described in Section \ref{sect:sysfft}.
As the accuracy of these objects is crucial, and their computation costs are relatively low in comparison to later operations, it is desirable to generate high-resolution lookup tables for $A_{\alpha i j}$ and $B_{\alpha i}$ and then interpolate using a high-order polynomial.

As discussed in Section \ref{sect:sysfft} the images of $G_i({\bf r})$ and $\Gamma({\bf r})$ are first zero-padded so that their spatial dimensions (in physical units) equal or exceed the maximum spatial extent of the input images that make up $I_i$.  It is noted that the \emph{WFIRST} design concept PSF of Figure \ref{fig:PSF} is supplied at relatively high resolution ($n_{\textrm{PSF}} \simeq 11$: see Section \ref{sect:sysfft}).   For rotated input frames, each $G_i({\bf r})$ must be rotated accordingly to the common coordinate system in which $\Gamma({\bf r})$ is defined. Each image $G_i({\bf r})$ is rotated using Lagrange polynomial interpolation at order $n_{\textrm{poly}} = 7$.  Then $\tilde{G}_i({\bf u})$ and $\tilde{\Gamma}({\bf r})$ are calculated using the FFT algorithm.  Limited only by the memory available (2GB in a 32-bit architecture), we may then add \emph{further} zero-padding to the bandlimited $\tilde{G}_i({\bf u})$ and $\tilde{\Gamma}({\bf r})$.  We choose to zero-pad each such that their linear dimension increases by a further factor $n_{\textrm{pad}}$ compared to the starting size of $\tilde{G}_i({\bf u})$ and $\tilde{\Gamma}({\bf r})$ (which were already padded once in real space, up to the size of the input images). In this example, and throughout this Paper, we choose $n_{\textrm{pad}} = 3$ (see Table \ref{tab:params}).  

The inverse FFT is then used to build the lookup tables described by equations \eqref{eq:aft} \& \eqref{eq:bft} which, following the zero-padding, have $n_{\textrm{PSF}} \times n_{\textrm{pad}}$ entries along the length of each input pixel in each dimension.  Finally, these tables are interpolated using a polynomial at order $n_{\textrm{poly}} = 7$  to estimate values of $A_{\alpha i j}$ and $B_{\alpha i}$ at the locations of the vectors ${\bf r}_j - {\bf r}_i$ and ${\bf r}_i - {\bf R}_{\alpha}$, respectively.  The resultant matrices can be seen in Figures \ref{fig:arand} \& \ref{fig:brand}.
\begin{figure}
\plotone{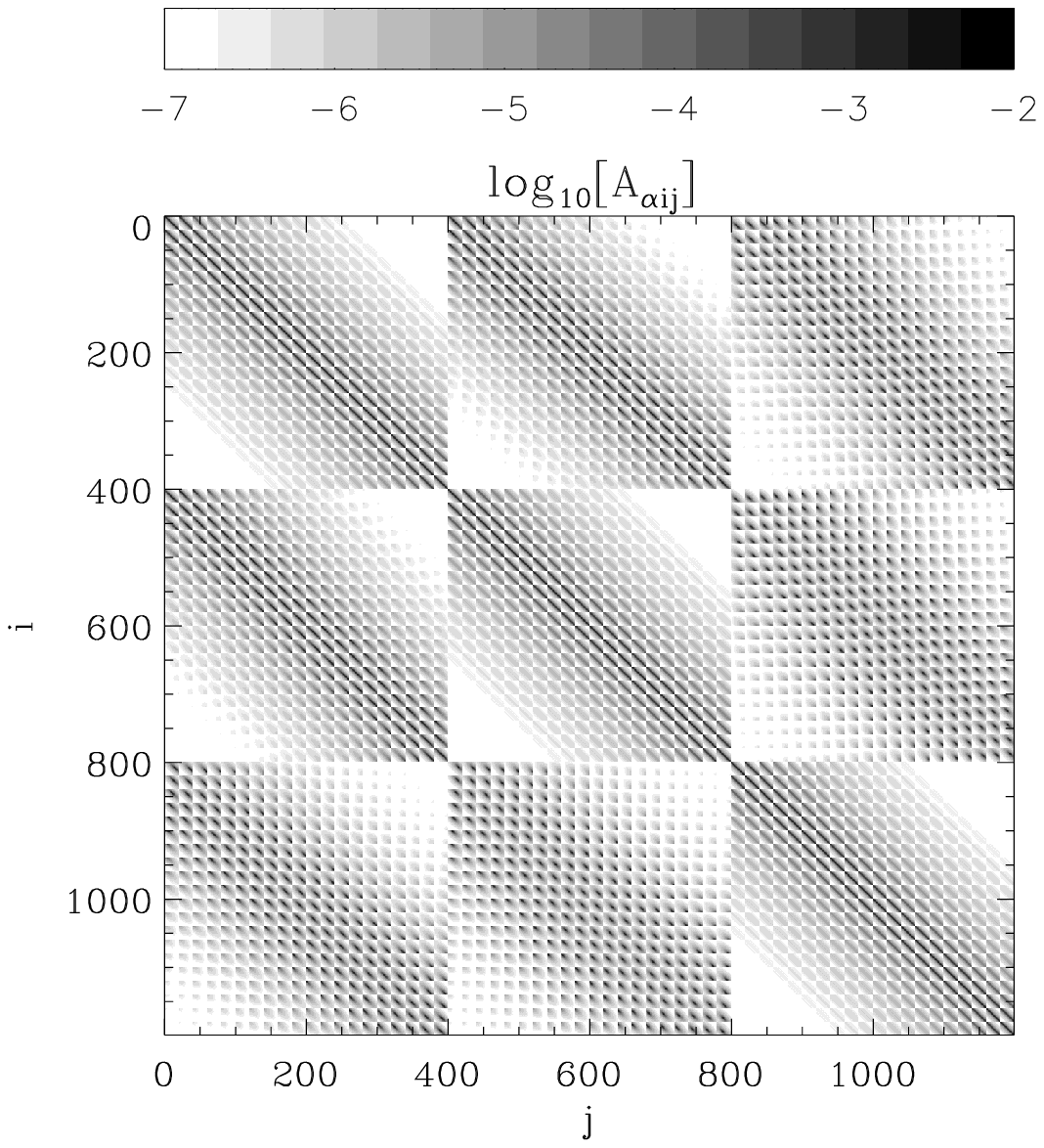}
\caption{System matrix $A_{\alpha ij}$ for the image configuration of Section \ref{sect:example} and Figure \ref{fig:rand}. \label{fig:arand}}
\end{figure}
\begin{figure}
\epsscale{0.8}
\plotone{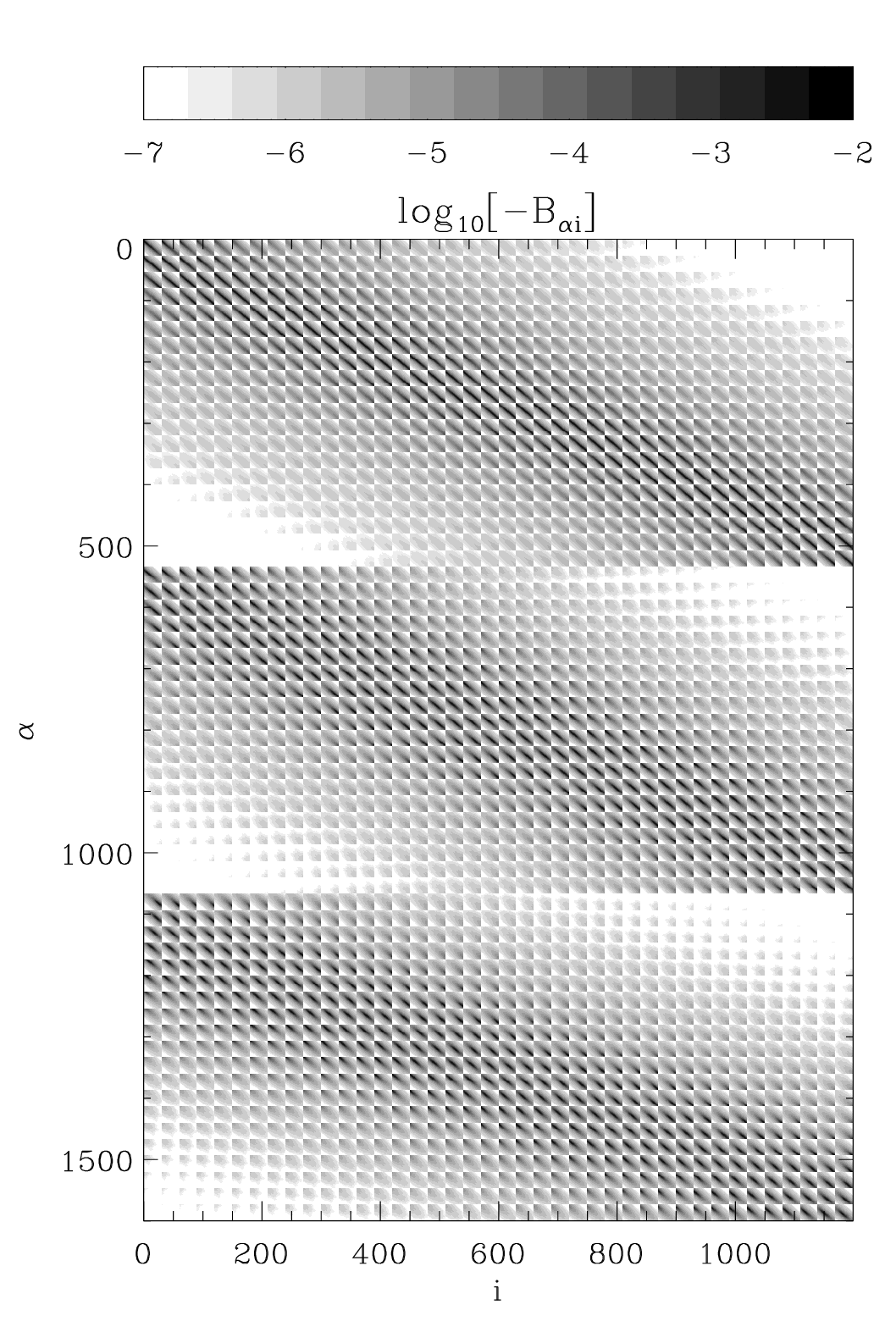}
\caption{System matrix $B_{\alpha i}$ for the image configuration of Section \ref{sect:example} and Figure \ref{fig:rand}. \label{fig:brand}}
\end{figure}

Once $A_{\alpha i j}$ and $B_{\alpha i}$ are calculated, the final stages in the preparation for the solution of $T_{\alpha i}$ are the eigendecomposition of the $A_{\alpha i j}$ matrix to give $\lambda_i$ and $Q_{ij}$, and the subsequent calculation of the projection matrix $P_{\alpha i}$.  In Figures \ref{fig:evalsrand} \& \ref{fig:evecsrand} we plot the eigenvalues and lowest-order eigenvectors for $A_{\alpha i j}$ in this example.  The condition number $\aleph$ (see, e.g., \citealp{cheneykincaid98})  for a symmetric, positive-definite matrix such as $A_{\alpha i j}$ is simply expressed in terms of the eigenvalues as
\begin{equation}
\aleph\left( A_{\alpha i j} \right) = \frac{\max[{\lambda_i}]}{\min[{\lambda_i}]}.
\end{equation}
A system is said to be singular if the condition number is infinite, and typically ill-conditioned if $\log_{10}(\aleph) \gtrsim$ the numerical precision of at which each matrix element is stored.  For the illustrative example in this Section, the condition number is found to be $\aleph\left( A_{\alpha i j} \right) = 8.369 \times 10^6$, which implies the system is adequately well-conditioned if using double precision arithmetic with machine epsilon $\varepsilon = 2^{-53} = 1.11 \times 10^{-16}$.  In practice the matrix that must be inverted is $A_{\alpha ij } + \kappa_{\alpha} N_{ij}$ so, given appropriate choices of $\kappa_{\alpha}$, the method may proceed even when $\aleph$ is large.  Nonetheless,  calculating $\aleph$ in this way provides both a useful check on the eigenvalues $\lambda_i$ (which must all be positive since $A_{\alpha i j }$ is a symmetric positive definite matrix), and an idea of when solutions for $T_{\alpha i}$ might become unstable for small $\kappa_{\alpha}$.

The projection matrix $P_{\alpha i}$ is then constructed from the eigenvectors $Q_{ij}$ and the matrix $B_{\alpha i}$ via matrix multiplication as shown in equation \eqref{eq:Pai}.  At this stage the problem is then fully set up, and we are ready to begin the solution for $T_{\alpha i }$.
\begin{figure}
\epsscale{1.0}
\plotone{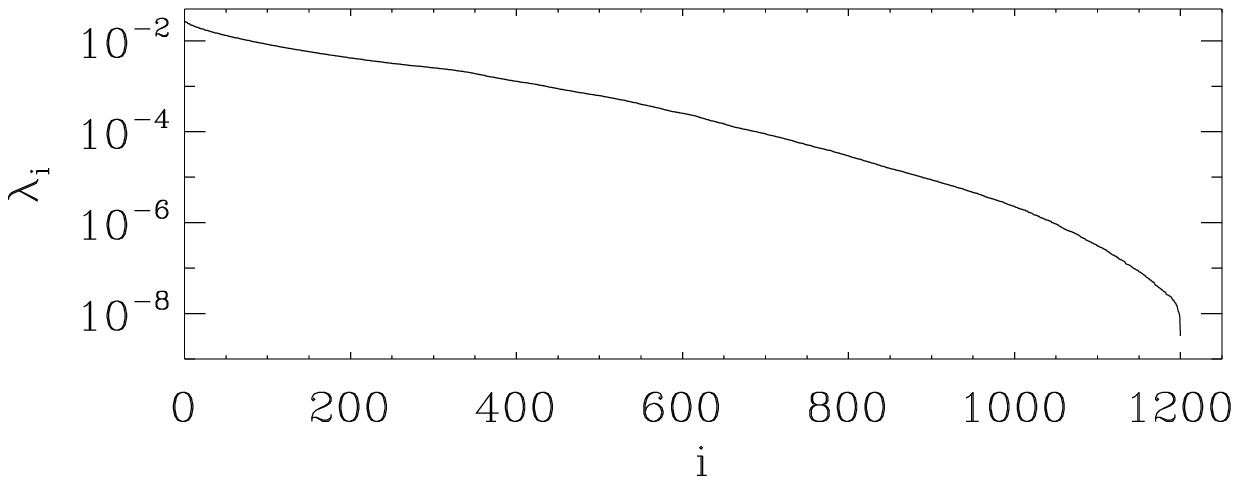}
\caption{Eigenvalues of the $A_{\alpha i j}$ matrix shown in Figure \ref{fig:arand}. The matrix in this example has a condition number of $\aleph = 8.369 \times 10^6$.\label{fig:evalsrand}}
\end{figure}
\begin{figure}
\plotone{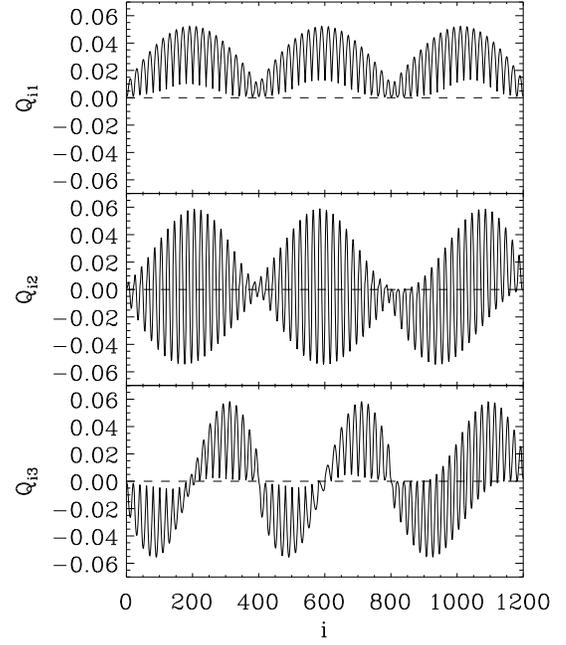}
\caption{The first three eigenvectors of the $A_{\alpha i j}$ matrix shown in Figure \ref{fig:arand}.\label{fig:evecsrand}}
\end{figure}

\subsection{Output Images}\label{sect:randout}
The solution of equation \eqref{eq:Tai} for $T_{\alpha i}$ as described in Section \ref{sect:bisect} proceeds in two separate stages. First, values of $\kappa_{\alpha}$ for each output pixel are independently determined so as to attempt to keep $U_{\alpha}$ or $\Sigma_{\alpha \alpha}$ within the pre-set limits. Second, and at far greater relative computational cost, these values are substituted into equation \eqref{eq:Tproj} to calculate $T_{\alpha i}$.

To determine $\kappa_{\alpha}$, it is necessary to complete the set of choices described in Section \ref{sect:outline}.  For the purposes of this example, we will demonstrate the solution of $\kappa_{\alpha}$ both by setting requirements upon $\Sigma_{\alpha \alpha}$ and by setting requirements on $U_{\alpha}$.  For both cases, we search within the broad range $[\kappa_{\alpha}^{\textrm{min}}, \kappa_{\alpha}^{\textrm{max}}] = [1.11  C_{\alpha}\times 10^{-16},  9.01 C_{\alpha} \times 10^{15}] $ (see Table \ref{tab:params}), effectively being limited only by the machine precision $\varepsilon$.  We also allow a total of $n_{\textrm{bis}} = 53$ interval bisections to find $\kappa_{\alpha}$.  Computational costs are negligibly affected by choices for these parameters.

In Figure \ref{fig:randbyS} we plot results from the \textsc{Imcom} code having required $\Sigma_{\alpha \alpha}^{\textrm{max}} - \Delta \Sigma_{\alpha \alpha}^{\textrm{max}} \le  \Sigma_{\alpha \alpha} < \Sigma_{\alpha \alpha}^{\textrm{max}} $, where $\Sigma_{\alpha \alpha}^{\textrm{max}} = 1$ and 
$\Delta \Sigma_{\alpha \alpha}^{\textrm{max}} = 10^{-2}$.   As the input noise covariance is in this case specified as a unitary, diagonal $N_{ij} = \mathbb{I}_n$, this corresponds to requiring that the output image $H_{\alpha}$ maintain the same level of noise variance as compared to the input image pixels. Figure \ref{fig:randbyS}a shows the value of $\kappa_{\alpha}$ selected at each output position ${\bf R}_{\alpha}$ in order to fulfill these requirements on $\Sigma_{\alpha \alpha}$, and Figure \ref{fig:randbyS}b shows the variation in the resultant leakage objective $U_{\alpha}$ across the output image.  This latter is plotted in its natural units of $C_{\alpha}$ (see Appendix \ref{app:asym}, equations \ref{eq:Utendinf} \& \ref{eq:Utendzero}). 
Degradation in image fidelity, manifesting as increased $U_{\alpha}$, can be clearly seen in the corner regions of the output in Figure \ref{fig:randbyS}b where input pixel coverage is reduced, but also within central locations where chance alignments of the input pixel pattern lead to encircled regions of sparse sampling.  
\begin{figure*}
\begin{center}
\plottwo{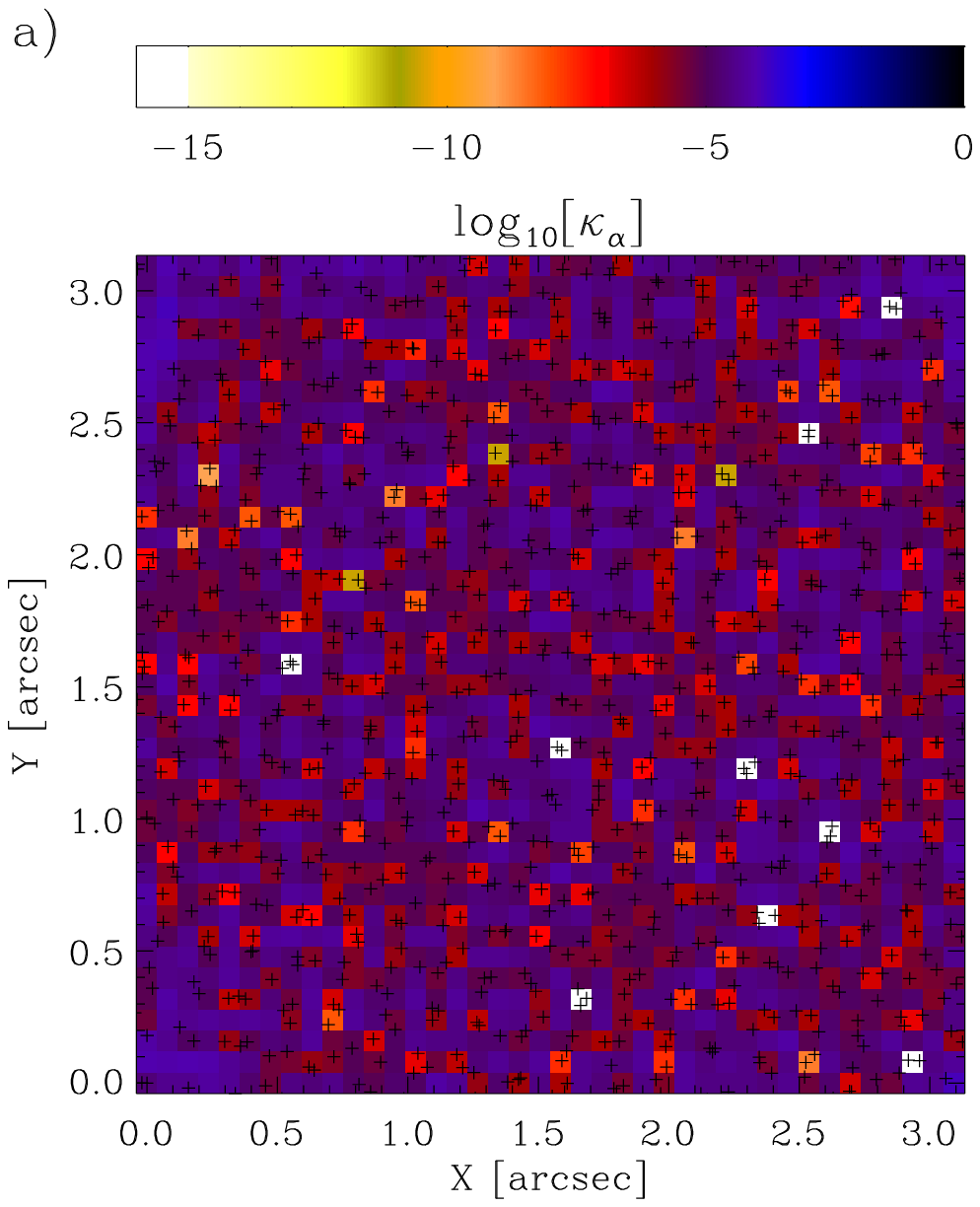}{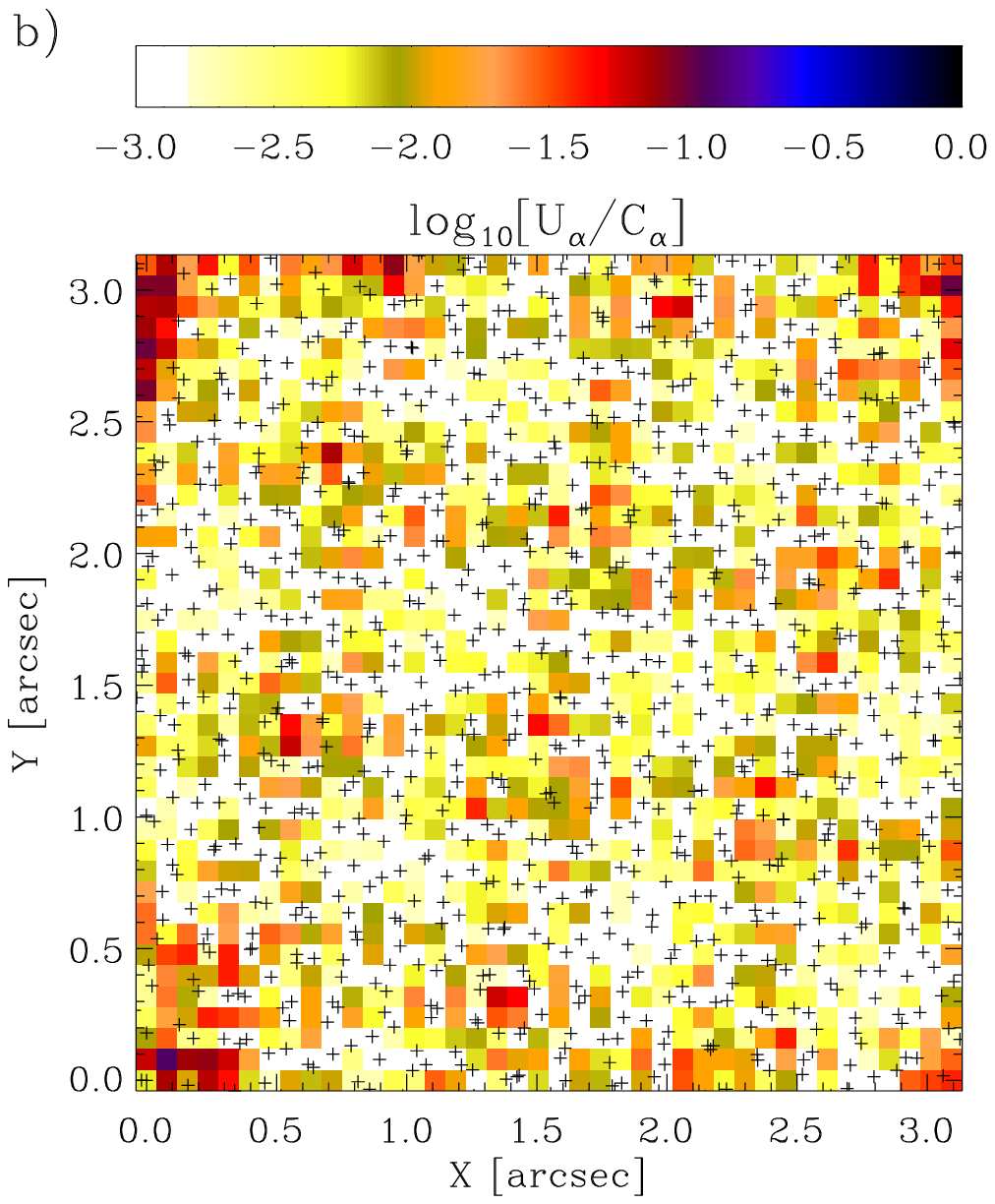}
\plottwo{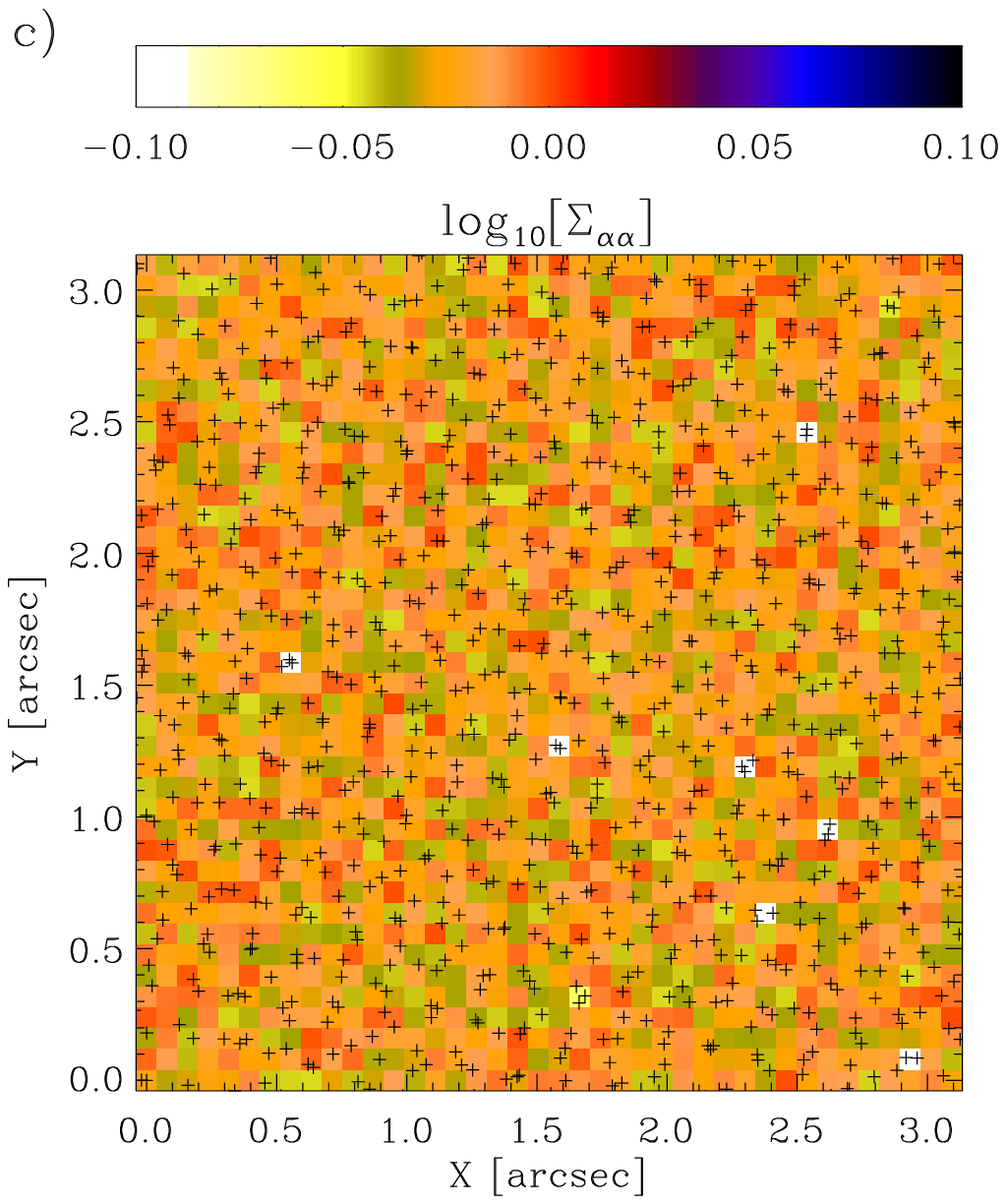}{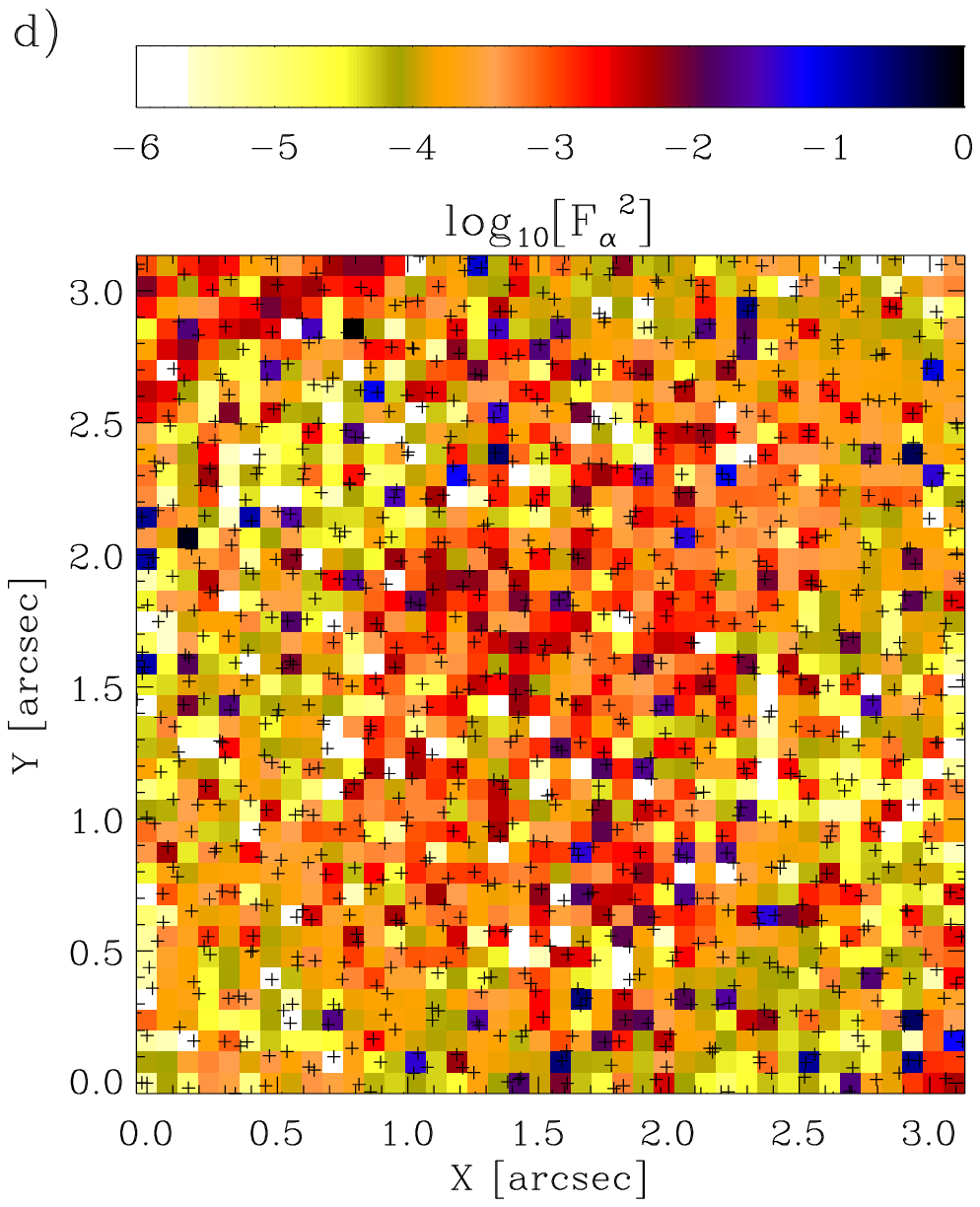}
\end{center}
\caption{Map of $\kappa_{\alpha}$ solution (a),  output leakage objective $U_{\alpha}$ (b),
noise variance $\Sigma_{\alpha \alpha}$ (c), and corresponding $F^2_{\alpha}$ (d)
for the input choices $\Sigma_{\alpha \alpha}^{\textrm{max}} = 10^{-1}$,  $\Delta \Sigma_{\alpha \alpha}^{\textrm{max}} = 10^{-3}$ in the demonstration of Section \ref{sect:example}. The color scale in the map for $U_{\alpha}$ does not reflect the full range of these values, but is selected to most clearly illustrate the dependence of $U_{\alpha}$ on the input dither pattern ${\bf r}_i$. \label{fig:randbyS}}
\end{figure*}

Overall we see that, given the requirements on the noise properties, the randomly dithered, three exposure system produces output with $U_{\alpha} \sim 0.01C_{\alpha}$ or worse over a significant fraction of $H_{\alpha}$.  This is not the level of fidelity that will be required for a dark energy mission, but the demonstration illustrates how aspects of the input dither pattern impact the quality of the recovered output.

The corresponding noise variance map $\Sigma_{\alpha \alpha}$ can be seen in Figure \ref{fig:randbyS}c.  As shown in Appendix \ref{app:asym}, the output $\Sigma_{\alpha \alpha}$ can be made arbitrarily small by increasing $\kappa_{\alpha}$, but not necessarily arbitrarily large as $\kappa_{\alpha}$ decreases.  The map of $\Sigma_{\alpha \alpha}$ shows where occasional chance alignments of multiple input pixels produce noise that is \emph{lower} than $\Sigma_{\alpha \alpha}^{\textrm{max}} - \Delta \Sigma_{\alpha \alpha}^{\textrm{max}}$, and comparison with Figure \ref{fig:randbyS}a shows that these points indeed lie where $\kappa_{\alpha} = \kappa_{\alpha}^{\textrm{min}}$. 

In Figure \ref{fig:randbyS}d we compare the \textsc{Imcom}-reconstructed image $H_{\alpha}$ to an independently-generated target image $J_{\alpha}$, calculated using equation \eqref{eq:Ja} and the highly-oversampled $f({\bf r})$ as shown in Figure \ref{fig:fgal} (the code for generating $J_{\alpha}$ is in fact written in the \textsc{IDL} language and shares no routines with \textsc{Imcom}).   These two image vectors are used to calculate the squared fractional residual,  which we define as
\begin{equation}\label{eq:f2a}
F^2_{\alpha} = \frac{(H_{\alpha} - J_{\alpha})^2} {J_{\alpha}^2}.
\end{equation}
Values of $F^2_{\alpha}$ depend upon $f({\bf r})$ directly and so cannot themselves be used to locate an optimal $T_{\alpha i}$, but in simulated examples they provide a useful comparison: for a well-controlled system we typically expect $F^2_{\alpha} \lesssim U_{\alpha} / C_{\alpha}$ [although this may be violated in practice if $J_{\alpha} \simeq 0$ or $f({\bf r})$ contains large or abrupt intensity variations such as those caused by bright stars or cosmic ray impacts].  The variation in $F^2_{\alpha}$ across the output image plane for this example can be seen in Figure \ref{fig:randbyS}d. This shows that $F^2_{\alpha} \lesssim U_{\alpha} / C_{\alpha}$ across much of the output in this example but can vary sharply due to the large variations in $U_{\alpha}$, particularly towards the edge regions of the image where $f({\bf r})$ is smaller.

\begin{figure*}
\begin{center}
\plottwo{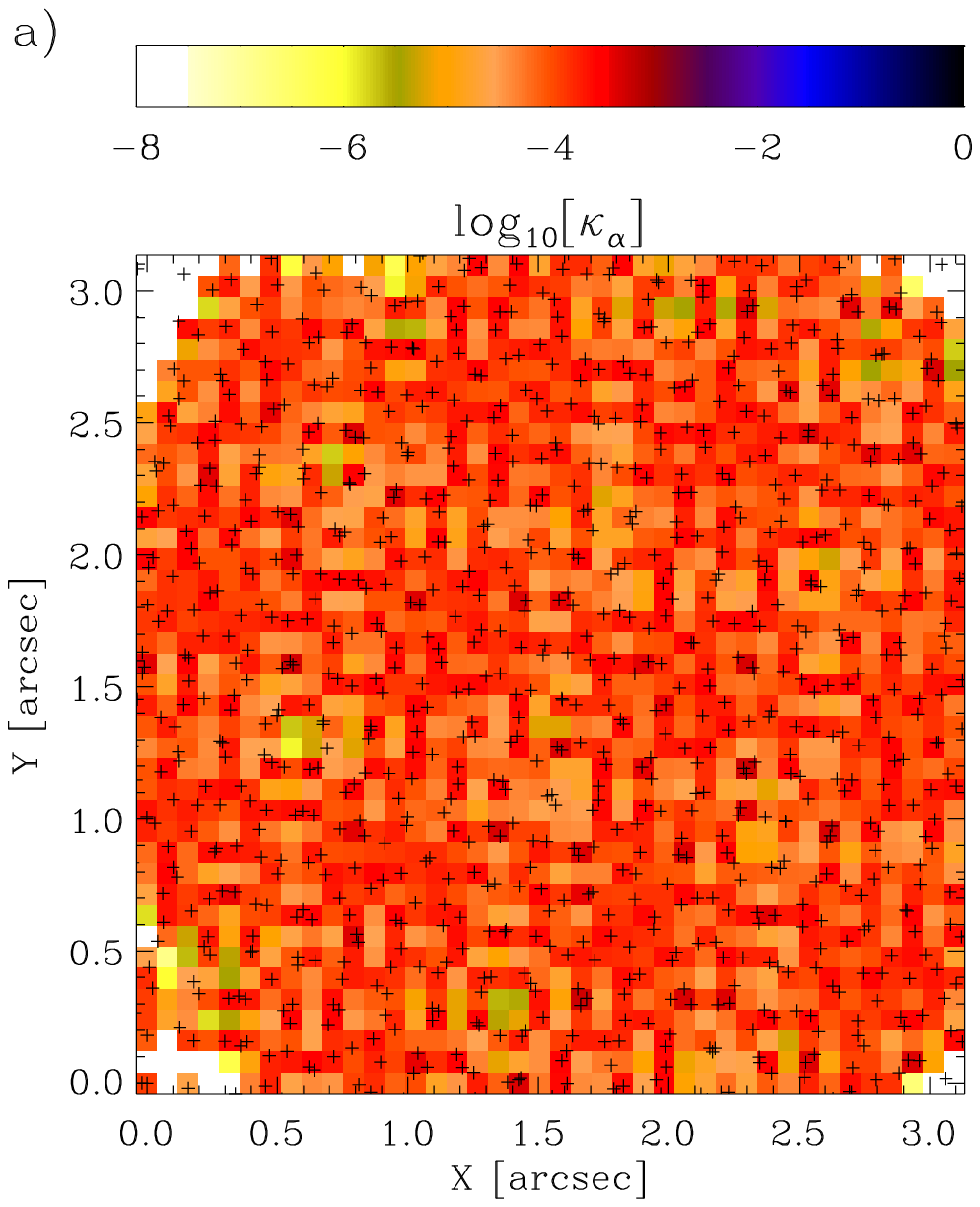}{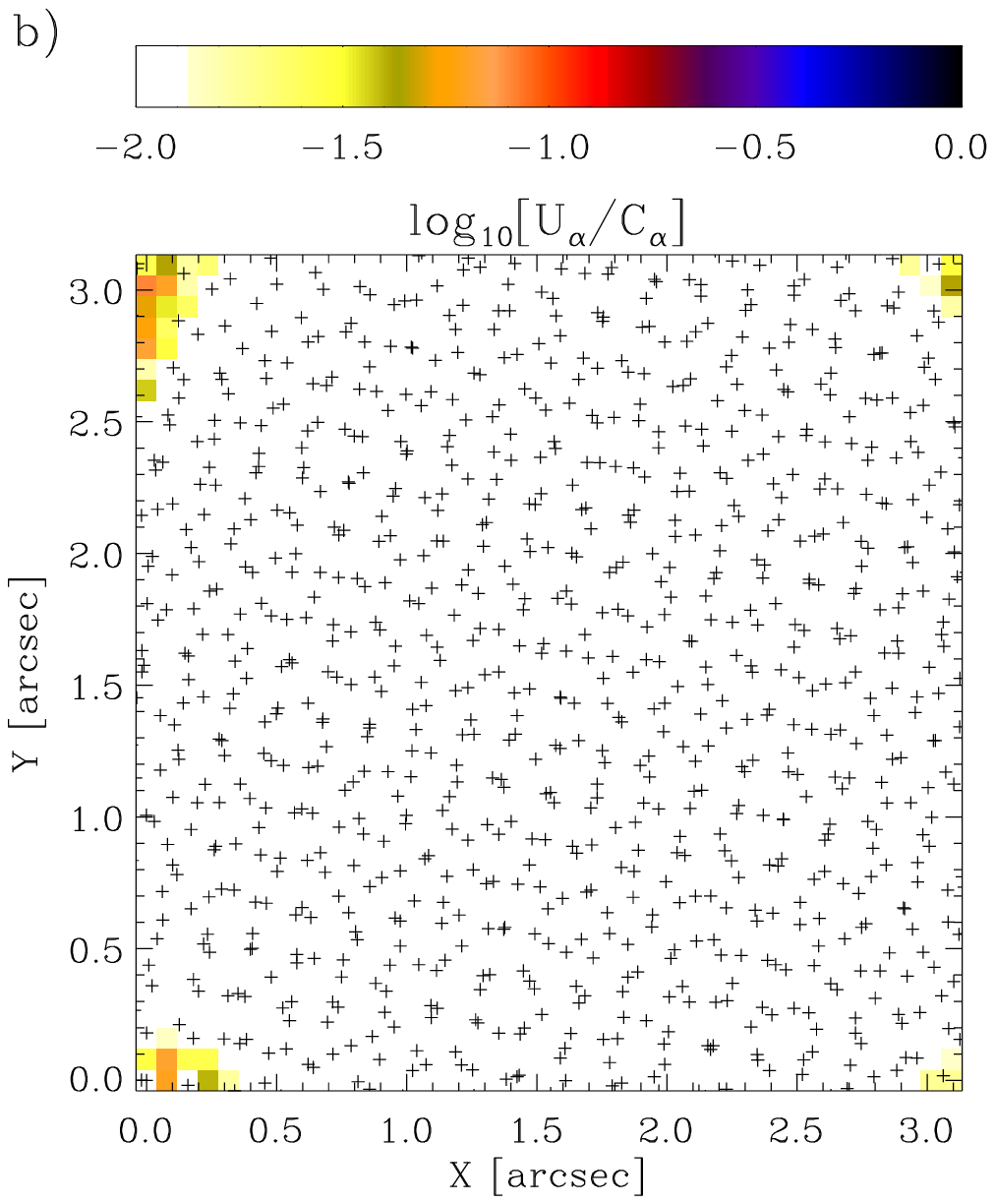}
\plottwo{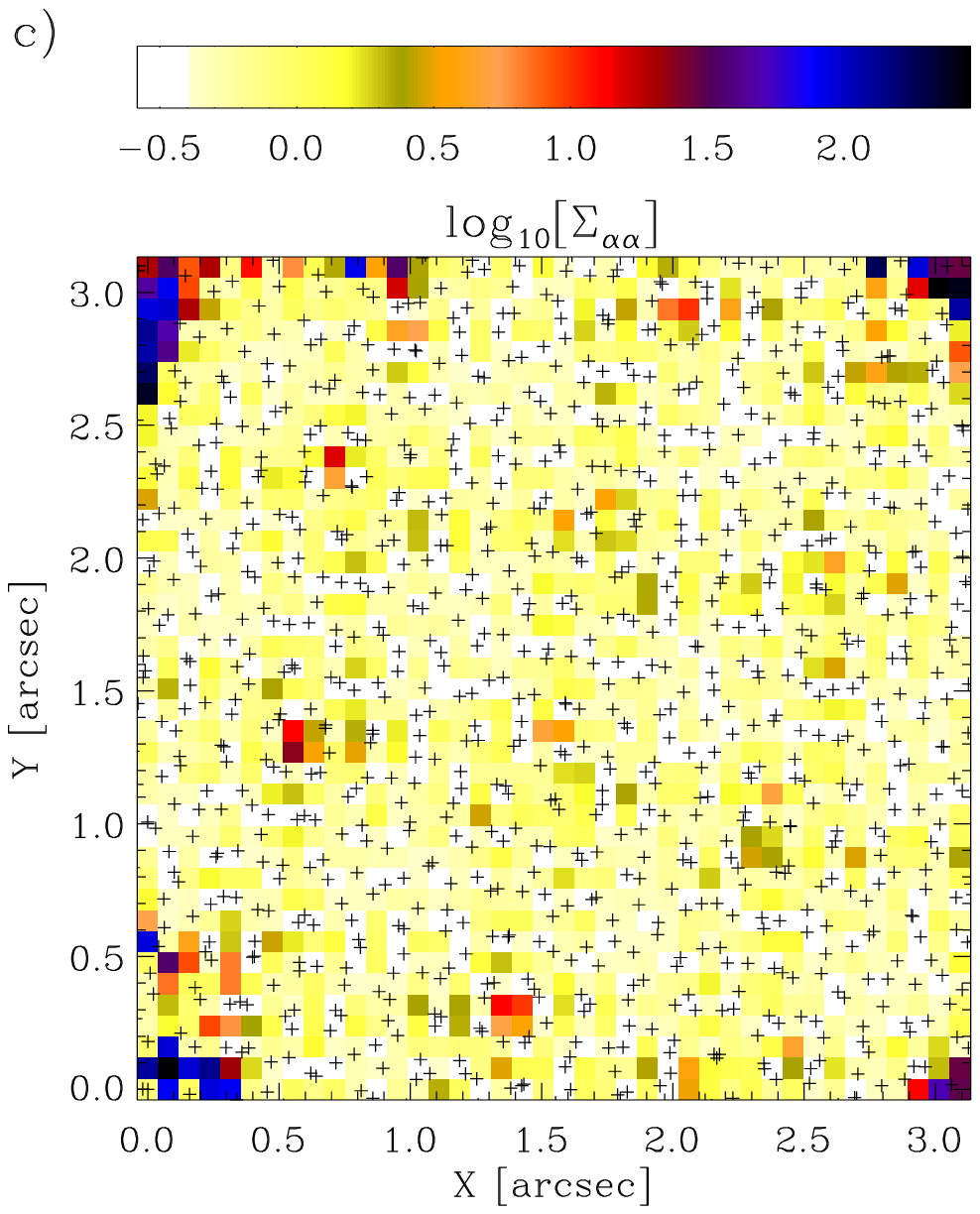}{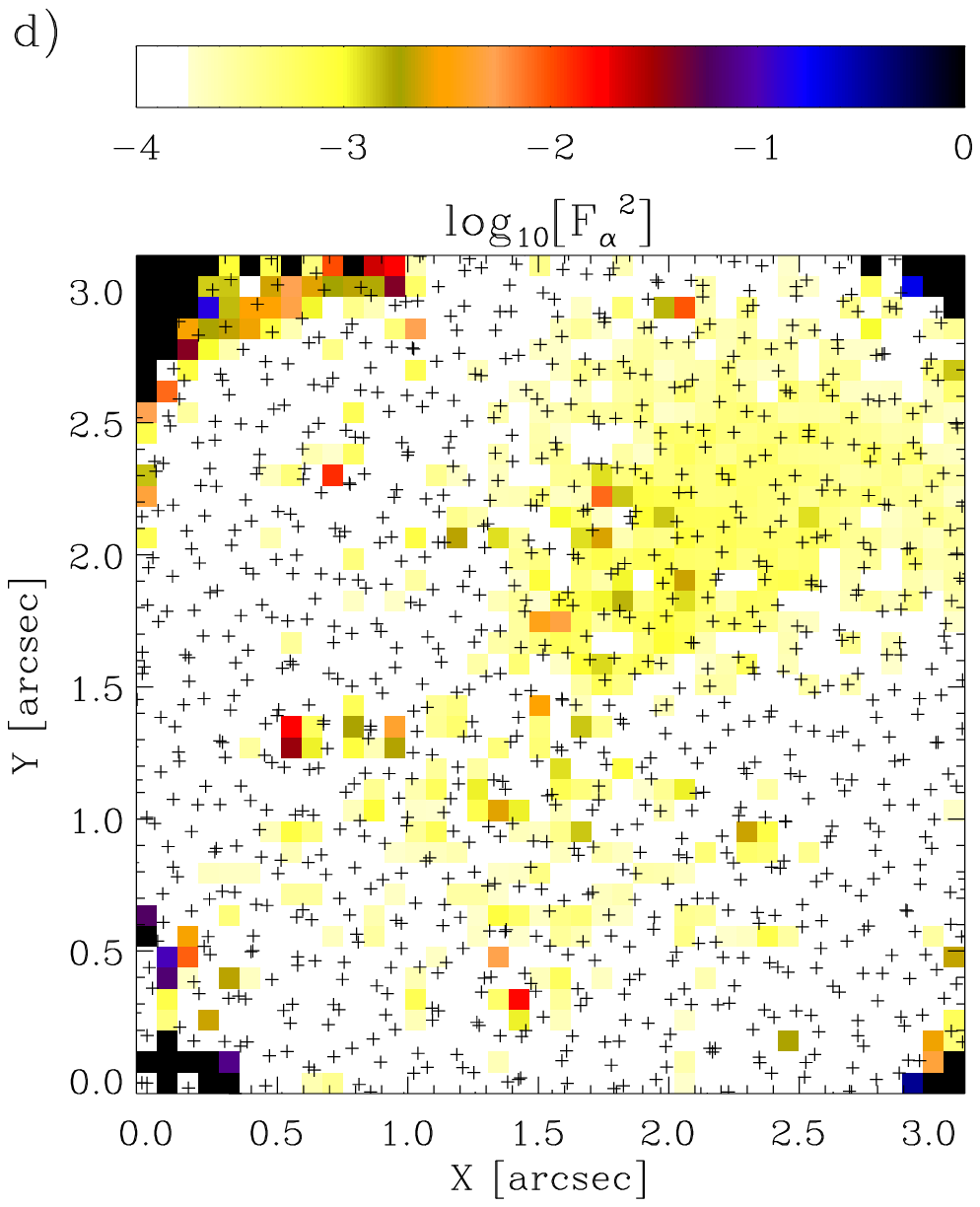}
\end{center}
\caption{Map of $\kappa_{\alpha}$ solution (a), output leakage objective $U_{\alpha}$ (b),
noise variance $\Sigma_{\alpha \alpha}$ (c), and corresponding $F^2_{\alpha}$ (d)
for the input choices $U_{\alpha}^{\textrm{max}} = 10^{-2}$,  $\Delta U_{\alpha}^{\textrm{max}} = 10^{-4}$ in the demonstration of Section \ref{sect:example}.  The failure to reach $U_{\alpha} \le 1 \times 10^{-2} C_{\alpha}$ in corner regions, even for $\kappa_{\alpha} = \kappa_{\alpha}^{\textrm{min}}$, can be clearly seen; the localization of peaks in the noise variance around encircled zones of reduced pixel coverage is also apparent. \label{fig:rand}}
\end{figure*}
We now examine a case in which the desired output image properties are specified in terms of the leakage objective $U_{\alpha}$ rather than the noise variance.  In Figure \ref{fig:rand} we plot results from the \textsc{Imcom} code having required $U_{\alpha}^{\textrm{max}} - \Delta U_{\alpha}^{\textrm{max}} \le  U_{\alpha} < U_{\alpha}^{\textrm{max}} $, where $U_{\alpha}^{\textrm{max}} = 10^{-2} C_{\alpha}$ and 
$\Delta U_{\alpha}^{\textrm{max}} = 10^{-4} C_{\alpha}$.  It can be seen from Figure \ref{fig:rand}b that in fact $U_{\alpha} > 10^{-2} C_{\alpha}$ in the sparsely-sampled corner regions of the output region.  Here in the corners the system has saturated at the lower limit for $U_{\alpha}$ within the broad range $[\kappa_{\alpha}^{\textrm{min}}, \kappa_{\alpha}^{\textrm{max}}]$, the practical minimum possible using double precision arithmetic and possibly approaching the theoretical minimum given by equation \eqref{eq:Utendzero}.

Figure \ref{fig:rand}c shows the output $\Sigma_{\alpha \alpha}$  for this solution: rapid variation is seen in the noise properties for this output image, with factors of $\sim 10$--100 greater variance than the input $N_{ij}$ seen in some regions.  These include the corner regions but also, as was the case for $U_{\alpha}$ in the previous example, encircled regions of sparse sampling that are reflected in the output $\Sigma_{\alpha \alpha}$.  There are also similar traces of these effects in the $F^2_{\alpha}$ map of Figure \ref{fig:rand}d, although these are most marked in the edge and corner regions where sparse input sampling combines with small values of $J_{\alpha}$ in the target image. 

Overall, it is seen that restricting $U_{\alpha} < 10^{-2} C_{\alpha}$ (a rather modest ambition if attempting precision scientific work) is difficult for the input image configuration and PSF of this example system.  While, as shown in Figure \ref{fig:randbyS}, it was possible to create an image with benign noise properties, this comes at a cost of significant $U_{\alpha}$ across much of the image.  Large values for $U_{\alpha}$ require significant leakage $L_{\alpha}({\bf R}_{\alpha} - {\bf r})$, and therefore imply a non-negligible, biasing residual $\langle Z_{\alpha} \rangle$.  The input sampling is too sparse to allow stable, unbiased reconstruction at this output resolution.  While this would have been apparent from the outset to someone with experience in stacking imaging data, the linear formalism allows it to be quantitatively diagnosed using $\Sigma_{\alpha \alpha}$ and $U_{\alpha}$ in a way that is independent of $f({\bf r})$.  This capability makes the method useful as a survey design tool.

In this Section we have attempted to demonstrate the flexibility and power of the linear image combination formalism developed in Section \ref{sect:linform} by tackling a difficult case.  For the three randomly-dithered exposure example, where the input sampling turns out to be insufficient for a fully-sampled, unbiased output, we have illustrated how the objective quantities $U_{\alpha}$ and $\Sigma_{\alpha \alpha}$ are used to understand the properties of $H_{\alpha}$.  In placing simple, user-specified constraints on these objective quantities we have shown how the \textsc{Imcom} implementation of Section \ref{sect:imp} can be used to efficiently explore the optimal trade-off between noise and fidelity, using a freely-varying $\kappa_{\alpha}$.  In trickier cases such as that presented here, there are further synthetic options that might be used to improve noise and bias in $H_{\alpha}$: specifying $\Gamma({\bf r})$ as a broader PSF is one possibility (see also Section \ref{sect:scalev}).

For the purposes of designing a dark energy mission, however, it is useful to address the problem in a slightly different manner.  Given a telescope PSF and an undersampling focal plane detector configuration, we may ask how many dithered exposures are needed, and in which pattern, to generate an unbiased, fully-sampled output $H_{\alpha}$.  This represents an important practical application of the techniques developed in Sections \ref{sect:linform} \& \ref{sect:imp}, and it is using this capacity as a design tool that we explore a range of dither patterns in the following Section.

\section{Tests of Multiple Observing Scenarios}\label{sect:wfirst}
In this Section we assess the ability of a number of dither pattern configurations to reconstruct fully-sampled images with a low level of leakage, a sure requirement for the precision photometry and shape measurement that will be necessary for the imaging component of a dark energy mission.   For the purposes of these tests we define low leakage as having $U_{\alpha} < 10^{-8} C_{\alpha}$, implying control of the output PSF properties to one part in $10^4$.  This ensures that uncontrolled changes to the PSF from linear image reconstruction will account for only a fraction of the total systematic error budget for a dark energy survey \citep{amararefregier08}.  We then examine the capability of the \emph{WFIRST} design concept described in Section \ref{sect:example}, with an optical PSF given by Figure \ref{fig:PSF} and a focal plane of H2RG detectors sampling at 0.18 arcsec, to generate fully-sampled outputs that are unbiased at this level of leakage.

The problem is set up as follows: We adopt $U_{\alpha}^{\textrm{max}} = 10^{-8} C_{\alpha}$ and $\Delta U_{\alpha}^{\textrm{max}} = 10^{-10} C_{\alpha}$ as our requirements on the properties of the output image $H_{\alpha}$, whereas other inputs such as the undersampling input pixel scale (0.18 arcsec), $G_i({\bf r})$, $\Gamma({\bf r})$, and the parameters of Table \ref{tab:params}, are all as described in Section \ref{sect:example} \emph{unless explicitly stated as otherwise}.  Using a realistic \emph{WFIRST} design concept in this way, we can begin to explore what survey design strategies may be needed to overcome undersampling for a dark energy mission of this sort.  We continue to set $N_{ij} = \mathbb{I}_n$ for the easy interpretation of $\Sigma_{\alpha \alpha}$ results. 

The output ${\bf R}_{\alpha}$ will again be placed at an interval of 0.079333 arcsec, the requirement for critical sampling.  In each following test we then vary the input pixel sampling information, adopting different dither patterns, numbers of exposures or detector coverage: in terms of the formalism, these variations are manifested as changes in ${\bf r}_i$ and $I_i$.  In this way we seek to show how the linear image combination formalism can be used to inform survey design, and make recommendations about survey dithering strategies for this \emph{WFIRST} design concept.

\subsection{The $2 \times 2$ Ideal Dither}\label{sect:2x2}
\begin{figure*}
\begin{center}
\plottwo{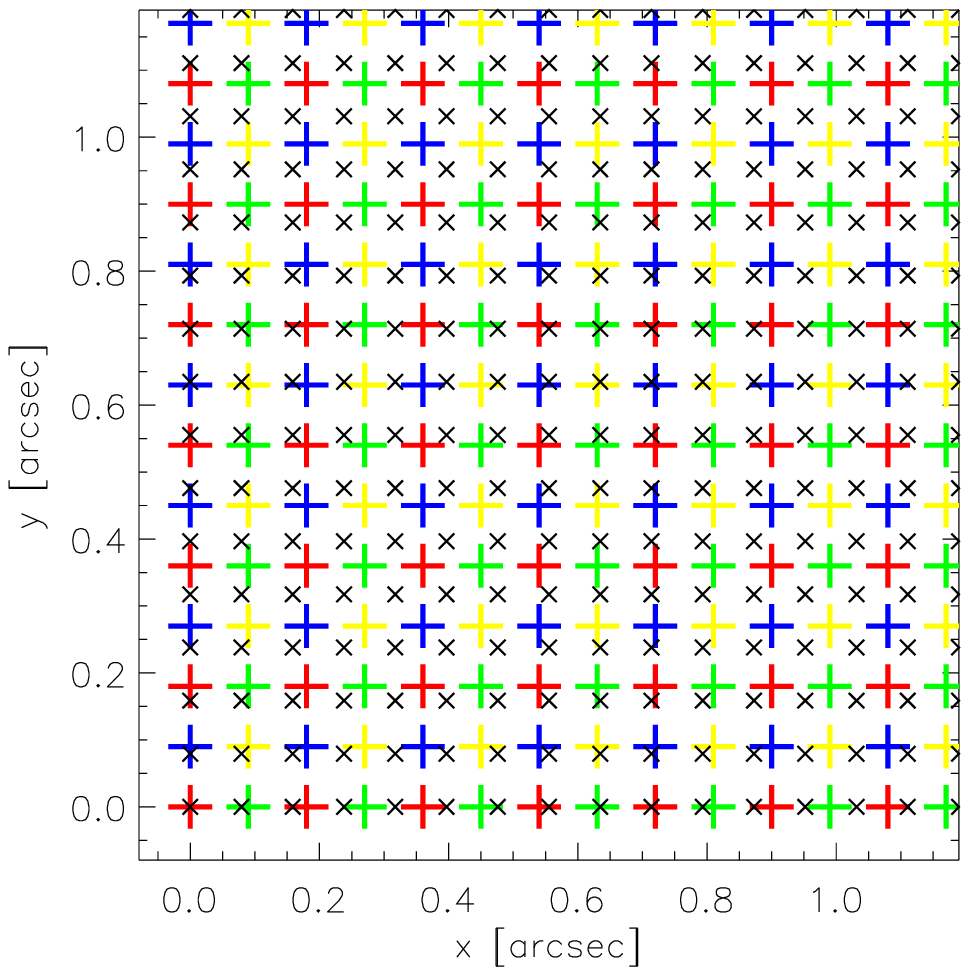}{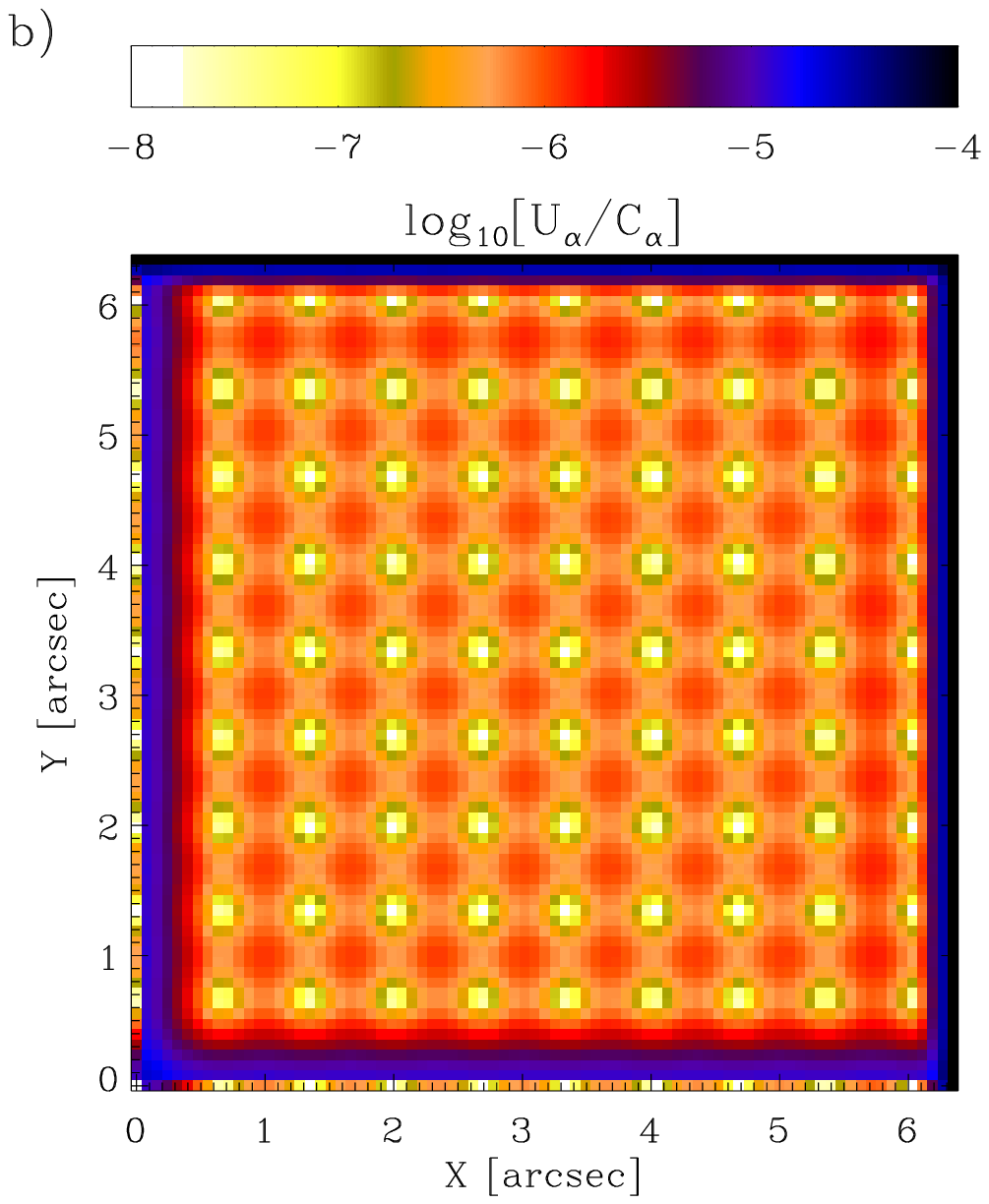}
\plottwo{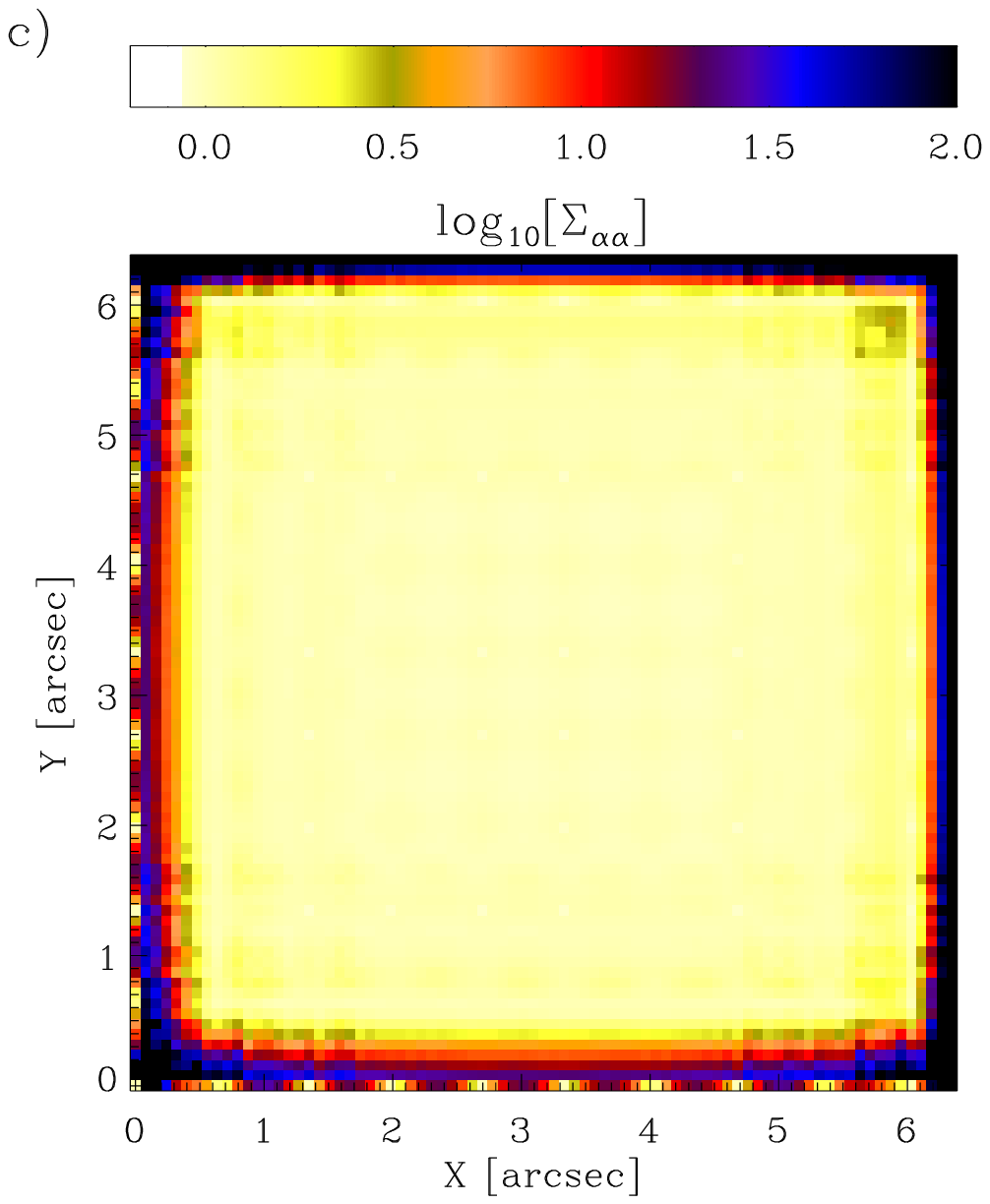}{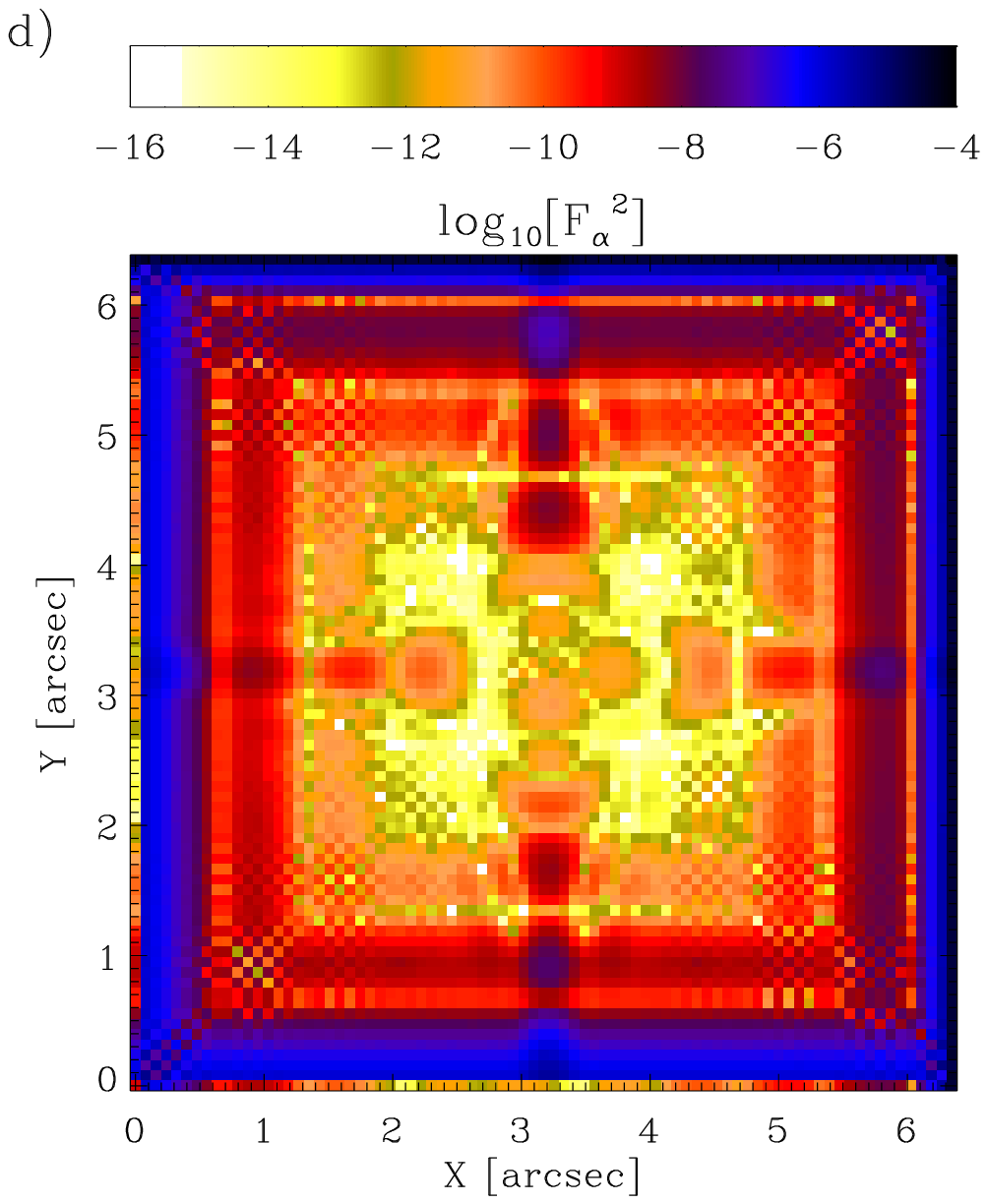}
\end{center}
\caption{Input pixel configuration (a) and leakage objective $U_{\alpha}$ (b), noise variance $\Sigma_{\alpha \alpha}$ (c), and squared fractional residual $F^2_{\alpha}$ (d) for the $2 \times 2$ ideal dither (see Section \ref{sect:2x2}). The color range chosen for $\Sigma_{\alpha \alpha}$ does not reflect the full range of values in the output, but highlights variations in the image center.}\label{fig:2x2}
\end{figure*}
In Figure \ref{fig:2x2} we plot results for four $36 \times 36$ pixel$^2$ input exposures, configured in a $2 \times 2$ ideal dither pattern.  The upper-left panel, Figure \ref{fig:2x2}a, illustrates this dithered configuration of input exposures (colored points) along with output sampling locations (black crosses), focused in upon a small sub-section of the total image area for clarity.  Figure \ref{fig:2x2}b shows the minimum leakage objective $U_{\alpha}$ that may achieved at fully-sampled resolution for this dither pattern; as can be seen, the threshold value of $U_{\alpha}^{\textrm{max}} = 10^{-8} C_{\alpha}$ is often exceeded (it is reached only where output pixels closely align with input pixel locations, indicated by the regularly spaced minima in $U_{\alpha}$).

The map in Figure \ref{fig:2x2}c shows the output noise variance $\Sigma_{\alpha \alpha}$ for the $2 \times 2$ dithered image.  Interestingly, this is both reasonably stationary (i.e.\ approximately constant), and comparable in amplitude to the input noise variance, across a large area of the output.  We note that in this Figure, and in many subsequent plots of $\Sigma_{\alpha \alpha}$, we choose the color scale not to show the full dynamic range of $\Sigma_{\alpha \alpha}$ but to better highlight behavior in the central regions of the image.  

At the extreme edges of the output, where input information is most scarce, $\Sigma_{\alpha \alpha}$ may exceed $10^5$. The size of such noisy edge regions is primarily determined by the extent of the PSF $G_i({\bf r})$ within each contributing exposure.  However, as will be discussed in Section \ref{sect:resources}, these regions are not likely to be used in a realistic survey strategy: Figure \ref{fig:patches} illustrates that it is not necessary for these edge regions to fall within the output $H_{\alpha}$ at all.  Instead, by tessellating successive $H_{\alpha}$ that cover only the central regions of overlapping input patches, a contiguous output image can be constructed for which all pixels are sufficiently far from an input edge.  In this Section, however, these noisier edge regions are kept so as to demonstrate the effect.

Figure \ref{fig:2x2}d shows the map of $F^2_{\alpha}$ across the image. The cruciform pattern seen in $F^2_{\alpha}$ is due to the light profile of the input $f({\bf r})$ seen in Figure \ref{fig:fgal}: this fact was verified by varying the centroid position of $f({\bf r})$, which was seen lead to equivalent changes in the position of the cruciform.  It can be seen that $F^2_{\alpha} < U_{\alpha} / C_{\alpha}$ across much of the output for this semi-realistic galaxy image.

\subsection{The $\sqrt{5} \times \sqrt{5}$ Ideal Dither}\label{sect:rt5xrt5}
\begin{figure*}
\begin{center}
\plottwo{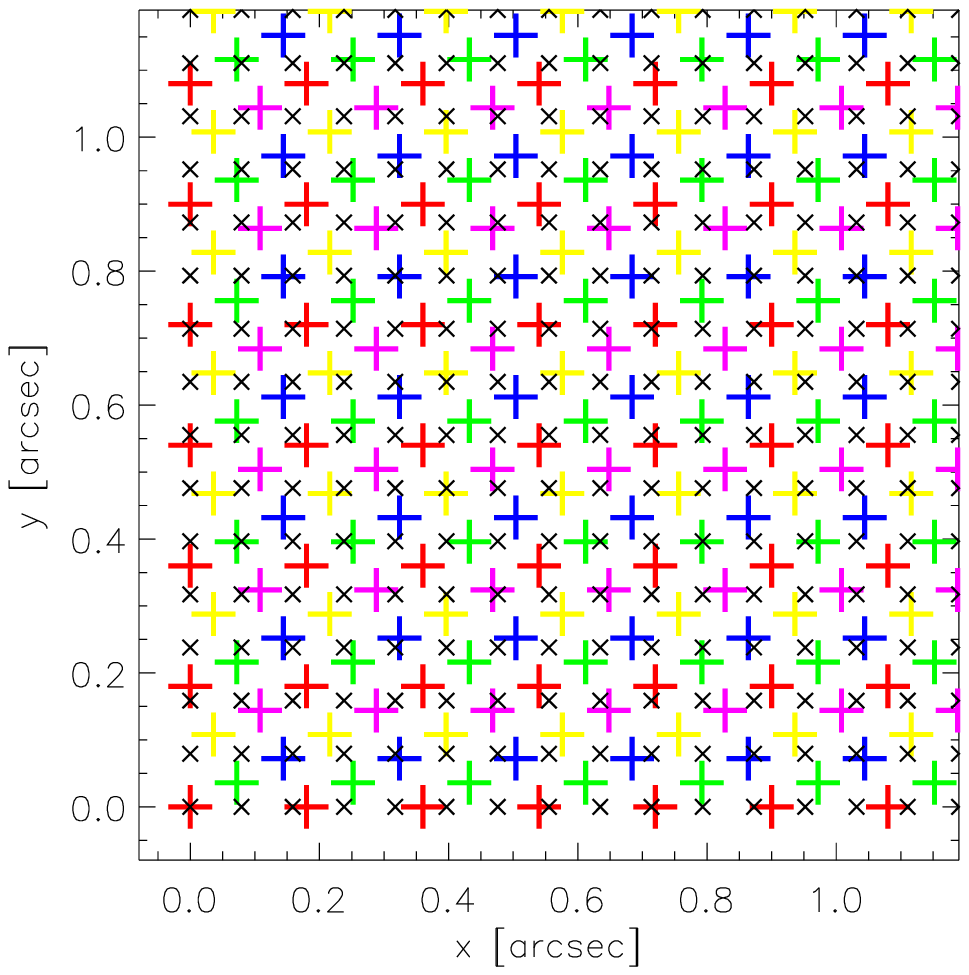}{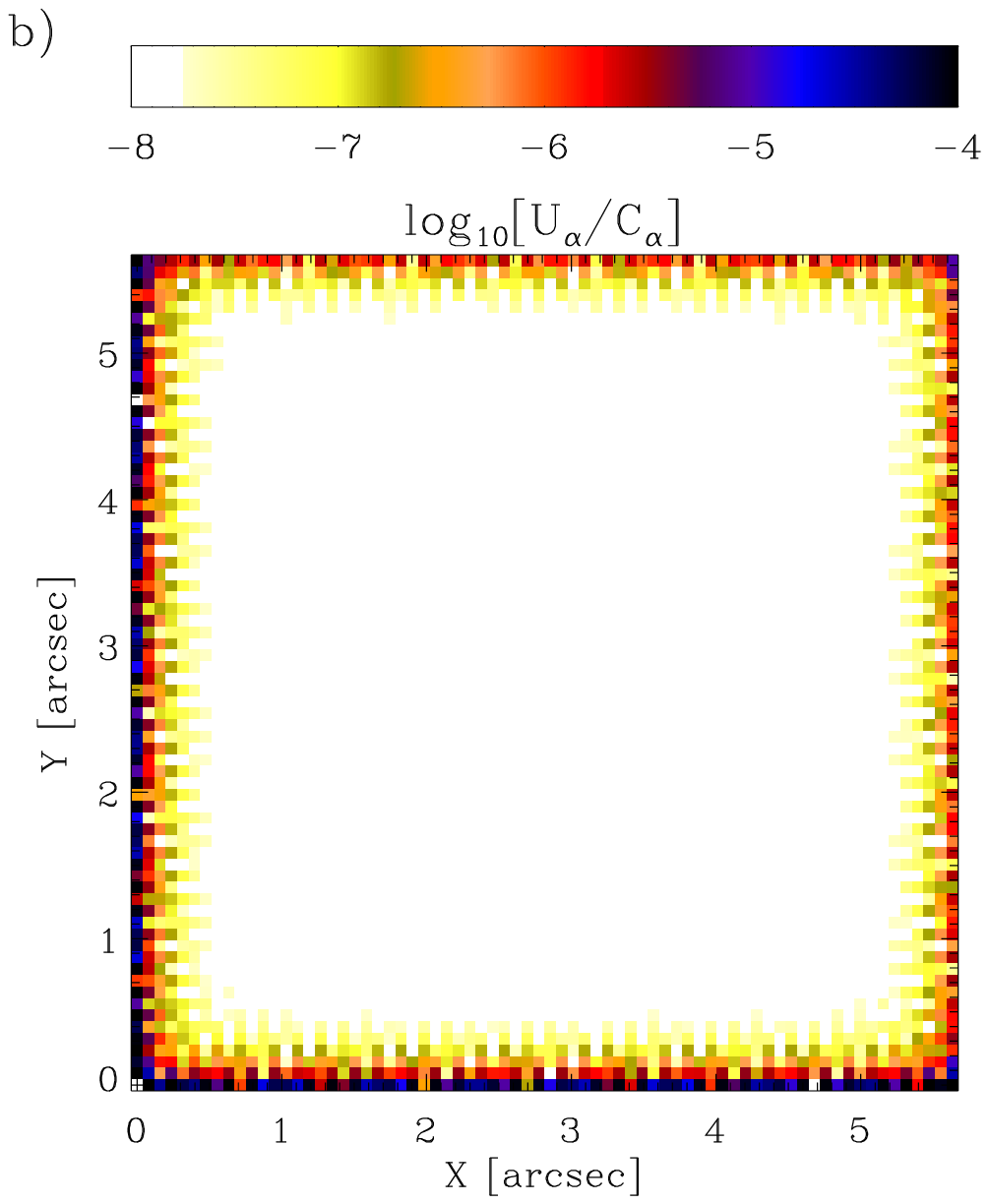}
\plottwo{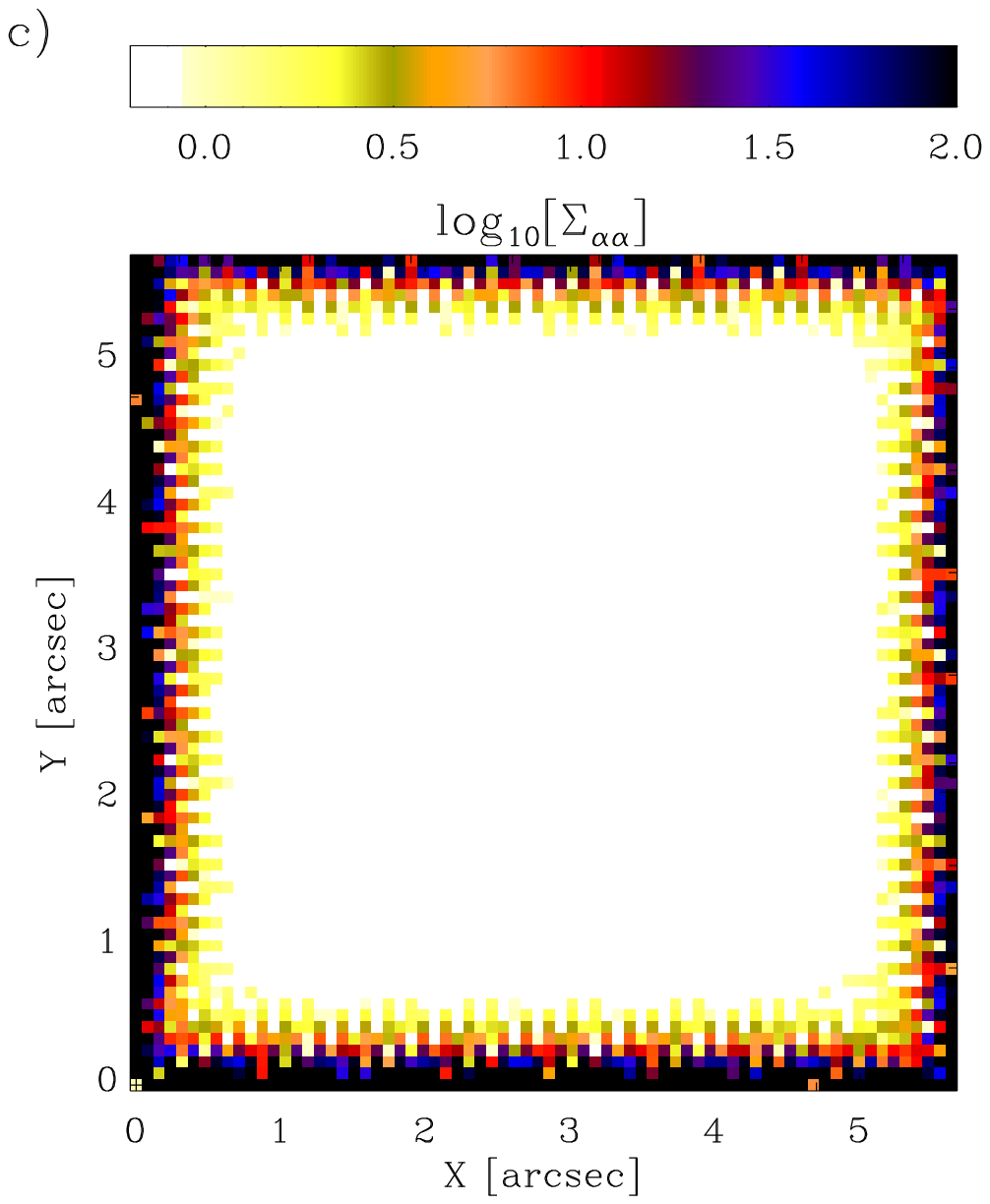}{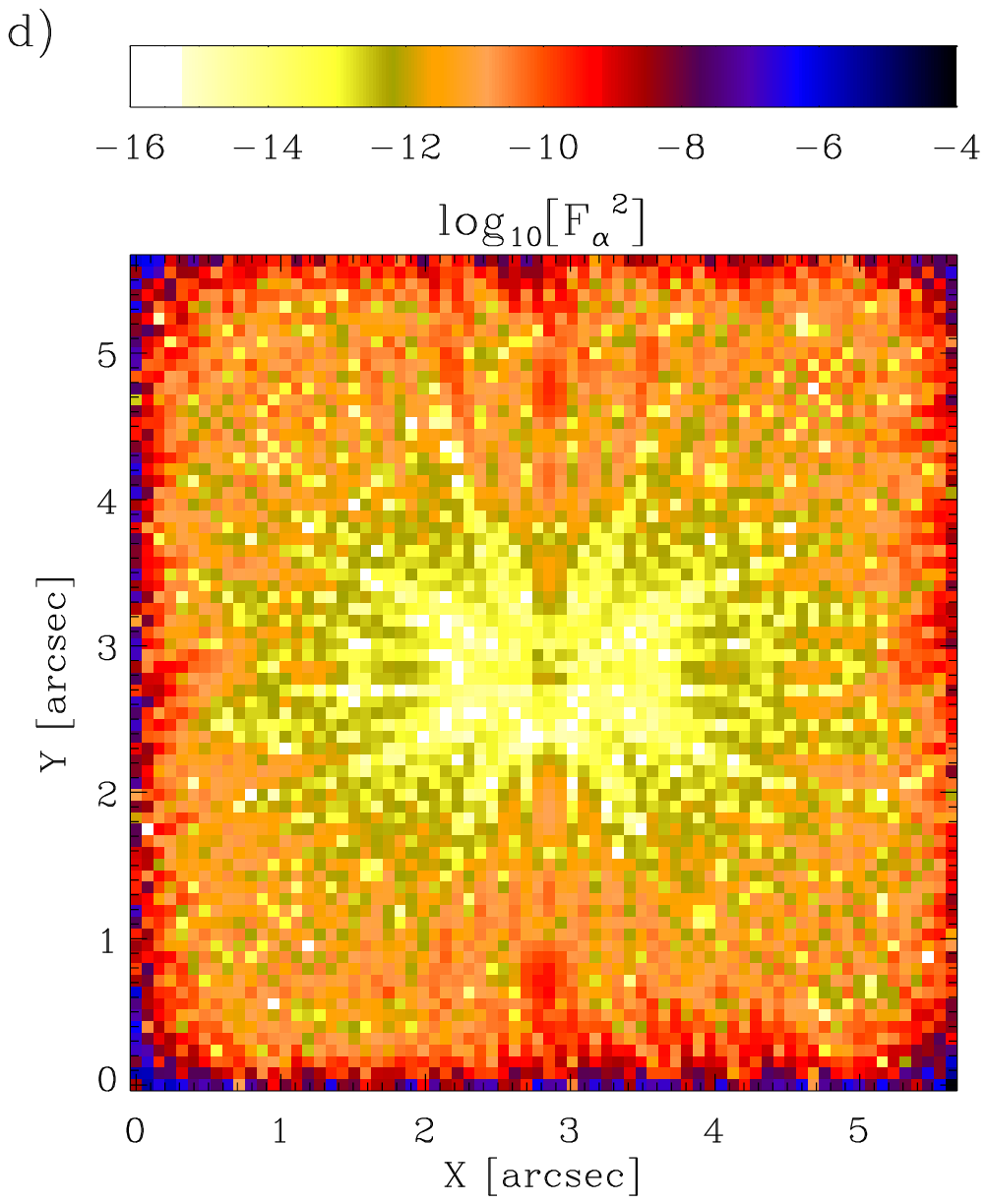}
\end{center}
\caption{Input pixel configuration (a), leakage objective $U_{\alpha}$ (b), noise variance $\Sigma_{\alpha \alpha}$ (c) and squared fractional residual $F^2_{\alpha}$ (d)  for the $\sqrt{5} \times \sqrt{5}$ ideal dither (see Section \ref{sect:rt5xrt5}). The color range chosen for $\Sigma_{\alpha \alpha}$ does not reflect the full range of values in the output, but highlights variations in the image center.}\label{fig:rt5xrt5}
\end{figure*}
In Figure \ref{fig:rt5xrt5} we plot results for five $32 \times 32$ pixel$^2$ input exposures, configured in a $\sqrt{5} \times \sqrt{5}$ ideal dither pattern.  The upper-left panel, Figure \ref{fig:rt5xrt5}a, illustrates this configuration of input exposures for a small sub-section of the total image area.  This is an interesting dither pattern which arranges five exposures per unit cell in a slanted, but regular, grid configuration.  The pattern has chirality: a reflection about the line $y = x$ produces a different, but functionally equivalent, set of dithers.

\begin{figure*}
\begin{center}
\plottwo{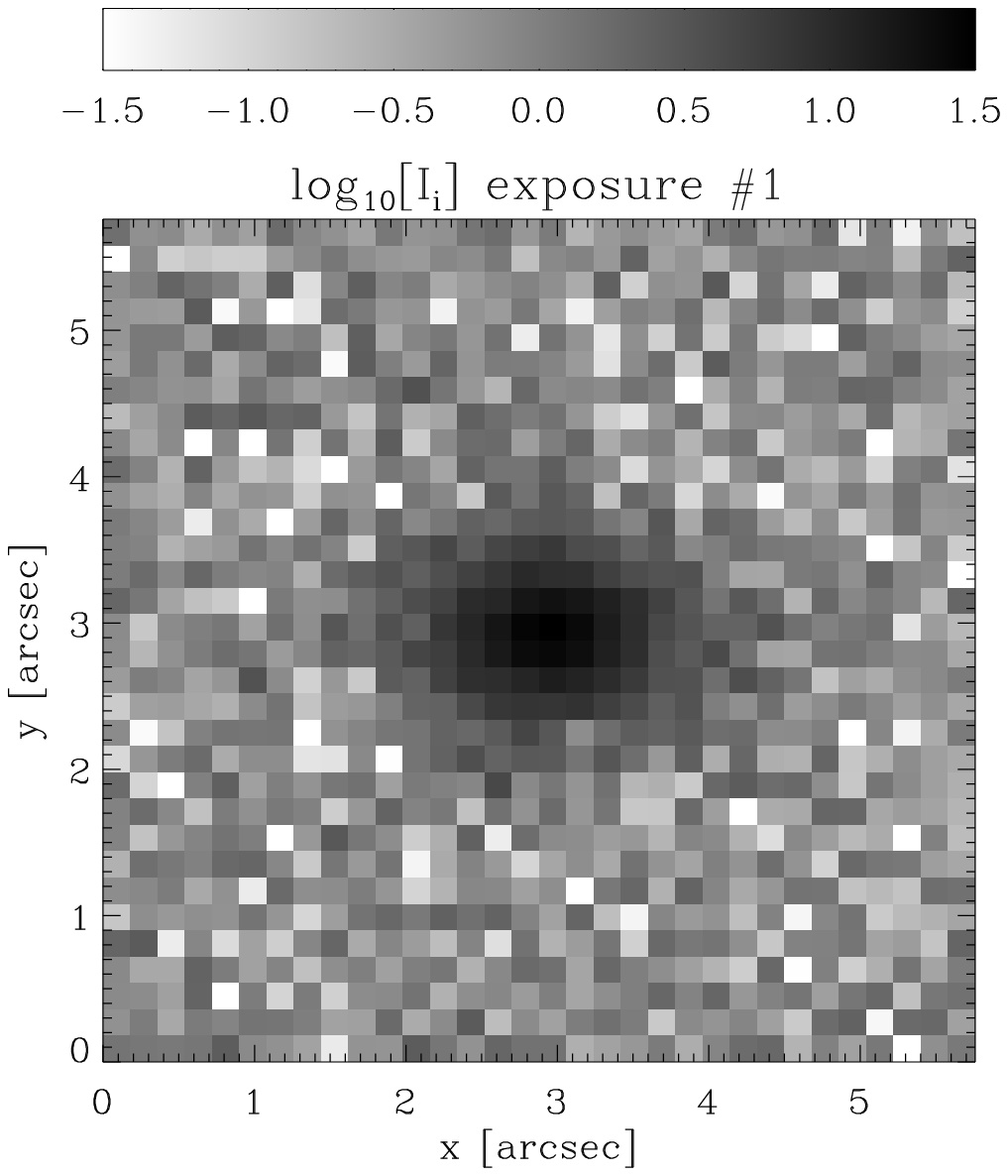}{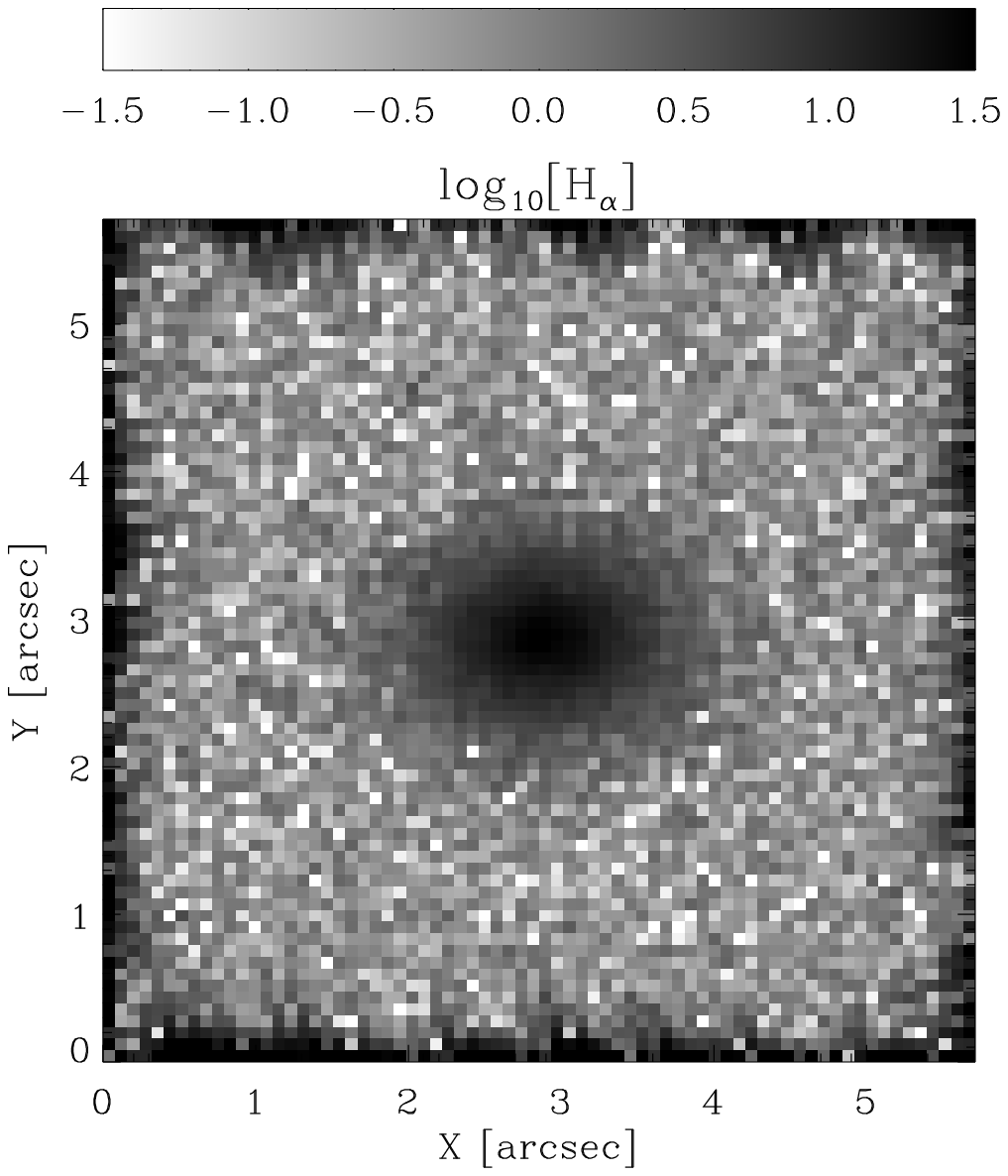}
\end{center}
\caption{Left panel: One of five \emph{noisy} input images dithered using the $\sqrt{5} \times \sqrt{5}$ ideal dither pattern of Section \ref{sect:rt5xrt5}, constructed by convolving $f({\bf r})$ with $G_i({\bf r})$ and adding a stochastic noise $\eta_i$ to each pixel $I_i$.  The noise added was drawn from a Gaussian distribution with unit variance and zero mean.  Right panel: The combined output image $H_{\alpha}$.  Both images show the logarithm of the absolute value of each $I_i$ and $H_{\alpha}$, to most clearly depict the central, edge, and background noise-dominated regions of each image. }\label{fig:rt5xrt5noise}
\end{figure*}
Figure \ref{fig:rt5xrt5}b shows the $U_{\alpha}$ achieved for this $\sqrt{5} \times \sqrt{5}$ dither pattern: across the broad central region of the output image the requirement $U_{\alpha} < 10^{-8} C_{\alpha}$ is now successfully met.  Figures \ref{fig:rt5xrt5}c \& d show the noise variance $\Sigma_{\alpha \alpha}$ and squared fractional residual $F^2_{\alpha}$. The former is seen to be approximately stationary, and comparable to the input noise variance, across the central regions of the output.  The values of $F^2_{\alpha}$ again form a vaguely-cruciform pattern that moves to align with the centroid of $f({\bf r})$ when this is shifted.  Overall, the $\sqrt{5} \times \sqrt{5}$ dither pattern is able to successfully reconstruct a fully-sampled output image for this \emph{WFIRST} design concept while preserving the input convolution kernel $G_i({\bf r})$ at a level $U_{\alpha} < 10^{-8} C_{\alpha}$.  A question remains as to whether a pattern such as this can be used to produce fully-sampled data in the presence of image pixel losses due to, e.g., cosmic rays, hot pixels, and other defects, or in the presence of variations in the focal plane plate scale: these issues are explored in Sections \ref{sect:rt5xrt5badpix} \& \ref{sect:scalev}.

For illustrative purposes, we show in Figure \ref{fig:rt5xrt5noise} an example of an input exposure and the \textsc{Imcom}-generated output for the $\sqrt{5} \times \sqrt{5}$ ideal dither, where actual stochastic noise $\eta_i$ was added to the input pixels $I_i$ before processing.  The noise added was Gaussian with unit variance.  The output $H_{\alpha}$ clearly shows the effects of increased $\Sigma_{\alpha \alpha}$ in edge regions, but demonstrates the successful image reconstruction at the image center.

\subsection{The 6 Exposure Random Dither}\label{sect:ran6}
\begin{figure*}
\begin{center}
\plottwo{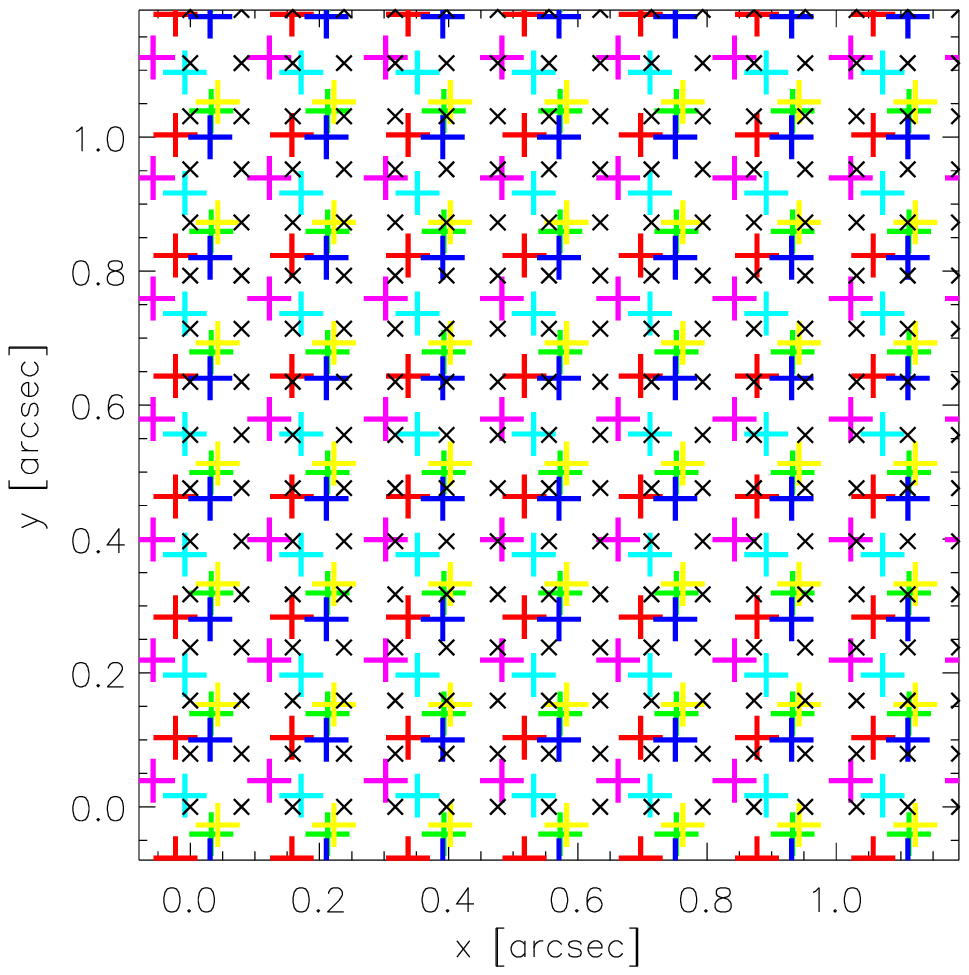}{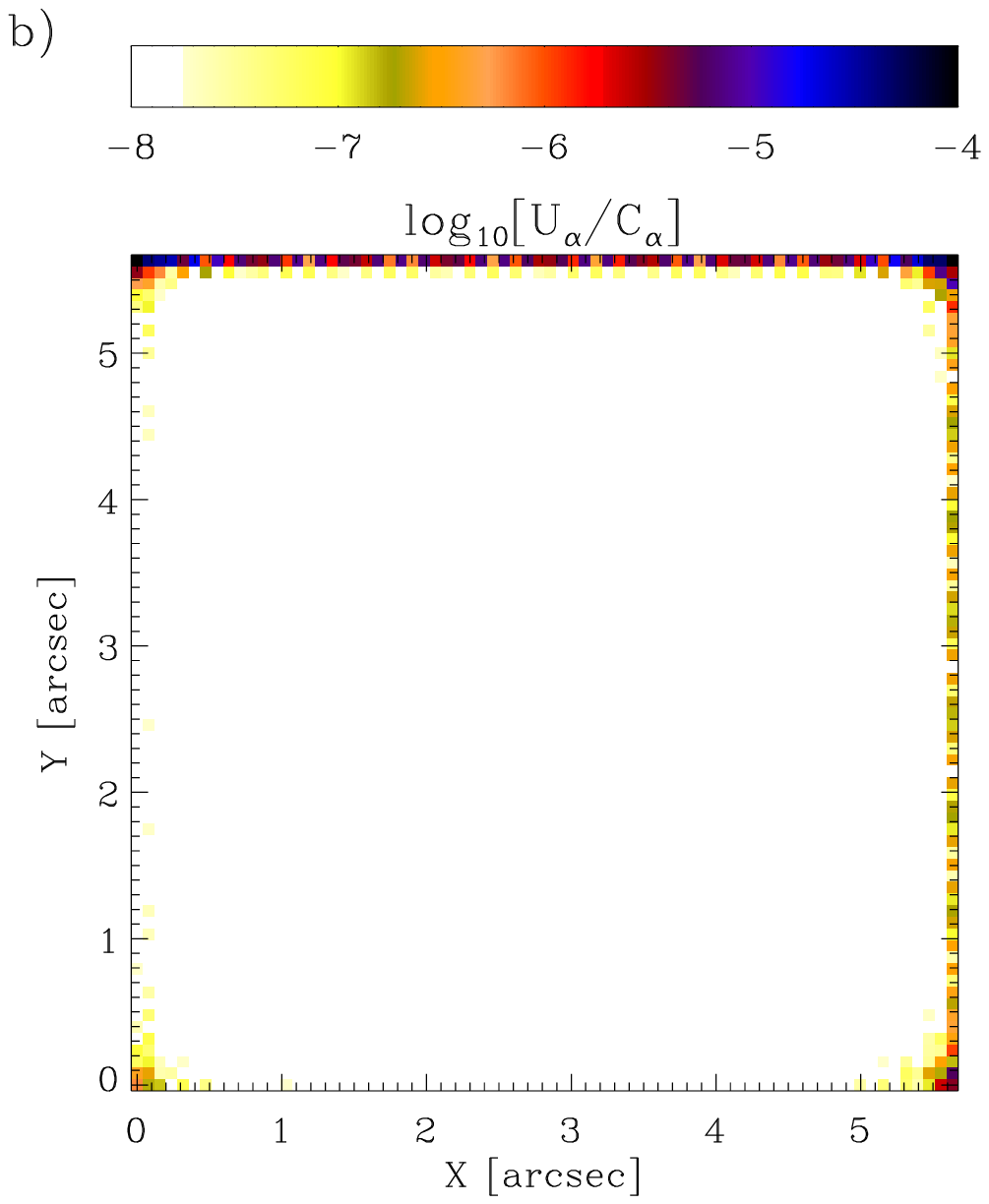}
\plottwo{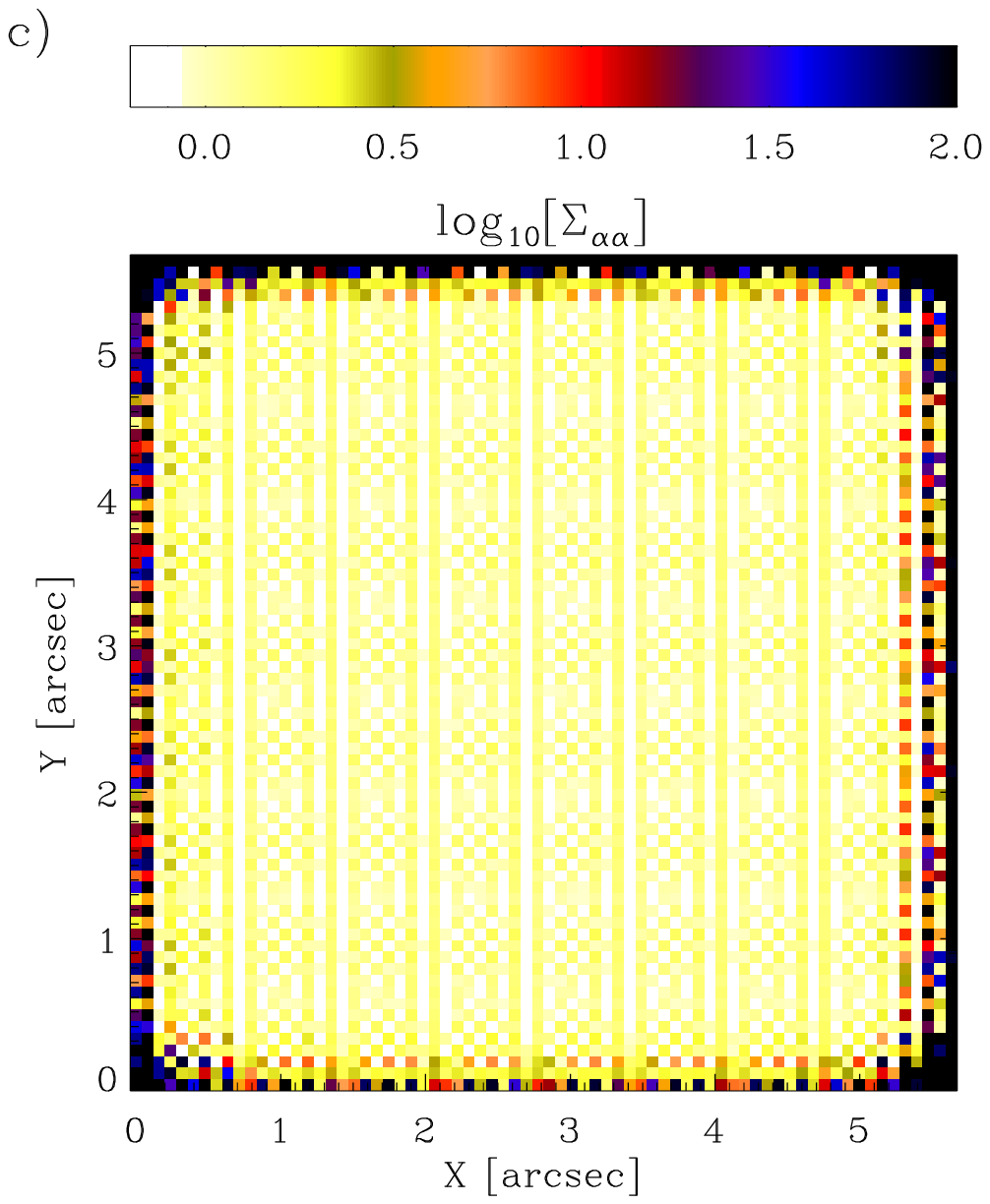}{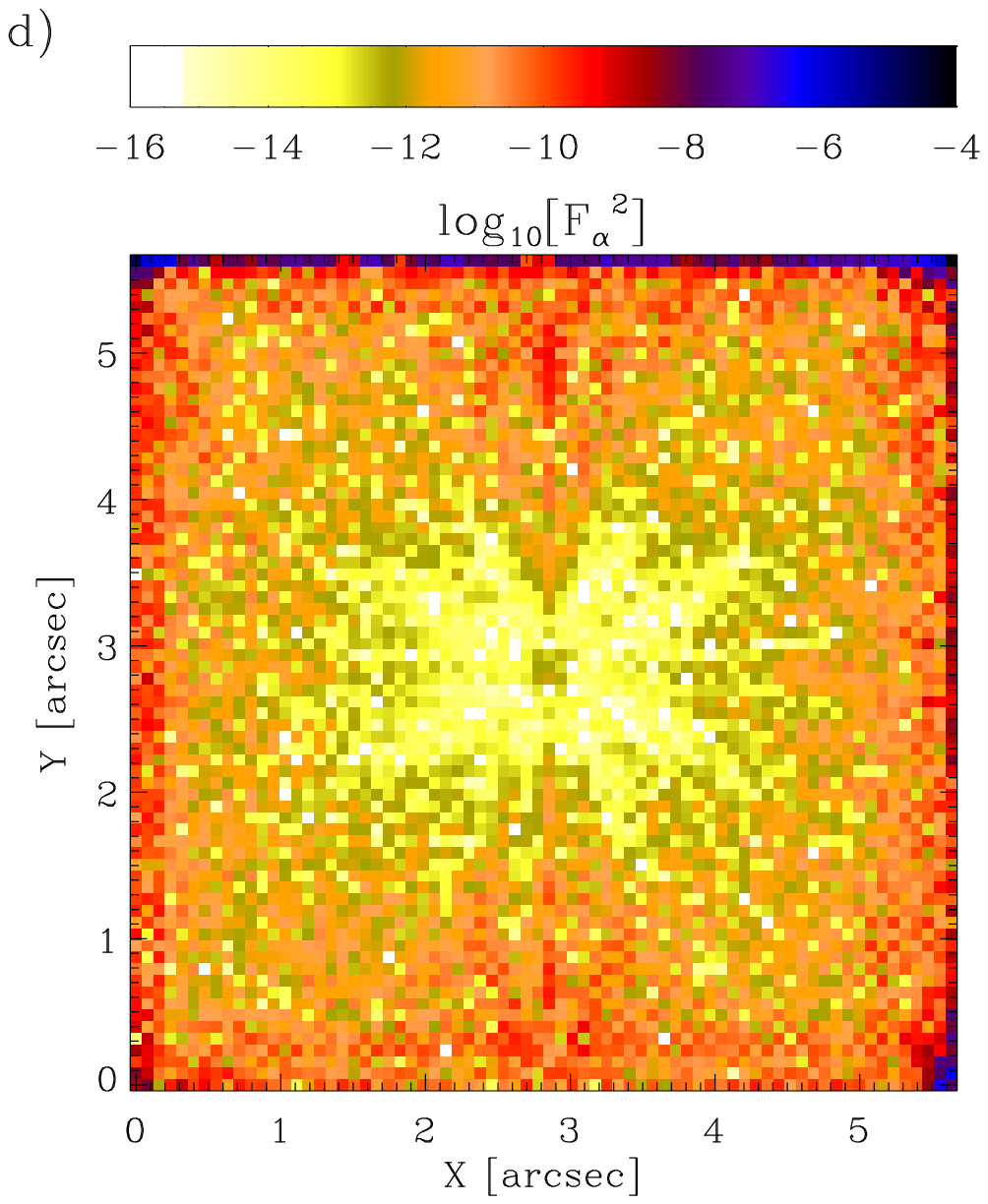}
\end{center}
\caption{Input pixel configuration (a), leakage objective $U_{\alpha}$ (b),
 noise variance $\Sigma_{\alpha \alpha}$ (c) and squared fractional residual $F^2_{\alpha}$ (d) for the 6 exposure random dither (see Section \ref{sect:ran6}). The color range chosen for $\Sigma_{\alpha \alpha}$ does not reflect the full range of values in the output, but highlights variations in the image center.}\label{fig:ran6}
\end{figure*}
In Figure \ref{fig:ran6}  we plot results for six $32 \times 32$ pixel$^2$ input exposures, configured in a randomly offset dither pattern; unlike in Section \ref{sect:example} there is no random roll in this example.  Figure \ref{fig:ran6}a illustrates this configuration of input exposures for a small sub-section of the total image area, along with the fully-sampling output locations.
The 5 exposure random dither was also tested, but produced poor results with $U_{\alpha} > 10^{-8} C_{\alpha}$ across much of the output in a small suite of randomly offset trial configurations.  One such test, not atypical of the small set of trials, produced $\Sigma_{\alpha \alpha} > 10$ for 65\% of the total output and $\Sigma_{\alpha \alpha} > 100$ for 22\% of the total output.

However, the leakage objective for the 6 exposure random dither in Figure \ref{fig:ran6}b shows good agreement with stated requirements across a large fraction of the output sample locations.  The $\Sigma_{\alpha \alpha}$ variance in Figure \ref{fig:ran6}b is also now significantly smaller than that which was seen in the five exposure dithers: an impressive impact from only a single additional exposure.   As was done for the 5 exposure dithers, these results were repeated internally for a small set of random 6 exposure dither patterns, and the results shown are typical of the results achieved.  It should be noted that the noise $\Sigma_{\alpha \alpha}$ is somewhat non-stationary, varying with ${\bf R}_{\alpha}$ across the output. 

As has been seen, good results have been achieved for a controlled reconstruction of fully-sampled images using certain configurations of input \emph{WFIRST}-like images.  The $\sqrt{5} \times \sqrt{5}$ ideal dither performed well, as did a single realization of a random dither of six exposures (at the cost of slightly inferior noise properties).  While the results shown for the random patterns were typical of a small internal test set of such patterns, the repeatability and range of these random dither results should be explored in more detail in subsequent work. 
Another investigation, which is within the scope of this Paper, is the robustness of linear image combination under identified image or pixel defects.  This is now explored.

\subsection{Cosmic Ray Impacts and Bad Pixels}\label{sect:rt5xrt5badpix}
We perform a very simple test of linear image combination under the presence of known image defects, such as might be caused by cosmic rays, dead pixels, and similar phenomena.  Taking the input images for the $\sqrt{5} \times \sqrt{5}$ ideal dither of Section \ref{sect:rt5xrt5}, we randomly select $5\%$ of input pixels to be flagged as not for use in the reconstruction of the input images.  This equates simply to a contraction of the $A_{\alpha i j}$ and $B_{\alpha i}$ by removal of the afflicted pixel rows and columns.
\begin{figure*}
\begin{center}
\plottwo{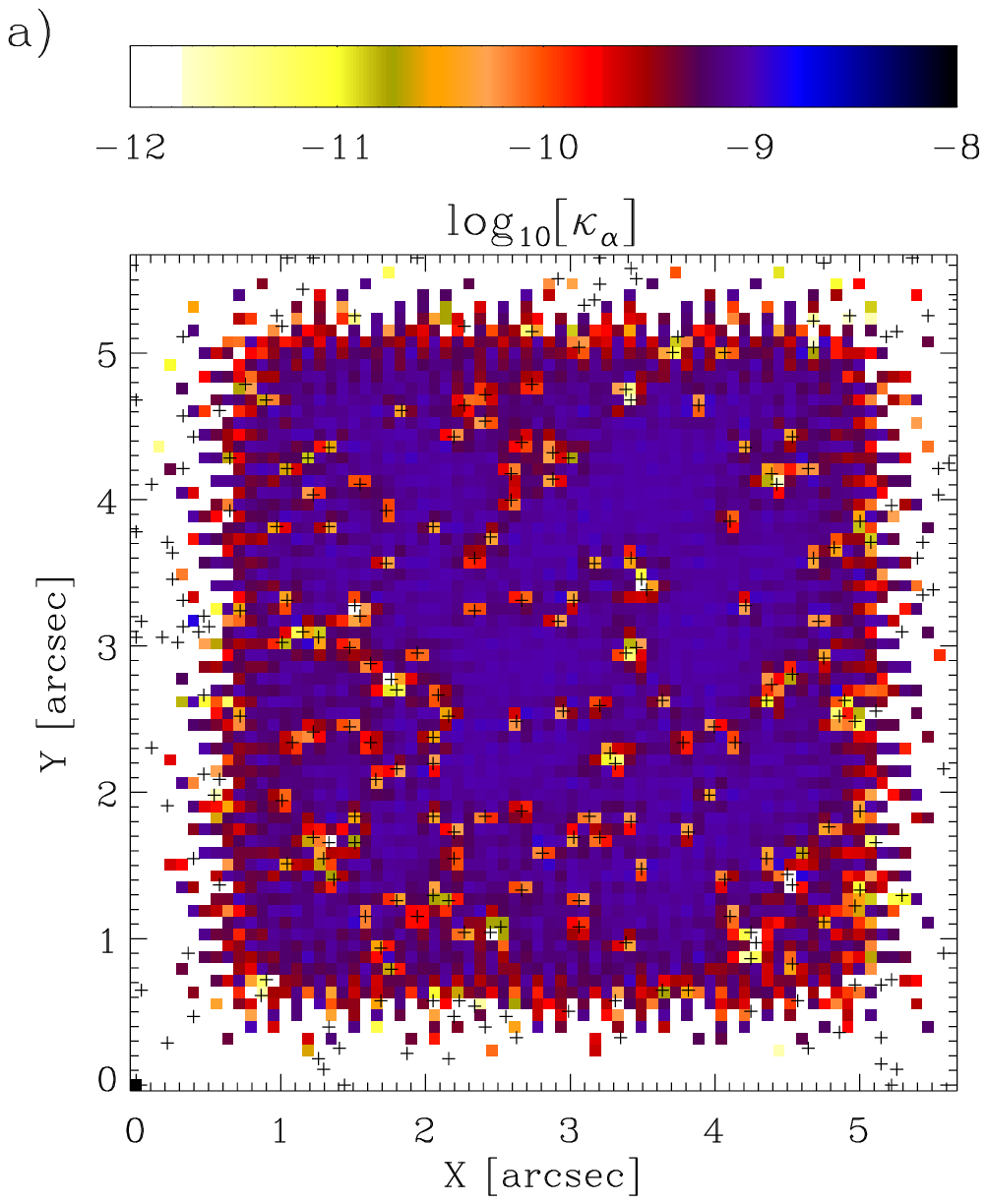}{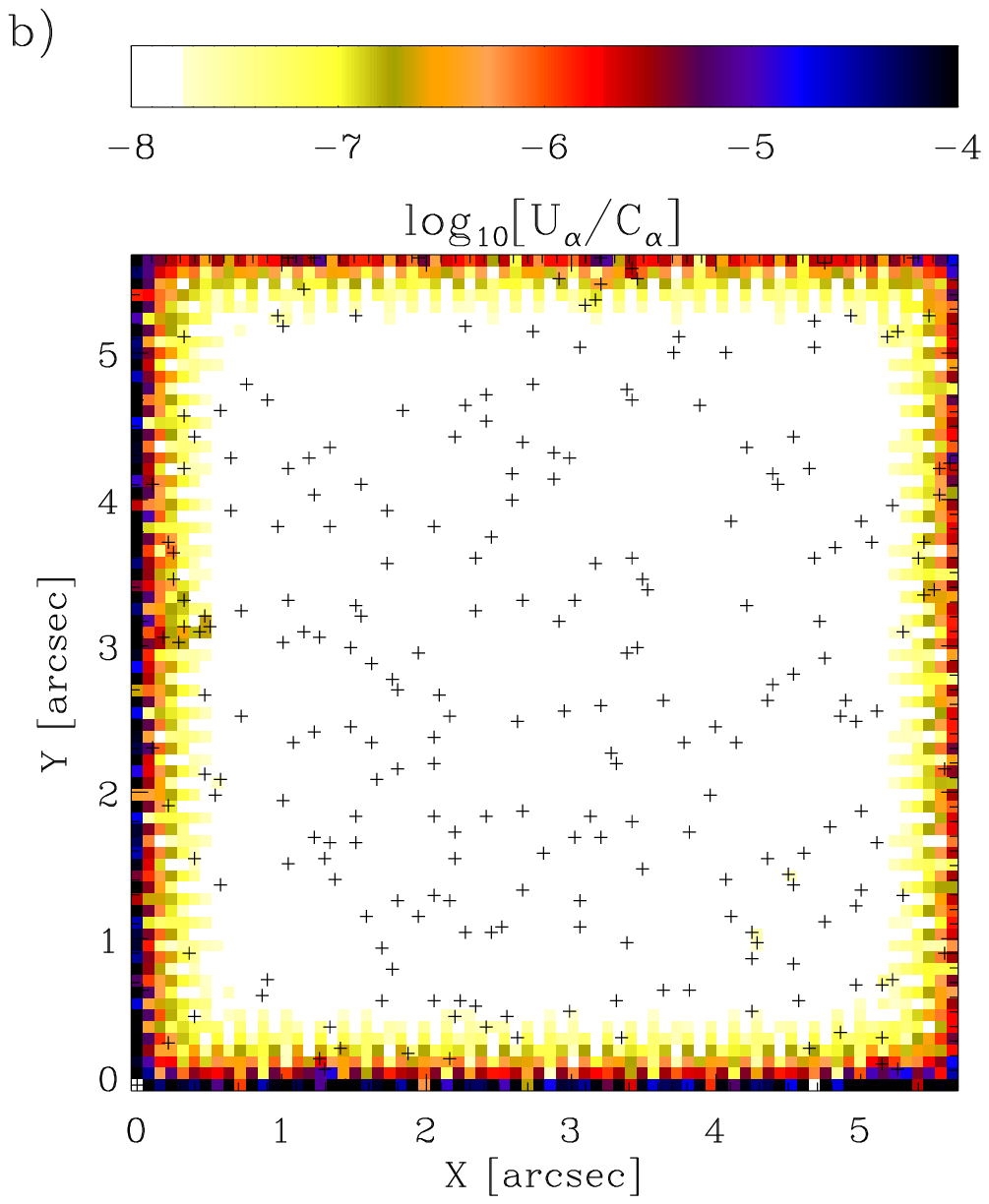}
\plottwo{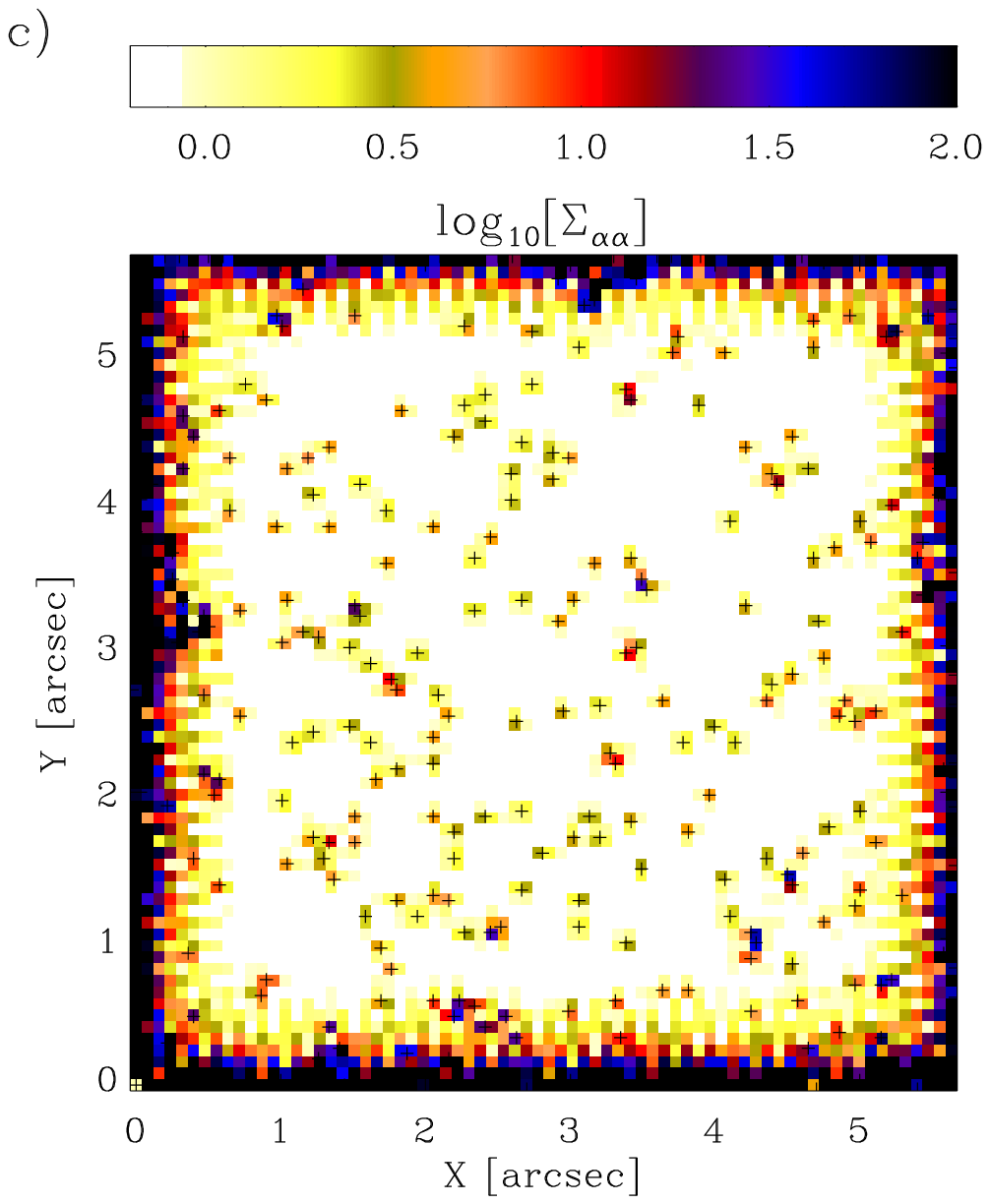}{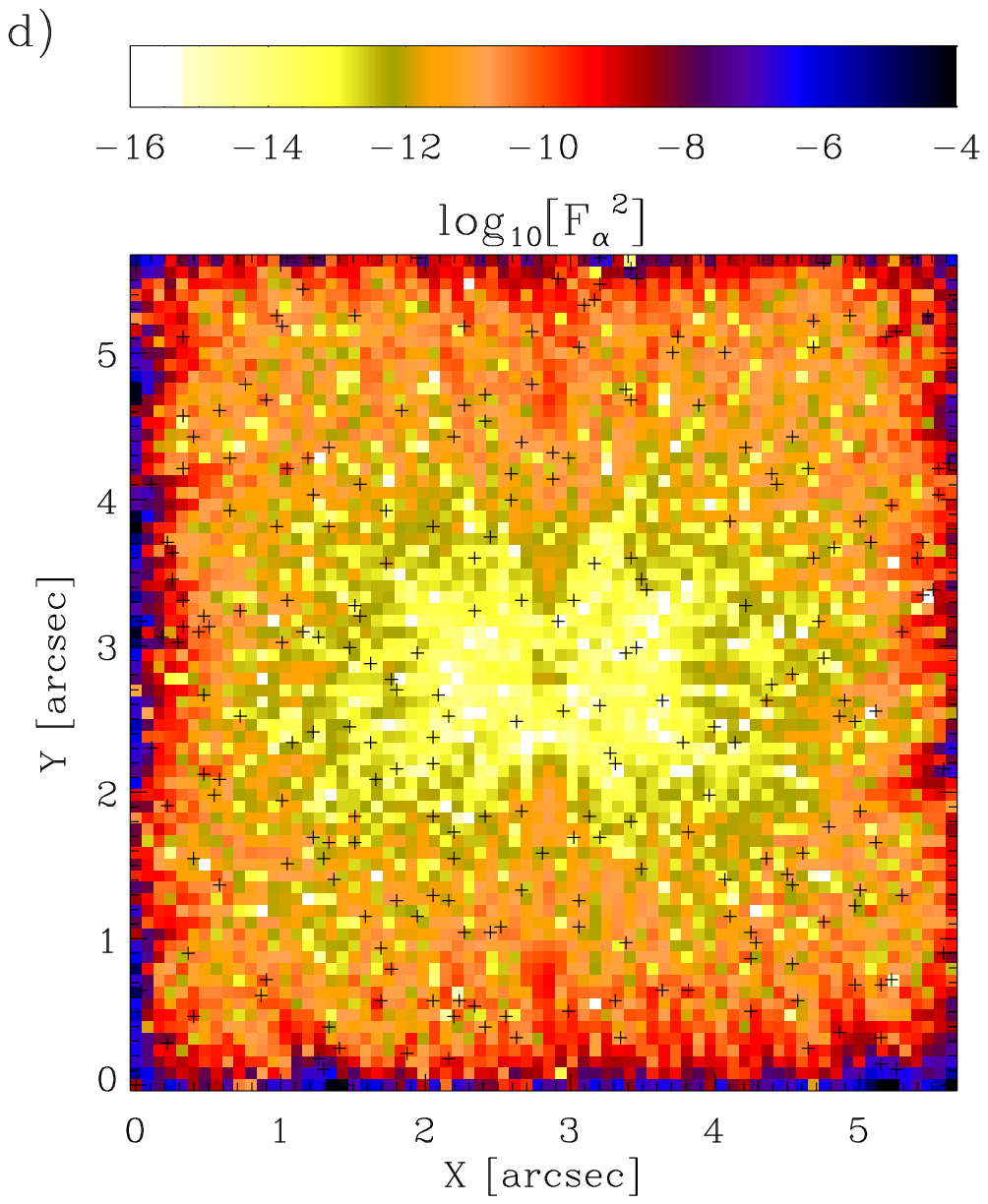}
\end{center}
\caption{Map of $\kappa_{\alpha}$ solution (a), leakage objective $U_{\alpha}$ (b), noise variance $\Sigma_{\alpha \alpha}$ (c), and squared fractional residual $F^2_{\alpha}$ (d) for the $\sqrt{5} \times \sqrt{5}$ ideal dither after losing a randomly-selected 5\% fraction of input pixels (See Section \ref{sect:rt5xrt5badpix}).  Input pixels that were not used are indicated using small black crosses. The color range chosen for $\Sigma_{\alpha \alpha}$ does not reflect the full range of values in the output, but highlights variations due to bad pixels in the image center.\label{fig:rt5xrt5badpix}}
\end{figure*}

In Figure \ref{fig:rt5xrt5badpix}a we plot the map of $\kappa_{\alpha}$ solution values along with, in Figure \ref{fig:rt5xrt5badpix}b, the leakage objective $U_{\alpha}$ that is maintained by the algorithm for this test.  Also shown are the locations of the $5\%$ of input pixels that were thrown out, as small black points.  Features in $\kappa_{\alpha}$ can be clearly seen where the \textsc{Imcom} algorithm has attempted to satisfy requirements on $U_{\alpha}$ where input image information is scarce. 

The increased noise variance in $\Sigma_{\alpha \alpha}$ for these bad pixel locations can be seen in Figure \ref{fig:rt5xrt5badpix}c.  Indeed, the noise can be large in regions of missing input information, particularly where pixel removals are clustered.  This information is represented in the output covariance $\Sigma_{\alpha \beta}$ given by equation \eqref{eq:Cr}, and is available to the user for weighting or fitting purposes if necessary.  If missing pixels are expected to be clustered (such as around a single cosmic ray hit) the rapid increase in noise for such events may represent a cause for concern that should be investigated using simulations of greater realism.   However, the level to which $U_{\alpha}$ (and indeed $F^2_{\alpha}$: Figure \ref{fig:rt5xrt5badpix}d) may be controlled is perhaps more important for many applications, and here results are encouraging.

\begin{figure*}
\begin{center}
\plottwo{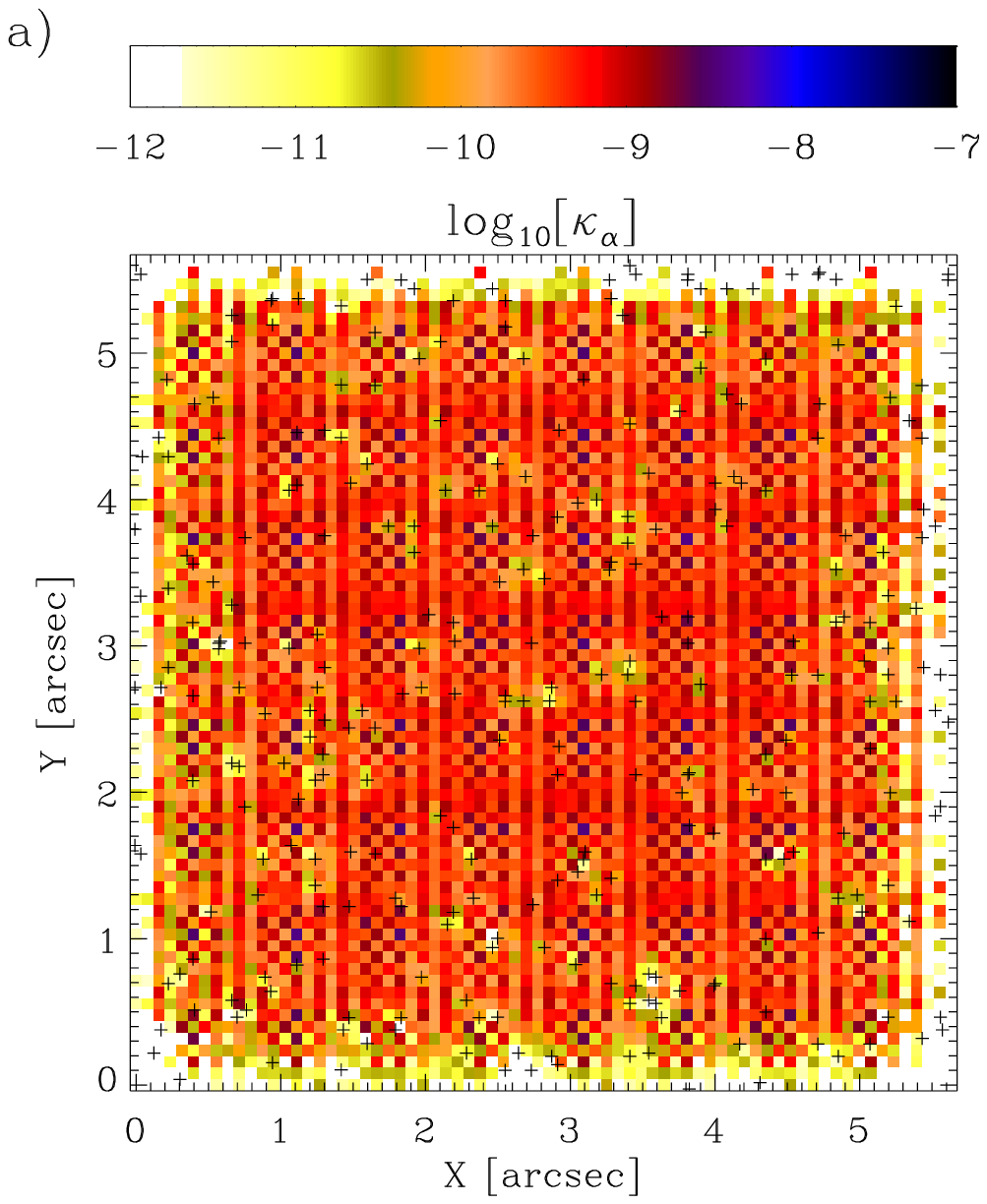}{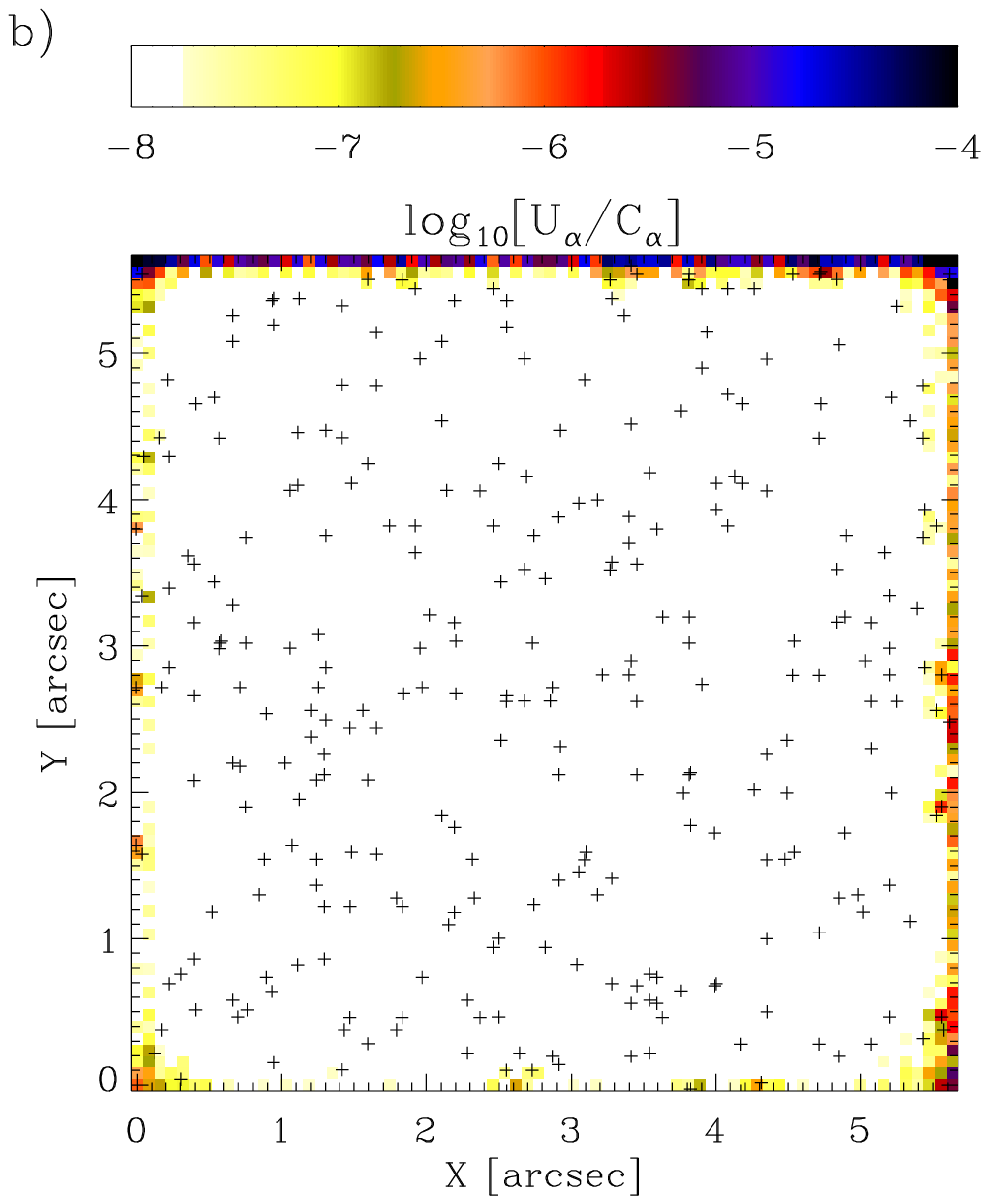}
\plottwo{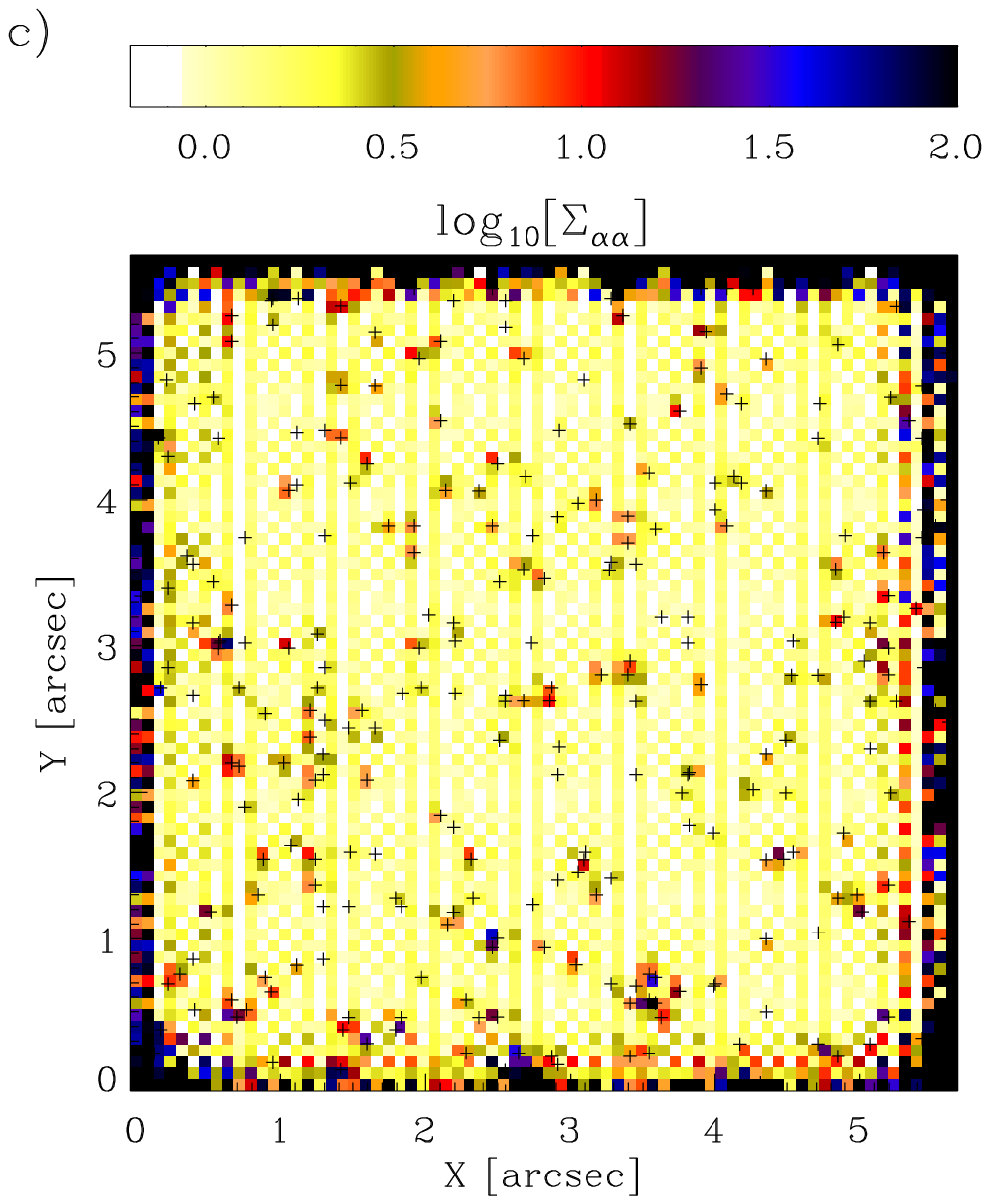}{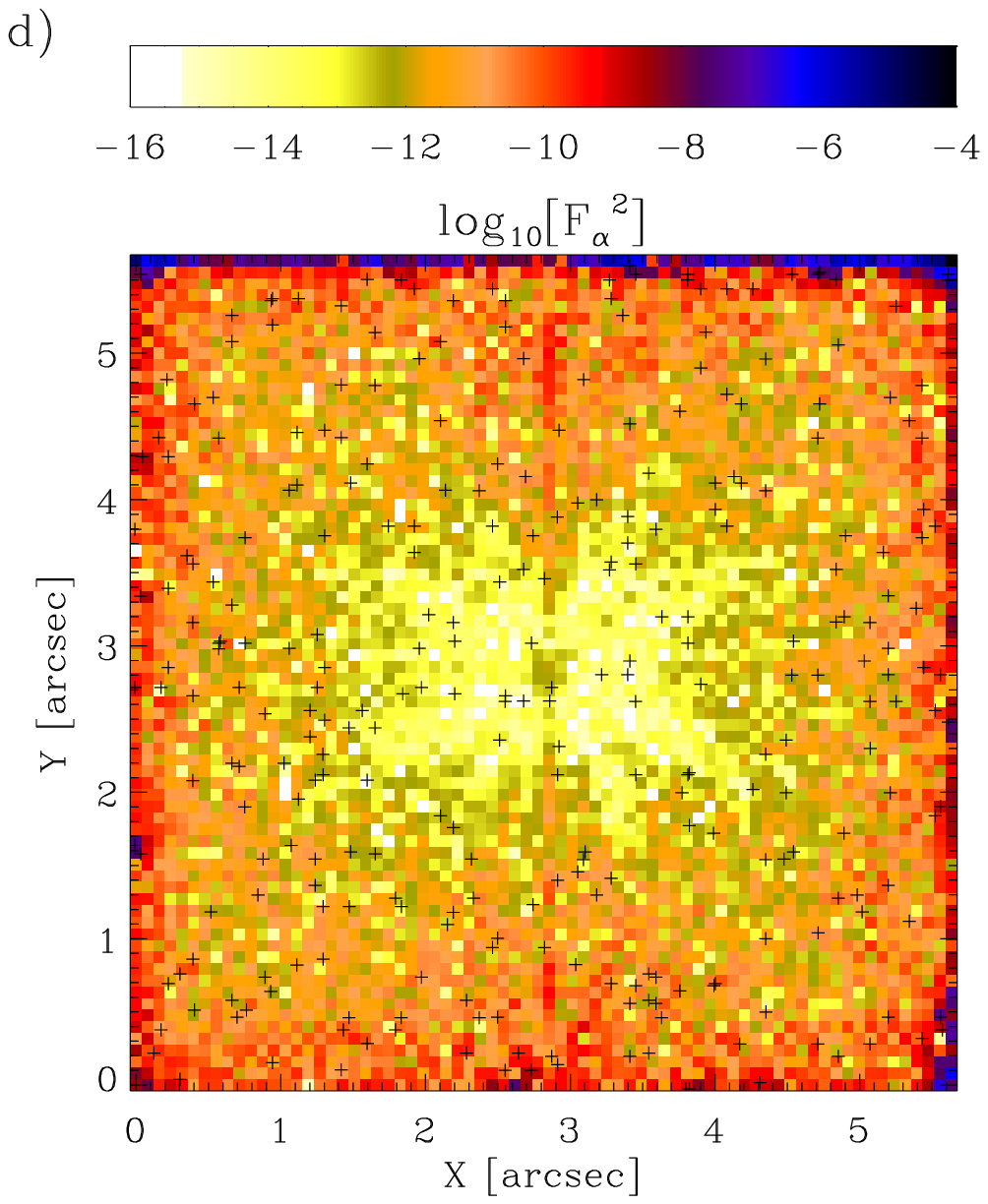}
\end{center}
\caption{Map of $\kappa_{\alpha}$ solution (a), leakage objective $U_{\alpha}$ (b), noise variance $\Sigma_{\alpha \alpha}$ (c), and squared fractional residual $F^2_{\alpha}$ (d) for the 6 exposure random dither after losing a randomly-selected 5\% fraction of input pixels (See Section \ref{sect:rt5xrt5badpix}).  Input pixels that were not used are indicated using small black crosses. The color range chosen for $\Sigma_{\alpha \alpha}$ does not reflect the full range of values in the output, but highlights variations due to bad pixels in the image center.}\label{fig:ran6badpix}
\end{figure*}
In Figure \ref{fig:ran6badpix} we show equivalent results after randomly removing $5\%$ of input pixels for the 6 exposure random dither discussed in Section \ref{sect:ran6}.  Once again, results for $U_{\alpha}$ are robust despite bad pixel losses, with $\Sigma_{\alpha \alpha}$ able to absorb uncertainty due to these small, localized instances of missing information.  As for the $\sqrt{5} \times \sqrt{5}$ ideal dither, $\Sigma_{\alpha \alpha}$ may become large where missing pixels are clustered and this behavior should be investigated further.

It has been shown that the \emph{WFIRST} design concept is able to produce output that is relatively robust to $5\%$ pixel losses from random locations in the input images $I_i$, both for the $\sqrt{5} \times \sqrt{5}$ ideal dither and a 6 exposure random dither configuration.  Tolerances on $U_{\alpha}$ may be maintained at the cost of increased $\Sigma_{\alpha \alpha}$ in regions of missing information: qualitatively this is the desired behavior.  However, there are other practical issues that may be encountered for certain survey strategies, one of which is the possibility of significant plate scale variations between exposures when combining wide angle-dithered images.  We now investigate linear image combination in such circumstances.

\subsection{Focal Plane Plate Scale Variations}\label{sect:scalev}
Unavoidable geometric field distortions, due to optical aberration and the alignment of each detector array in the focal plane assembly, are manifested as variations in the plate scale (angular pixel scale) across a telescope field of view.  These geometric distortions may be determined using precision astrometry, but it is not clear what impact they may have on a linear image reconstruction from multiple dithers.  Typically the distortion is smoothly varying, so that close pairs of pixels on the detector plane show smaller relative scale variations on average than widely-separated pairs.  With wide-angle slew dither strategies being considered for the \emph{WFIRST} imaging survey, the behavior of linear image combination when facing plate scale variations in the input images is an important design consideration.

We investigate the effect of plate scale variations using the 6 exposure random dither pattern discussed in Section \ref{sect:ran6}.  For each input exposure we vary the relative fractional plate scale by a random amount drawn from a Gaussian distribution of width $1\%$.  The variation is applied to all pixels in each given exposure, and is the same in both $x$ and $y$ directions: the effect is to uniformly expand or contract the pixel grid.  The charge diffusion and pixel response component of each $G_i({\bf r})$ is appropriately modified to reflect the new input pixel scale, and $\Gamma({\bf r})$ is chosen as the $G_i({\bf r})$ corresponding to exposure with the largest pixels.

In the upper panels of Figure \ref{fig:ran6scalev} we plot the $U_{\alpha}$ and $\Sigma_{\alpha \alpha}$ that can be achieved for this dither configuration, having stipulated $U_{\alpha}^{\textrm{max}} = 10^{-8} C_{\alpha}$ and $\Delta U_{\alpha}^{\textrm{max}} = 10^{-10} C_{\alpha}$ as usual.   It is clear that the impact of $\sim 1\%$ plate scale variations is severe: $\Sigma_{\alpha \alpha}$ shows significant increases as compared to the results of Section \ref{sect:ran6}, whereas degradation $U_{\alpha}$ is more modest but perceptible.  An increase in noise variance by a factor $\gtrsim 10$ compared to the input pixels, as seen here over much of the output, is not a tolerable cost for fully-sampling a dark energy imaging survey.  The success of linear image combination is clearly sensitive to small variations in the input plate scale, and a mitigating strategy must be found.

\begin{figure*}
\begin{center}
\plottwo{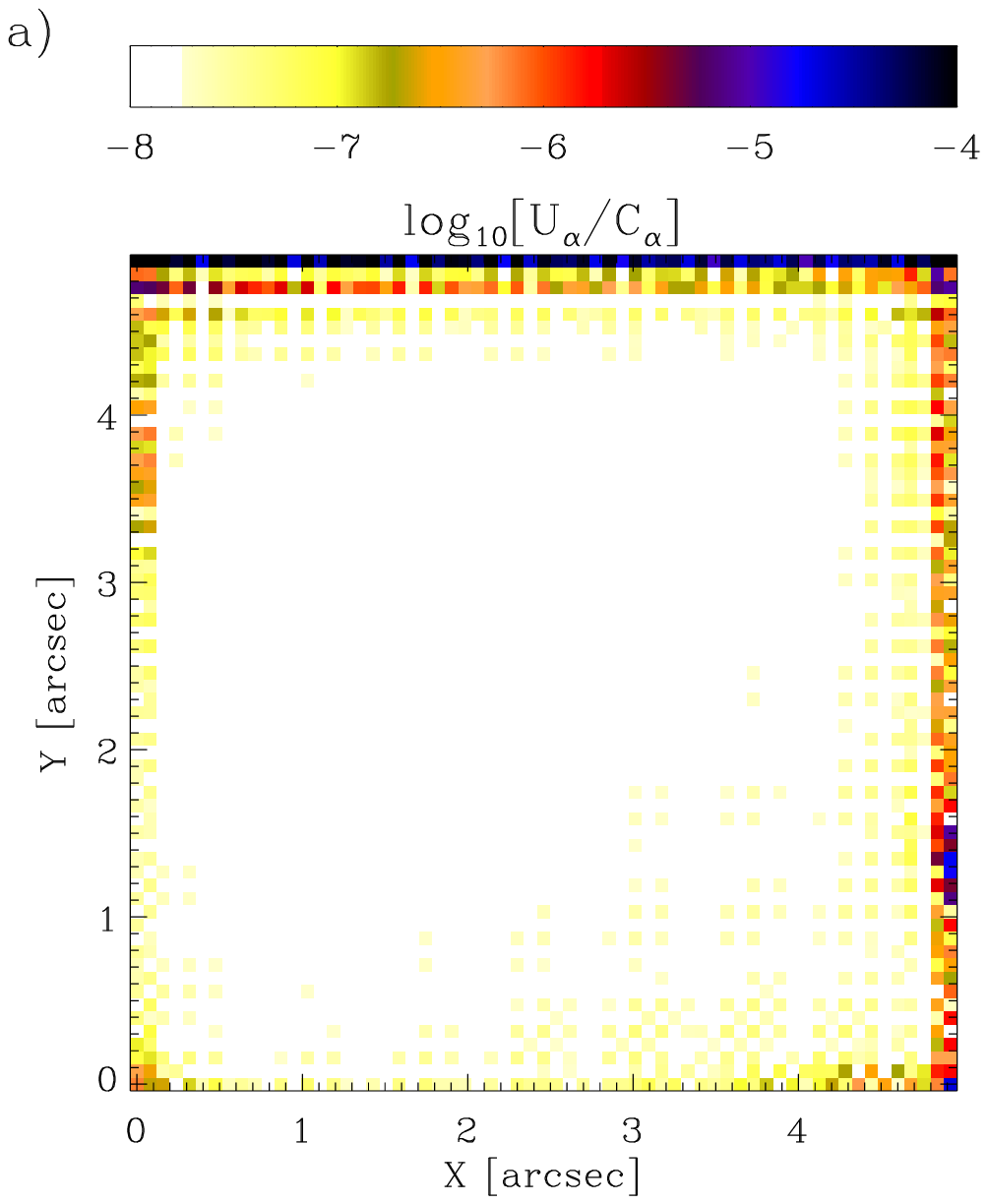}{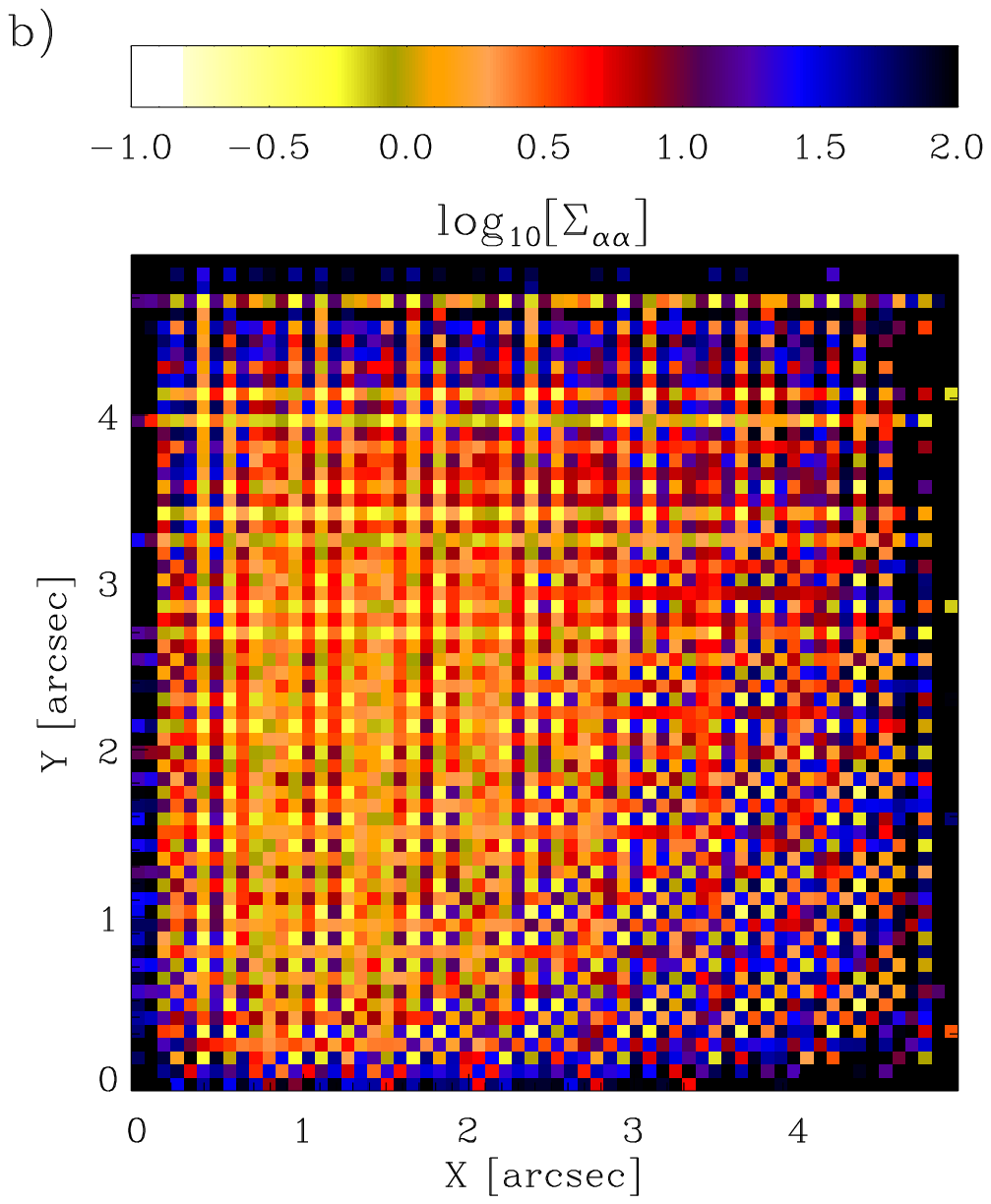}
\plottwo{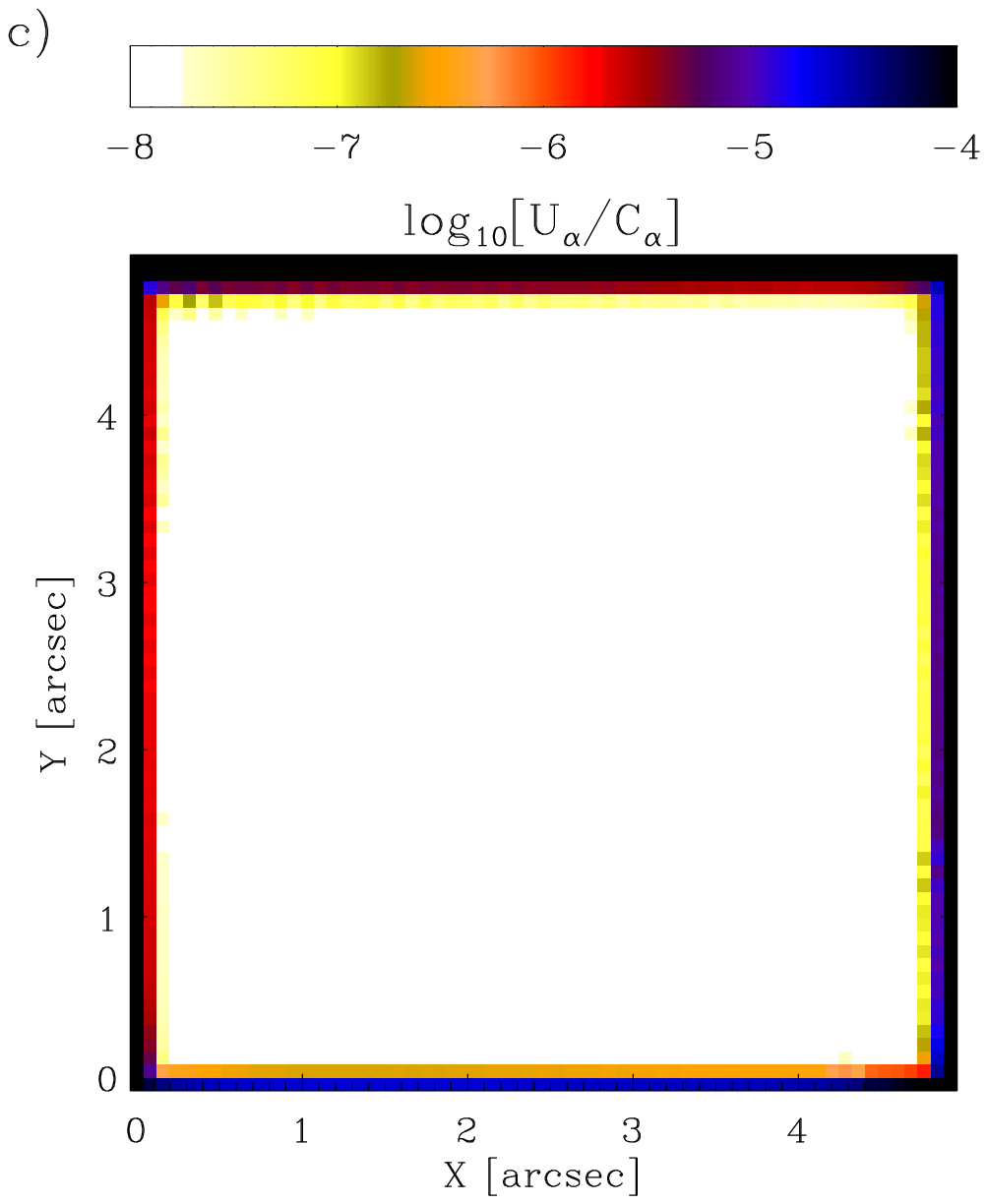}{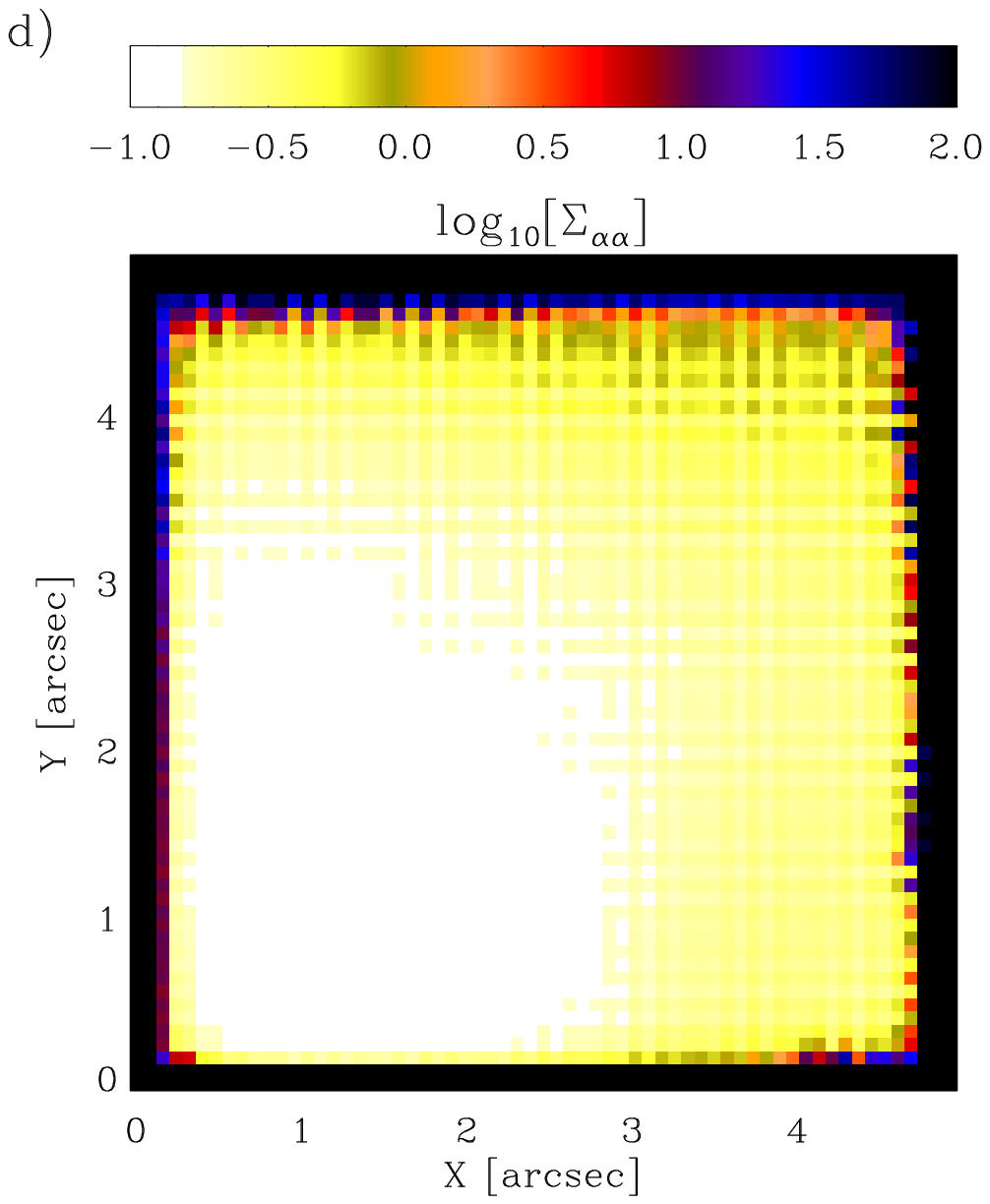}
\end{center}
\caption{Upper panels: Leakage objective $U_{\alpha}$ (a), and noise variance $\Sigma_{\alpha \alpha}$ (b), for a 6 exposure random dither after adding $\sim$1\% variations in the pixel scale for each input exposure.Lower panels: $U_{\alpha}$ (c), and noise variance $\Sigma_{\alpha \alpha}$ (d), for the same input pixel configuration, but adding a Gaussian smoothing of $\sigma = 0.09$ arcsec (0.5 pixel) to the desired output PSF $\Gamma({\bf r})$ (See Section \ref{sect:scalev}). Plotted color scales do not necessarily reflect the full range of $U_{\alpha}$ or $\Sigma_{\alpha \alpha}$, but are chosen to emphasize the central regions of greatest interest. }\label{fig:ran6scalev}
\end{figure*}
One such mitigating strategy is to increase the number of dithered exposures used in the reconstruction.  This was attempted, adding additional exposures to the 6 exposure random dither, up to a total of nine.  Results were modest, showing only marginal improvements in $\Sigma_{\alpha \alpha}$ as extra dithers were added.

Another option to mitigate the effects of scale variation is to exploit the user's freedom in setting the desired output PSF $\Gamma({\bf r})$.
In the lower panels of Figure \ref{fig:ran6scalev} we plot results using the same input pixel configuration as in the upper panels, but having added an additional Gaussian smoothing with $\sigma = 0.09$ arcsec (0.5 input pixels) to $\Gamma({\bf r})$.  Dramatic improvements to the noise properties $\Sigma_{\alpha \alpha}$ can be seen. $U_{\alpha}$ is also improved enough to bring it below $10^{-8} C_{\alpha}$ for a far greater proportion of the output area (see Figure \ref{fig:ran6scalev}c).

We see, therefore, that the damaging impact of plate scale variations can be suppressed to some extent by the addition of synthetic smoothing to the desired PSF $\Gamma({\bf r})$.  Considering the effect upon the MTF $\tilde{\Gamma}({\bf u})$, this can be seen simply as a suppression of the high-frequency modes that are most challenging to recover in the presence of plate scale variations.  The choice of optimal smoothing filter will, in general, depend closely upon the nature of the plate scale variations encountered between input exposures, and on the dither strategy employed.  The successful linear reconstruction of input images has been shown to be sensitive to this effect: further study aimed at finding optimal mitigating strategies, using more realistic simulations, will be needed.

This concludes our first set of trials for the optimal linear image combination formalism.  These early results suggest that the technique has merit as a design tool and can be used to explore competing dither strategies for generating oversampled output.  It has been used to identify encouraging robustness of linear reconstruction to the presence of randomly-located, missing input pixels, but has also highlighted a sensitivity to plate scale variation that should be explored further.

Clearly, however, an ultimate goal for the method is also  its use in constructing the oversampled images from actual all-sky imaging data supplied by a mission such as \emph{WFIRST}.  Therefore, we now turn to a discussion of whether such an approach represents a computationally feasible (i.e.\ affordable) option for analyzing data from such a mission.

\section{Practical Considerations \& Computational Costs}\label{sect:resources}
The linear image combination formalism provides a natural framework for constructing oversampled images, but an equally natural question begs itself: can the method be applied to a $10\,000$ deg$^2$ imaging survey (such as proposed for \emph{WFIRST}) within an acceptable timescale, and  using reasonable computing resources?  To answer this question, we must begin by estimating the total number of floating point operations required to apply the algorithm to a survey of this area.

As discussed in Section \ref{sect:imp}, the solution for any given patch of sky requires $O(n^3 + 2n^2 m)$ floating point operations, neglecting terms that are lower order in $n$ or $m$.  Therefore, the optimum data processing strategy is simply to split any imaging survey into as many small patches as possible: this is clear, since $n$ for each patch is proportional to its area.  Reduced memory costs and the ease of coarse-grained parallelization are other advantages of processing many small patches of sky. In Figure \ref{fig:patches} we illustrate one such patch schematically, noting the choice of different size input and output regions to avoid edge effects.  The output regions (solid line) for such patches can in principle be laid down in a tessellating pattern} to cover an entire survey area, with the larger input regions (dashed line) therefore overlapping to provide approximately uniform coverage free from the edge effects discussed in Section \ref{sect:2x2}.
\begin{figure}
\epsscale{0.75}
\plotone{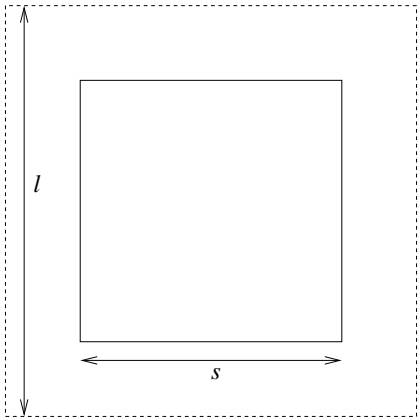}
\caption{One approach for efficiently processing an entire dark energy survey is to split it into small patches as shown here. In this approach, input pixels $I_i$ across multiple exposures are drawn from within a square of dimensions $l \times l$ (dashed line). From these, an output image $H_{\alpha}$ is reconstructed within a smaller, contained square of dimensions $s \times s$ (solid line).  Successive patches of sky may be aligned so that the smaller output regions tessellate to cover the entire survey. \label{fig:patches}}
\epsscale{1.0}
\end{figure}

We will now consider the computational cost of such a procedure.  Using the labels of Figure \ref{fig:patches}, and taking the same \emph{WFIRST} input pixel size and output sampling rate (0.18 and 0.079333 arcsec, respectively) as Sections \ref{sect:example} \& \ref{sect:wfirst}, let us consider patches of outer side $l = 25 \times 0.18 ~ \textrm{arcsec} = 4.5$ arcsec and inner side $s = 38 \times 0.079333 ~ \textrm{arcsec} = 3.014$ arcsec.  This is a small patch, and thus an efficient means to cover a large survey, but is of sufficient size relative to the PSF to avoid significant degradation in $H_{\alpha}$ due to edge effects (see, e.g., Section \ref{sect:2x2}).  A $10\,000$ deg$^2$ survey requires $1.4 \times 10^{10}$ such patches for full coverage.  Assuming five dithered input exposures gives $n = 3.125 \times 10^3$ and $m = 1.444 \times 10^3$ for each patch, resulting in order $n^3 + 2n^2 m \simeq 5.8 \times 10^{10}$ floating point operations per patch.  To process all patches independently to cover the total survey area thus requires $\sim 10^{21}$ floating point operations, a considerable computational cost.

However, because of the extreme parallelizability of the problem the use of supercomputing resources is a realistic option.  The RoadRunner\footnote{http://www.lanl.gov/roadrunner/} supercomputer at Los Alamos National Laboratory is capable of a sustained rate of 1 PFLOPS ($10^{15}$ Floating Point Operations Per Second) in double precision arithmetic, and would therefore require approximately 12 days to complete the image combination necessary for a full dark energy survey.   While it must be stressed that this is very much a ``back of the envelope'' estimate, and subject to usual additional factors due to practical realities and data management, the resources required are conceivably within reach even today (if costly).  As of November 2010 this machine was rated as the seventh fastest on earth but, thanks in part to the advent of inexpensive Graphical Processing Units, the cost and availability of such computing power must be expected to change favorably over the next decade.  Moreover, the survey imaging data also require considerable time simply to be collected by the telescope. This fact makes it unlikely that optimal linear image combination would represent a rate-determining step for a dark energy mission, even given more modest access to supercomputing resources.

Finally, this discussion has assumed that the linear combination method described in this Paper would be performed repeatedly and independently for each and every patch of sky.  However, departing from this general but `brute force' approach there are doubtless many gains in efficiency to be made when processing imaging data for a dark energy survey.  Not least, image combination efforts might be focused in regions where significant concentrations of light have been detected on a coarser grid, avoiding empty regions of the sky.
Also, there are likely to be close similarities between the PSF and input pixel configurations for nearby patches: this might allow results from costly stages in the calculation (e.g.\ the eigendecomposition of $A_{\alpha i j}$) to be reused multiple times.  
Further optimizing the implementation of the linear image combination method would make an interesting topic for future work, particularly as plans for a space-based imaging survey move closer towards reality.

\section{Conclusions}\label{sect:conc}
It has been the intention of this Paper to demonstrate a simple, yet general, formalism for the linear combination of undersampled astronomical images to produce oversampled output.  Image reconstruction based on this formalism allows the explicit control of any changes to the effective image PSF $G_i({\bf r})$, and noise variance $\Sigma_{\alpha \alpha}$, due to the combination process.
Such an approach is likely to be a useful addition to a growing set of image analysis tools being built up to face unprecedented challenges in astronomical inference: challenges arising primarily in the bid to understand dark energy.

Despite the numerical emphasis of the approach, we have shown how significant computational savings may be made via efficient implementation, with further savings being possible.  It is planned to make the prototype Fortran 95 code \textsc{Imcom}, used to generate the results of this Paper, available to the public; this will be either in its current form or as part of a more developed, scaled-up software package for use with greater volumes of data.  Even using the current prototype implementation, it has been shown that the method is a feasible means of processing all-sky data from a dark energy mission.

Within the linear framework presented, we have demonstrated some interesting results for the \emph{WFIRST} design concept adopted in Section \ref{sect:example} and explored in more detail in Section \ref{sect:wfirst}.  First, we have found that this design allows the controlled reconstruction of fully-sampled images if five input dithers (in a $\sqrt{5} \times \sqrt{5}$ ideal pattern) or six input dithers (randomly offset) are available.  We also found that the $\sqrt{5} \times \sqrt{5}$ ideal dither and 6 exposure random dither configurations proved reasonably robust in $U_{\alpha}$ despite suffering $5\%$ bad input pixel losses in random locations. Largely, these defects were able to be absorbed into an increase in the noise variance $\Sigma_{\alpha \alpha}$.   Whether this conclusion is robust when the bad pixel patterns are less random remains to be seen: the contiguous spatial extent of defects such as cosmic rays is likely to be a complicating factor for the control of both $U_{\alpha}$ and $\Sigma_{\alpha \alpha}$ in the output, and merits further investigation. 

Another result, discussed in Section \ref{sect:scalev}, was the sensitivity of the linear image combination method to variations in the pixel plate scale between exposures.  Plate scale variations of the order $\sim 1\%$ were seen to significantly degrade $U_{\alpha}$ and $\Sigma_{\alpha \alpha}$ for 6 exposure random dither.  It was found that adding an additional Gaussian smoothing kernel of standard deviation $\sigma = 0.09$ arcsec (0.5 input pixels) to the desired output PSF $\Gamma({\bf r})$ mitigated this degradation significantly.  The dependence of the strength of this effect upon input image parameters such as the plate scale variation, input PSF, and dither strategy will need to be investigated more thoroughly than was possible in this Paper.  Coping efficiently with plate scale variation clearly represents a non-trivial challenge when using optimal linear techniques to combine multiple input images.

In the examples tested in Section \ref{sect:wfirst} it was also found that the stationarity of the noise characterized by $\Sigma_{\alpha \alpha}$ (also a desirable property if $U_{\alpha}$ may be kept low) seems to be best preserved for the ideal dither patterns, even while $U_{\alpha}$ may show strong spatial variation (e.g.\ Section \ref{sect:2x2}).  Conversely, setting limits on $U_{\alpha}$ for the random dither patterns causes significant spatial variation in $\Sigma_{\alpha \alpha}$.  This effect is also something to be investigated further when designing dither strategies for a dark energy survey mission.  Correlated, non-stationary noise introduces noise rectification biases that must be calculated and removed when performing accurate photometry and shape measurement.  Precise calibration of such biases most likely proceeds best via simulations or deep (high signal-to-noise) training data, although analytic methods exist to determine the leading order terms \citep{kaiser00,bernsteinjarvis02}.

For precisely such reasons, it is suggested that inference regarding galaxy shapes might best occur at the level of raw pixel images (e.g.\ \citealp{milleretal07,kitchingetal08}), while image stacks or other post-processed science tools would be used to provide crucial ancillary information such as fully-sampled PSF images and precise object centroid estimates.  The discussion of this question is appropriate and timely.  Whatever the eventual conclusions of such a debate, there is undoubted value in a method for combining multiple undersampled images to generate oversampled output that offers control over any changes to the PSF, while simultaneously suppressing noise where possible.  The formalism presented in this Paper, being linear, is the simplest approach that meets these objectives.  

There remain some important unanswered questions regarding the use of the technique for real data.  As mentioned early in Section \ref{sect:linform}, the input PSF $G_i({\bf r})$ must be known, modeled or approximated before the use of the linear image combination formalism.  Any model of $G_i({\bf r})$ which is sensitive to aliasing in images of stars from the input $I_i$ risks imparting similar defects to the output image $H_{\alpha}$.  This would be a problem if attempting non-parametric reconstruction of the PSF directly from stellar images, for example.  However, even in this case, a mitigating strategy may be found.  If $G_i({\bf r})$ does not vary rapidly across the instrumental field of view then images of different stars within a small region of a given exposure will approximate multiple images of the same PSF, but with a variety of centroid offsets (assuming variations due to star color are small, and ignoring varying flux which can be normalized).  Fully-sampled images of an ``average'' star for each small region might thus be reconstructed by applying the linear image combination formalism in combination with an accurate estimation of the individual stellar centroids. 

In practice, any model for $G_i({\bf r})$ can only be an estimate of the true convolving kernel, and the impact of this uncertainty on estimates of key observables (e.g.\ \citealp{paulinetal08,rowe10}) must be quantified in context of optimized linear image combination.  A related question that must be addressed is the effect of color variation in $G_i({\bf r})$ for images observed using broad-band spectral filters.  Furthermore, and beyond its importance when modelling $G_i({\bf r})$, the need for accurate stellar centroid estimation also arises when considering astrometric registration for each of the individual exposures that make up $I_i$.  This registration is needed before the images may be combined, and any model of ${\bf r}_i$ can only approximate the true astrometric solution.  The impact of realistic levels of uncertainty in ${\bf r}_i$ upon the quality of the output image $H_{\alpha}$ must be investigated.  Work to explore some of these important questions is already underway in the laboratory, where they are equally relevant in the precise characterization of detector technology for a dark energy mission.  Happily, the formalism presented here provides a useful framework for exploring these issues.

\acknowledgments

The authors would like to thank Tod Lauer and Gary Bernstein (particularly regarding the use of lookup tables for calculating the system matrices $A_{\alpha ij}$ and $B_{\alpha i}$) for useful discussion and suggestions, and the anonymous referee for useful comments that improved the manuscript.  This work was performed in part at the Jet Propulsion Laboratory, operated by the California Institute of Technology under a contract for the National Aeronautics and Space Administration (NASA).  BR has been supported by the NASA \emph{WFIRST} Project Office.
CH is supported by the Department of Energy (DE- FG03-02-ER40701), the National Science Foundation (AST-0807337), and the David \& Lucile Packard Foundation.





\appendix

\section{Asymptotic Behavior of Output Image Properties}\label{app:asym}
The solution to the system of equations derived by minimizing $W_\alpha$ in equation \eqref{eq:obj2}, which is given by equation \eqref{eq:tai}, may be written succinctly as
\begin{equation}\label{eq:Tsuc}
{\bf T} = -\frac{1}{2} \left( {\bf A}_{\alpha} + \kappa {\bf N} \right)^{-1} {\bf B}_\alpha.
\end{equation}
Using this result with equation \eqref{eq:Cr} we may write the noise variance in each output pixel as
\begin{eqnarray}
\Sigma_{\alpha\alpha} & = & \sum_{ij} T_{\alpha i} T_{\alpha j} N_{ij} \\
& = & \frac14 {\bf B}_\alpha^T \left({\bf A}_\alpha + \kappa {\bf N} \right)^{-1} {\bf N} \left({\bf A}_\alpha + \kappa {\bf N}\right)^{-1} {\bf B}_\alpha  \label{eq:SAA}.
\end{eqnarray}
The leakage objective $U_{\alpha}$ at each pixel may similarly be written
\begin{equation}
U_{\alpha} = \frac14 {\bf B}_\alpha^T \left({\bf A}_\alpha + \kappa {\bf N} \right)^{-1} {\bf A_\alpha} \left({\bf A}_\alpha + \kappa {\bf N}\right)^{-1} {\bf B}_\alpha -\frac{1}{2} 
{\bf B}_\alpha^T \left({\bf A}_\alpha + \kappa {\bf N} \right)^{-1} {\bf B_\alpha} + C_\alpha. \label{eq:UAA}
\end{equation}
Noting the simple $\kappa$ dependence in the expressions above, we see immediately that
\begin{equation}
\frac{\partial\Sigma_{\alpha\alpha}}{\partial\kappa}  =  -
\frac12 {\bf B}_\alpha^T \left({\bf A}_\alpha + \kappa {\bf N} \right)^{-1} {\bf N} \left({\bf A}_\alpha + \kappa {\bf N}\right)^{-1} {\bf N} \left({\bf A}_\alpha + \kappa {\bf N} \right)^{-1}  {\bf B}_\alpha < 0 ,
\end{equation}
and
\begin{equation}\label{eq:dUdk}
\frac{\partial U_{\alpha}}{\partial\kappa}  = 
\frac12 {\bf B}_\alpha^T \left({\bf A}_\alpha + \kappa {\bf N} \right)^{-1} {\bf N} 
\left[ \mathbb{I}_n - \left({\bf A}_\alpha + \kappa {\bf N}\right)^{-1}{\bf A_\alpha} \right] 
\left({\bf A}_\alpha + \kappa {\bf N} \right)^{-1}  {\bf B}_\alpha > 0 ,
\end{equation}
where $\mathbb{I}_n$ is the $n \times n$ identity matrix.


These results can be interpreted in terms of the asymptotic behavior of the output image properties across the full range of $\kappa$.  Taking first the limit $\kappa \rightarrow \infty$, we see that 
\begin{equation}
\lim \limits_{\kappa\rightarrow\infty} \left\{ \Sigma_{\alpha\alpha} \right\} \rightarrow 0^+ , \: \: \: \: \: \:  \: \: \:  \: \: \:  \: \: \:  \: \: \: \lim \limits_{\kappa\rightarrow\infty} \left\{ \frac{ \partial \Sigma_{\alpha\alpha}}{\partial \kappa} \right\} \rightarrow 0^- .
\end{equation}
As $\kappa$ becomes large the system tends convergently to an output image of zero  noise variance (or covariance).  This result can in fact be trivially seen from equation \eqref{eq:Tsuc}, as ${\bf T} \rightarrow 0$ in the limit $\kappa \rightarrow \infty$.  The output image $H_\alpha$ likewise tends to zero everywhere, and the leakage and leakage objective tend as
\begin{equation}
\lim \limits_{\kappa\rightarrow\infty} \left\{ L_{\alpha}({\bf r}) \right\} \rightarrow  - \Gamma ({\bf r}),
\: \: \: \: \: \:  \: \: \:  \: \: \:  \: \: \:  \: \: \:  \lim \limits_{\kappa\rightarrow\infty} \left\{ U_\alpha \right\} \rightarrow C_{\alpha}, \: \: \: \: \: \:  \: \: \:  \: \: \:  \: \: \:  \: \: \: \lim \limits_{\kappa\rightarrow \infty} \left\{ \frac{ \partial U_{\alpha}}{\partial \kappa} \right\} \rightarrow 0^+. \label{eq:Utendinf}
\end{equation}
It follows hence that values of the leakage objective $U_\alpha$ are most naturally quoted in units of $C_\alpha$, converging upon unity in the limit of least image fidelity.

Turning to opposing limit $\kappa \rightarrow 0$, we find that
\begin{equation}
\lim \limits_{\kappa\rightarrow 0} \left\{ U_{\alpha} \right\} \rightarrow C_\alpha - \frac{1}{4} {\bf B}^T_{\alpha} {\bf A}^{-1}_{\alpha} {\bf B}_{\alpha} , \: \: \: \: \: \:  \: \: \:  \: \: \:  \: \: \:  \: \: \: \lim \limits_{\kappa\rightarrow 0} \left\{ \frac{ \partial U_{\alpha}}{\partial \kappa} \right\} \rightarrow 0^+ . \label{eq:Utendzero}
\end{equation}
While we can see, therefore, that the system converges stably to this minimum value there is no guarantee that $U_\alpha \rightarrow 0$ as $\kappa \rightarrow 0$, and \emph{therefore no guarantee of an unbiased linear solution for} $T_{\alpha i }$.  The noise variance exhibits the following limiting behavior
\begin{equation}
\lim \limits_{\kappa\rightarrow 0} \left\{ \Sigma_{\alpha\alpha} \right\} \rightarrow \frac14 {\bf B}_\alpha^T {\bf A}_\alpha^{-1} {\bf N} {\bf A}_\alpha^{-1} {\bf B}_\alpha , \: \: \: \: \: \:  \: \: \:  \: \: \:  \: \: \:  \: \: \: \lim \limits_{\kappa\rightarrow 0} \left\{ \frac{ \partial \Sigma_{\alpha\alpha}}{\partial \kappa} \right\} \rightarrow - \frac12 {\bf B}_\alpha^T {\bf A}_\alpha^{-1} {\bf N} {\bf A}_\alpha^{-1} {\bf N} {\bf A}_\alpha^{-1} {\bf B}_\alpha.
\end{equation}
Depending on $ A_{\alpha i j}$ and $B_{\alpha i}$, these variances may be extremely large resulting in noisy output images.  In practice a compromise between noise and fidelity will be desired.




\bibliographystyle{apj}
\bibliography{btprmnras_letter}

\end{document}